\documentclass{aa}   %

\usepackage{color}

\usepackage{orcidlink}
\usepackage{upgreek}    

\begin{document}

\title{Central flashes during stellar occultations}
      \subtitle{Effects of diffraction, interferences, and stellar diameter}

\author{
B. Sicardy$^{\orcidlink{0000-0003-1995-0842}}$\inst{1}
\and
L. Dettwiller$^{\orcidlink{0000-0003-4350-755X}}$\inst{2}
}

\institute{
Laboratoire Temps Espace (LTE), Observatoire de Paris, Universit\'e PSL, CNRS UMR 8255, Sorbonne Universit\'e, LNE, 61 Av. de l'Observatoire, F75014 Paris, France \\
\email{Bruno.Sicardy@obspm.fr}
\and
Universit\'e Jean Monnet Saint-Etienne, CNRS, Institut d Optique Graduate School,
Laboratoire Hubert Curien UMR 5516, F-42023 Saint-Etienne, France
}
 
\date{Received May 19, 2025; accepted January 07, 2026 in \it Astronomy \& Astrophysics}


  \abstract
   {Central flashes may occur during stellar occultations by objects in the Solar System.}
   {Catalog diffraction effects on the flash with point-like stars, monochromatic waves, and different cases of spherical transparent atmosphere; describe  the corrections due to  stellar diameters.}
   {To describe diffraction, we used the Huygens principle, the Sommerfeld lemma, and the stationary phase method, and we treated the effects of finite stellar diameter using Clausius' theorem.}
   {For point-like stars, the central flash shape is that of the classical Poisson spot, but with a greater height. For tenuous atmospheres that cannot focus the stellar rays at the shadow center, the flash is amplified by the factor $(R_0/r_0)^2$ compared to the Poisson spot, where $R_0$ and $r_0$ are the object and the shadow radii, respectively. For denser atmospheres that can focus the rays at the shadow center, the flash peaks at $2\pi^2(R_{\rm CF}/\lambda_{\rm F})^2 \phi_\perp(0)$, where $R_{\rm CF}$ is the central flash layer radius, $\lambda_{\rm F}$ is the Fresnel scale, and $\phi_\perp(0)$ is the flux that would be observed at the shadow center without focusing. For isothermal atmospheres with scale height $H$, the height is $2\pi^2 (R_{\rm CF} H)/\lambda_{\rm F}^2$. Fringes surrounding the central flash are  separated by $\lambda_{\rm P}= \lambda_{\rm F}^2/R_{\rm CF}$, which is related to the separation between the primary and secondary stellar images. For a projected stellar diameter $D_* \gg \lambda_{\rm P}$, the flash is described by complete elliptic integrals, and has a full width at half maximum of $1.14 D_*$ and a peak value of $8H/D_*$.}
   {For Earth-based occultations by Pluto and Triton observed in the visible with point-like stars, diffraction causes flashes with very large heights of $\sim$10$^4$-10$^5$, spread over a very small meter-sized region in the shadow plane. In practice, the flash is usually smoothed by the stellar diameter, but still reaches high values of $\sim$50 and $\sim$200 during Pluto and Triton occultations, respectively. Diffraction dominates when using millimeter wavelengths or longer. We discuss the effects of departure from sphericity, atmospheric waves, and stellar limb darkening.}
   
 
   \titlerunning{Central flashes: Effects of diffraction, interferences, and stellar diameter}       

   \maketitle

\nolinenumbers         
\section{Introduction}
\label{sec_intro}

During a stellar occultation by a body with an atmosphere, the latter acts like a lens that refracts the light rays coming from the star. For an observer at distance $\Delta$ from the occulter, and if the atmosphere is dense enough, there is a layer (called the central flash layer) with radius $R_{\rm CF}$ that focuses the rays toward the shadow center. 
The value of $R_{\rm CF}$ depends on the molecular refractivity and on the number density profile of the atmosphere. 
When approaching the shadow center, an observer records a surge of brightness known as the central flash.

One of the first observations of a central flash came from the occultation of $\epsilon$ Geminorum by the Martian upper atmosphere on 8 April 1976. A modeling of the flash was made in the framework of pure geometrical optics, i.e. without accounting for diffraction effects \citep{Elliot1977}. 
It included the calculation of the caustic created by an oblate atmosphere near the shadow center, and it also mentioned the effect of the finite apparent stellar diameter.

However, diffraction was considered more than a century ago as the technique of stellar occultation by bodies without atmospheres came into use. \cite{Eddington1909} was one of the first to describe the effect of Fresnel diffraction during occultations of stars by the lunar limb.
Diffraction fringes were observed three decades later during a lunar occultation of Regulus, see \cite{Arnulf1936} who gave an upper limit of 2.0 milli-arcsec for the apparent angular diameter of this star. 
With the advent of fast photometers, and starting with \cite{Whitford1947}, many stellar diameters were measured with this technique (see \cite{Ridgway1977,Ridgway1979}, and the review by \cite{White1987}, who mentions 124 stellar diameter measurements). 
Besides stellar diameters, lunar occultation also provided another important parameter, the local limb slope \citep{Evans1970}.

After the discovery of rings around Uranus and Neptune, diffraction caused by narrow semitransparent rings (or small opaque objects) was formalized by \cite{Roques1987}.
Concerning occultations by bodies with an atmosphere, considerations of some effects due to both diffraction and stellar diameter dates back to at least \cite{Fabry1929}. \cite{French1976} studied the Fresnel diffraction at the edge of a shadow produced by bodies with tenuous isothermal atmospheres. The role of diffraction in stellar scintillation, due to local atmospheric inhomogeneities of the occulter, was analyzed by \cite{Cooray2003}, and applied to the sharp variations in signal (spikes) associated with ray-crossing. \cite{Young2012} proposed an algorithm to generate global occultation light curves by objects with arbitrary atmospheres that accounts for both diffraction and refraction.

In a brief note published in Nature, \cite{Hubbard1977} considered the effect of diffraction on the central flash, and assumed a point-like monochromatic stellar source and a spherical, 
transparent,\footnote{Hereinafter, the adjective ``transparent'' refers not only to an atmosphere without absorption, but also without scattering.}
and isothermal atmosphere with scale height $H$. 
Hubbard's expression of the stellar flux\footnote{Proportional to the irradiance for a given telescope.} normalized to its value outside the occultation is, for a dense enough atmosphere and in the presence of diffraction, 
\begin{equation}
\phi_{\rm Diff}(r)= (2\pi)^2 \left( \frac{R_{\rm CF}H}{\lambda \Delta} \right) J_0^2 \left( \frac{2\pi R_{\rm CF}r}{\lambda \Delta} \right).
\label{eq_CF_Hubbard}
\end{equation}
Here $\lambda$ is the wavelength in the vacuum of the incoming wave, $r$ is the distance of the observer to the shadow center, counted in the shadow plane, i.e. the plane perpendicular to the star-observer line, and $J_0$ is the Bessel function of the first kind and $n=0$ order, more generally defined by\footnote{In this paper, the symbol := indicates  a definition or a notation.}
\begin{equation}
J_n (u) := \frac{1}{\pi} \int_0^\pi \cos (nt - u \sin t) \, {\rm d}t.
\label{eq_Bessel}
\end{equation}

The first zero of $J_0$ is reached for $u \approx 2.4$, which implies from Eq.~\ref{eq_CF_Hubbard} that the typical width of the flash in the shadow plane is $\sim \lambda \Delta/(2R_{\rm CF})$. This is the expected image size of a point-like source produced by a lens with focal length $\Delta$ and diameter $2R_{\rm CF}$, for a wave of wavelength $\lambda$.

\cite{Hubbard1977}'s short note does not provide details on the steps that lead to Eq.~\ref{eq_CF_Hubbard}. Because $J_0(0)=1$, it predicts a surge in flux that peaks at a value of $\phi(0)= (2\pi)^2 R_{\rm CF}H/(\lambda \Delta),$ which can be very large, as discussed later. We show later that this peak is surrounded by successive bright and dark circular fringes around the shadow center, with a fringe spacing of $\approx \lambda \Delta/(2R_{\rm CF})$.

One goal of this paper is to discuss the calculations that lead to Eq.~\ref{eq_CF_Hubbard} and to study the structure of the flash
in the idealized case of a perfectly spherical and transparent atmosphere. This is a necessary step before tackling more complicated problems, in particular the cases of distorted atmospheres. In the idealized case, 
we show that there is a continuous gradation between the classical Poisson spot produced by an airless body and the central flash produced by a spherical atmosphere.
We then examine the smoothing effect of the finite stellar diameter on the flash structure.
We also provide applications to Pluto and Triton occultations.
The main quantities used in this paper are listed in Table~\ref{tab_definitions}, while more specific definitions are given on a case-by-case basis.

\begin{table}[!t]
\caption{Definitions of main quantities.}
\label{tab_definitions}
\renewcommand{\arraystretch}{1.1}
\begin{tabular}{ll}
\hline
\hline
 Symbol & Quantity \\
\hline
$a$                                        & Amplitude of the received wave \\
$E$, $F$                                & Complete elliptic integrals of the second \\
                                               & and first kind, respectively (Eqs.~\ref{eq_G}) \\
$F_{\rm r}$                             & Fresnel function (e.g. Eqs.~\ref{eq_Fresnel_fringes},\ref{eq_Fresnel_func}) \\
$G(w)$                                    & $\left[E(w) - (1-w^2) F(w)\right]/w$ \\
$H$                                         & Scale height of isothermal atmosphere \\
$I$                                          & Radiance of the stellar disk (e.g. Eqs.~\ref{eq_phi_LD_gen},\ref{eq_limb_darkening}) \\
$J_n$                                      & Bessel function of first kind and \\
                                                & n$^{\rm th}$ order (Eq.~\ref{eq_Bessel}) \\
$K$                                         & Molecular refractivity \\
$n_{\rm g}$                              & Molecular number density \\
$r$                                           & Distance of observer to  shadow \\ 
                                                & center (Fig.~\ref{fig_tenuous_vs_dense}) \\
$R$                                         & Closest approach of stellar ray to body \\
                                                & center (Fig.~\ref{fig_tenuous_vs_dense}) \\
$R_{\rm CF}$                          & Radius of central flash layer \\
$r_*$, $D_*$                            & Apparent stellar radius and stellar \\ 
                                                & diameter projected at occulter distance \\
                                                                                              
$\Delta$                                 & Distance from observer to occulter \\
$\theta$                                  & Polar angle of ray at  closest approach \\
                                               & to body center (Fig.~\ref{fig_geo_occ}) \\
$\lambda$                              & Wavelength in vacuum \\
$\lambda_{\rm F}$                  & Fresnel scale (Eq.~\ref{eq_Fresnel_scale}), typical separations \\
                                                & of Fresnel fringes \\
$\lambda_{\rm P}$                  & Separation of Poisson fringes (Eq.~\ref{eq_inter_fringe})  \\
$\nu$                                       & Gas refractivity \\
$\varphi$                                & Phase of the received wave \\
$\phi(r)$                                  & Normalized stellar flux at $r$ in the limits of \\
                                                & geometrical optics and point-like star \\
$\phi_\perp(r)$                        & Same as $\phi(r)$, but ignoring the focusing by \\
                                               &limb curvature \\
$\chi$                                      & Angle intervening in the Rayleigh-Sommerfeld \\ 
                                               & inclination factor (Fig.~\ref{fig_geo_occ}) \\
$\tau(R)$                                 & Integrated line of sight refractivity for \\ 
                                                & a ray passing at $R$ \\
$\omega$                               & Bending of the stellar ray (Fig.~\ref{fig_tenuous_vs_dense})  \\
\hline
\end{tabular}
\tablefoot{Certain quantities listed above have indices 1 or 2 when they refer to the primary and secondary images, respectively.
Besides $\perp$, various indices are used for $\phi(r)$, depending on the cases under study.}
\end{table}

\section{Geometrical optics}
\label{sec_geom_optics}

Before tackling the problem with diffraction, we first summarize the basic results obtained in the framework of pure geometrical optics. Numerous papers have been devoted to this topic; see the review in \cite{Sicardy2022}, which provides the equations given in this section.

\begin{figure}[!h]
\centerline{\includegraphics[totalheight=90mm,trim=0 0 0 0]{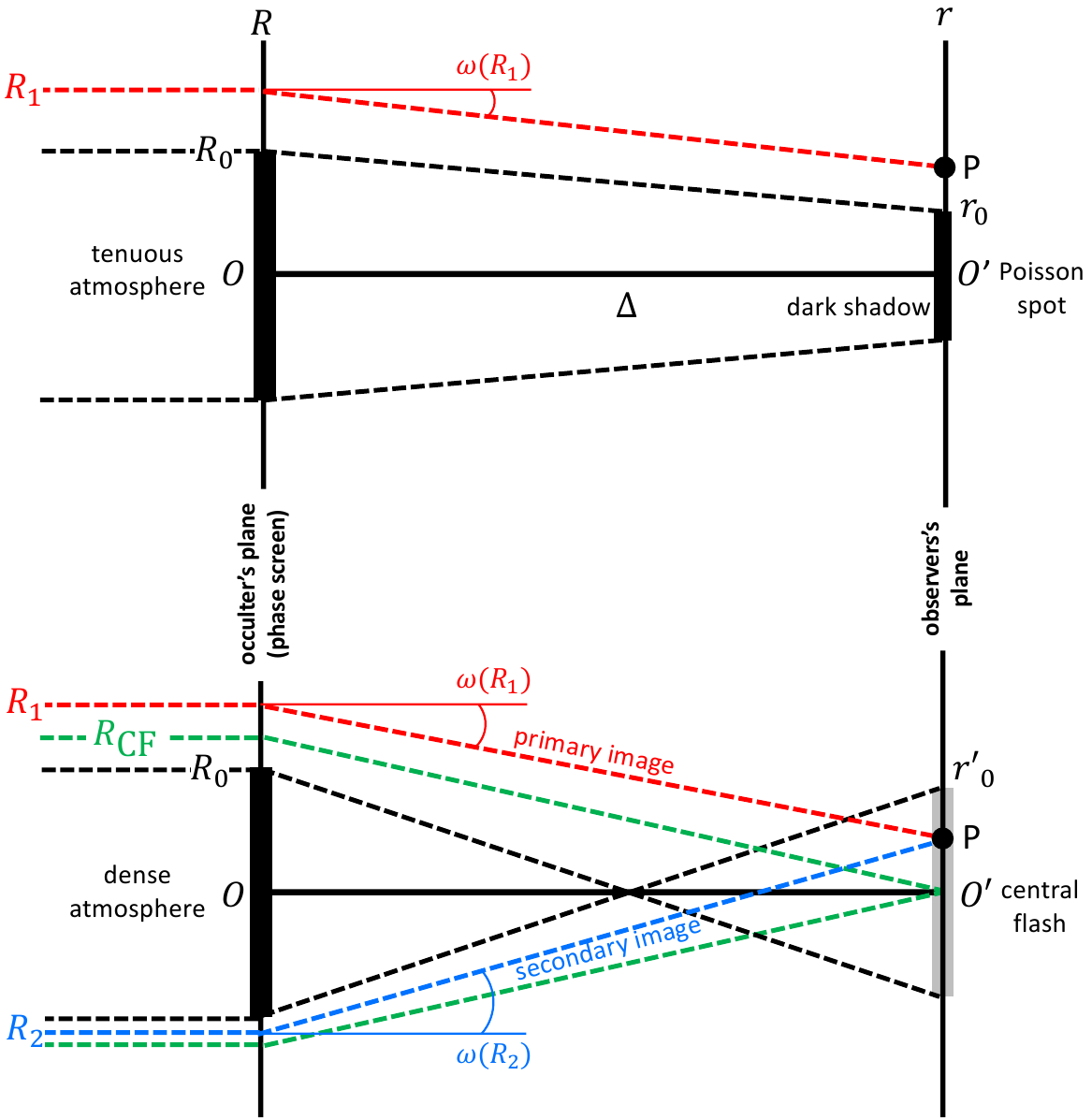}}
\caption{Principle of a stellar occultation by the atmosphere of an opaque spherical body with radius $R_0$, replaced here by a disk perpendicular to the figure.
\it Upper panel\rm: Tenuous atmosphere case, where the rays grazing the surface of the occulter at $R_0$ cannot converge toward  the shadow center, creating a dark shadow of radius $r_0$ (thick black line in the observer's plane) with a Poisson spot at its center in $O'$.
\it Lower panel\rm: Dense atmosphere case,  where there is a layer with radius $R_{\rm CF}$ that focuses the rays toward the shadow center, creating a central flash in $O'$. In the gray zone of radius $r'_0$, an observer at P receives a ray coming from the  primary (secondary) image plotted as the red (blue) dashed line.}
\label{fig_tenuous_vs_dense}
\end{figure}

We consider a spherical opaque object of radius $R_0$ surrounded by a perfectly spherical and transparent atmosphere, observed from a distance $\Delta \gg R_0$. As discussed in Sect.~\ref{sec_wave_optics} and illustrated in Fig~\ref{fig_tenuous_vs_dense}, the atmosphere can be approximated as a thin phase screen that slightly refracts a ray coming from infinity and passing at distance $R > R_0$ from the body center. 
This approximation stems from the fact that in the case of Earth-based observations, the bending angle due to refraction is generally less than 10$^{-5}$ radian (the apparent angular radius of the occulter), which means that the bending of the ray in the atmosphere is less than a meter or so, justifying the straight line approximation 
\citep{Sicardy2022}\footnote{
This approximation breaks down if the occulting body is close to the observer and/or if the  observation is conducted at radio wavelengths, which means that the rays go deep into the atmosphere; see for instance  the Mariner V radio occultation by Venus in October 1967 \citep{Fjeldbo1971}.}.
In the geometrical optics approach, the ray is deviated by an angle $\omega(R)$ (taken as positive), before reaching the observer P at distance $r$ from the shadow center. 
This angle depends on the profile of refractivity (refractive index minus one), $\nu,$ of the gas. The latter is in turn dependent on the molecular number density profile of the atmosphere, $n_{\rm g}$, through the relation $\nu \approx K n_{\rm g}$, where $K$ is the molecular refractivity that depends on $\lambda$.

Figure~\ref{fig_tenuous_vs_dense} illustrates our problem.
An occultation by a spherical object of radius $R_0$ is observed from far away ($\Delta \gg R_0$), so that the atmosphere causes a negligible deviation $\omega \ll 1$ of the light rays (straight-line approximation). As discussed by \cite{Young2012}, the atmosphere then acts as a phase screen that causes an increment of optical path $\tau(R)$ for the incoming wave, compared to the case where there is no atmosphere. 

Three cases are considered in this paper.
The first case occurs when no atmosphere is present around the body. We refer to this as the ``airless body case". It produces the classical Fresnel fringes at the edge of the shadow in the observer's plane, as well as the Poisson spot at its center. Although it is not the problem at stake here, this case serves as a starting point because both the Fresnel fringes and the Poisson spot survive in the presence of an atmosphere. 

The second case occurs when the atmosphere is not dense enough to refract the rays toward the shadow center. We refer to it as the ``tenuous atmosphere case". It means that the rays (of geometrical optics) grazing the limb at radius $R_0$ are not deviated enough to reach the shadow center. This creates a circular dark shadow\footnote{As opposed to just the shadow, in which an attenuated stellar flux can still be recorded.}
of radius $r_0$ with a Poisson spot at its center. Outside the dark shadow, an observer at P only sees one stellar image. 

The third case occurs when the atmosphere is dense enough for there to be a layer with radius $R_{\rm CF}$, called the flash layer, which focuses the stellar rays toward the shadow center at $r=0$. We refer to this as the ``dense atmosphere case".
In this case, a luminous ray can reach the observer at P by following two different possible paths, which correspond to the primary (or near-limb) and secondary (or far-limb) 
images\footnote{In a wide sense, because they are astigmatic. Moreover, we ignore the formation of multiple images (mirages) that  occur with strong local temperature gradients (e.g., \citealt{Stansberry1994}).} of the star, respectively.

The terms dense and tenuous are used here as descriptive terms 
and not absolute notions. For instance, Pluto and Triton have dense atmospheres in our nomenclature, while their surface pressures is of the order of 10~$\upmu$bar, which is by no means a high value. Moreover, the fact that an atmosphere is able or not to focus the stellar rays onto the shadow center depends on the observer distance, $\Delta$, and the wavelength, $\lambda$. Thus an atmosphere may be dense for some large values of $\Delta$ and tenuous for smaller values of $\Delta$, as seen in  Fig~\ref{fig_tenuous_vs_dense}.

Considering the dense atmosphere case, the closest-approach distances $R_1$ and $R_2$ are given as a function of $r$ through the implicit equations\footnote{In the rest of the paper and by convention, we replace $\approx$ by = when no new approximation is introduced.}
\begin{equation}
\begin{array}{l}
\displaystyle R_1(r) \approx r + \omega(R_1) \Delta {\rm~~and} \\ 
 \\
\displaystyle R_2(r) \approx -r + \omega(R_2) \Delta.
\end{array}
\label{eq_R1_R2}
\end{equation}
Because $d\nu/dR < 0$, and also $d\omega/dR < 0$, two rays passing at different distances $R$ diverge, which causes a decay of the stellar flux. Its value, normalized to the unocculted stellar flux, is obtained from the conservation of energy for each image, 
\begin{equation}
\phi_{\perp 1,2} (r) = \left| \frac{{\rm d}R_{1,2}(r)}{{\rm d}r} \right| = \frac{1}{\displaystyle 1 - \Delta \frac{{\rm d} \omega}{{\rm d}R}[R_{1,2}(r)]}.
\label{eq_flash_geo_optics_perp}
\end{equation}
The index $\bot$ indicates that this flux results from the differential refraction  of the rays perpendicular to an assumed rectilinear limb. 
Because ${\rm d}\omega/{\rm d}R < 0$, we have $\phi_\perp (r) < 1$, i.e. a decrease of flux.  
Differential refraction is the dominant effect during the occultation until very close to the shadow center, where the limb curvature causes a ray focusing that results in an increase of flux by a factor $R/r$. Thus, the flux received by a telescope centered at P from any one of the stellar images is
\begin{equation}
\phi_{1,2} (r) = \phi_{\perp,1,2} (r) \left( \frac{R_{1,2}}{r} \right).
\label{eq_phi_to_phi_perp}
\end{equation}
Near the shadow center, $R_1 \approx R_2 \approx R_{\rm CF}$ and the two stellar images send equal fluxes to the telescope, so that the central flash profile in the geometrical optics approximation is
\begin{equation}
\phi (r) \approx \phi_\perp (0) \left( \frac{2R_{\rm CF}}{r} \right).
\label{eq_flash_geo_optics}
\end{equation}

In the particular case of an isothermal atmosphere with a constant scale height $H \ll R$, the bending angle 
is\footnote{
Strictly speaking, due to the variation of gravity with $R$, we should write $H(R)$. As this dependence is weak, for the sake of brevity this dependence is ignored in this paper.
A more accurate expression of Eq.~\ref{eq_omega} includes a corrective factor $1+[(9-b)H/(8R)]$, where $b$ is an exponent of order unity that describes the variation of temperature with $R$ \citep{Young2009}. As $H \ll R$, this correction is ignored here.}
\begin{equation}
\omega(R) \approx \nu(R) \sqrt{\frac{2\pi R}{H}}.
\label{eq_omega}
\end{equation} 
As the atmospheric density decreases exponentially with $R$, so does approximately $\omega(R)$, and thus ${\rm d}\omega/{\rm d}R \approx -\omega/H$ and $1 - \Delta ({\rm d}\omega/{\rm d}R) \approx 1 + \omega\Delta/H$. Near the shadow center ($r \ll R_{\rm CF}$), we have $\omega \approx R_{\rm CF}/\Delta$, and considering that $H \ll R_{\rm CF}$, Eq.~\ref{eq_flash_geo_optics_perp} finally provides 
\begin{equation}
\phi_\perp (0) \approx \frac{H}{R_{\rm CF}}.
\label{eq_phi_perp_0}
\end{equation}
From Eq.~\ref{eq_flash_geo_optics}, we finally obtain 
\begin{equation}
\phi(r) =  2 \left( \frac{H}{R_{\rm CF}} \right)  \left(\frac{R_{\rm CF}}{r} \right) = \frac{2H}{r},
\label{eq_flash_geo_optics_iso}
\end{equation}
a result already derived by \cite{Young1977}. The factor $(H/R_{\rm CF})$ is the stellar flux that would come from each image without limb curvature, while $(R_{\rm CF}/r)$ accounts for the focusing of the rays caused by the limb curvature.

We note that the factor $R_{\rm CF}$ disappears from the expression of $\phi(r)$ in Eq.~\ref{eq_flash_geo_optics_iso}: the flash profile only depends on $H$ and $r$, and is independent of the object size and of the wavelength. 
We also note that the flux reaches the full unocculted stellar flux at $r = 2H$, and diverges to infinity at 
$r=0$ (this result can be retrieved if $\lambda$ approaches zero in Eq.~\ref{eq_CF_Hubbard}). 
This singularity disappears when diffraction or finite stellar diameter are taken into account.

\section{Wave optics, general formalism}
\label{sec_wave_optics}

\begin{figure}[!t]
\centerline{\includegraphics[totalheight=65mm,trim=0 0 0 0]{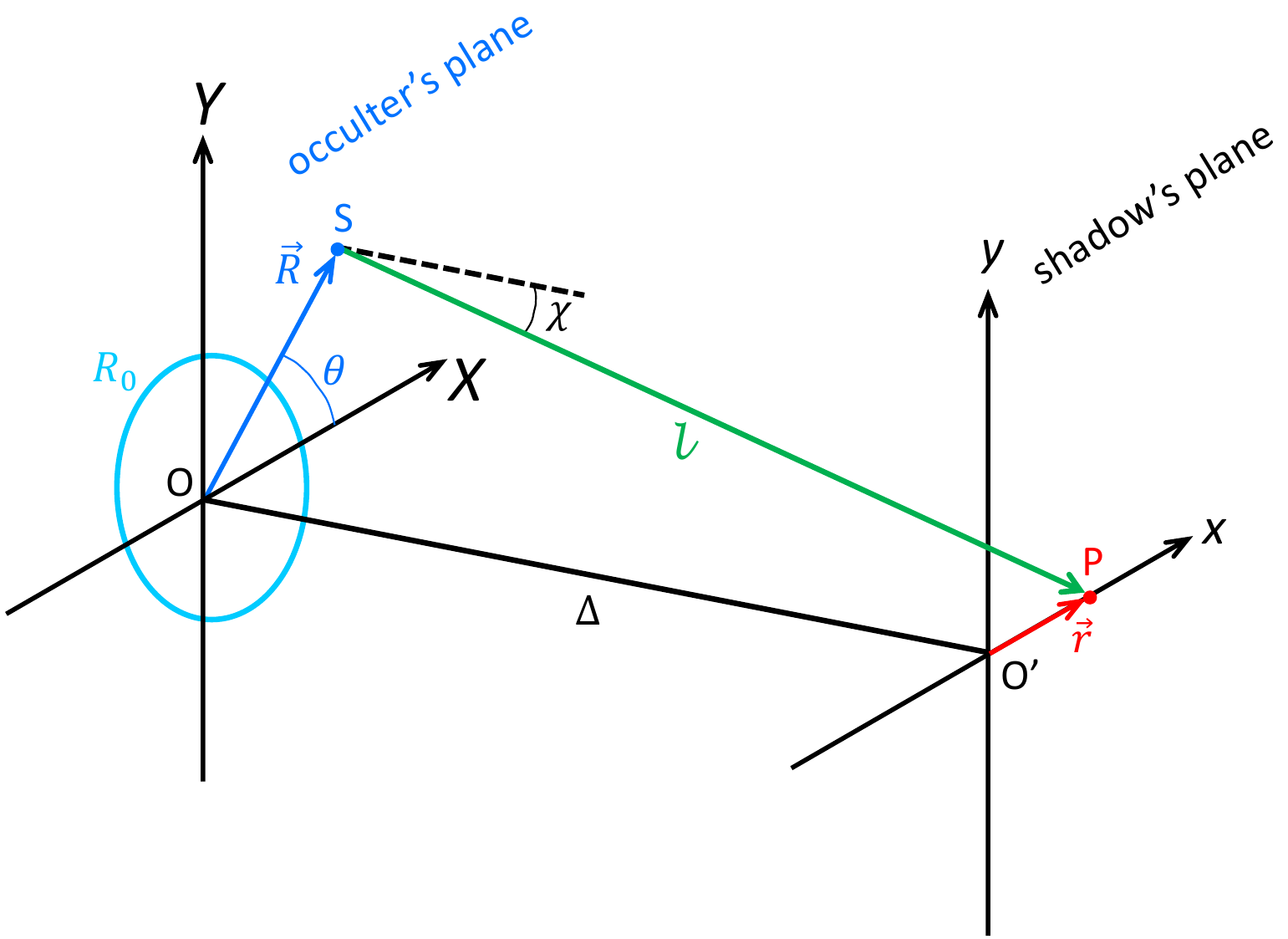}}
\caption{Quantities used in the wave optics calculations.
A star at infinity at the left of the figure along the $OO'$ direction sends a plane wave through the occulter atmosphere, considered as a thin phase screen $XOY$. 
A source point S in this screen, with polar coordinates $(R,\theta)$, emits an elementary wave toward the observer in the shadow plane at P, with an inclination angle $\chi$ with respect to the normal of the screen.
If the atmosphere is spherical, the point P can be placed without loss of generality along the  $O'x$ axis, at a distance $r=|x|$ from the shadow center.
The phase of the elementary wave received at P depends on both the phase shift induced by the atmosphere at S and by the distance $l$ traveled by the wave between S and P.}
\label{fig_geo_occ}
\end{figure}

We now consider the role of diffraction around the shadow center, assuming a point-like star and a monochromatic plane wave. The wave hits the object described in Sect.~\ref{sec_geom_optics} and illustrated in Fig.~\ref{fig_geo_occ}. A ray that passes at radius $R>R_0$ from the body center is affected by the atmosphere considered as a phase screen, which causes a phase change of
\begin{equation}
\varphi_{\rm a}(R) = k\tau(R) = \frac{2\pi}{\lambda} \tau(R),
\label{eq_phase_a}
\end{equation}
where $k=2\pi/\lambda$ is the wave number. 

The point source S in the atmosphere sends to P a secondary spherical wave along a segment of length $l$ (Fig~\ref{fig_geo_occ}). The phase change between S and P has a purely geometrical origin that amounts to $\varphi_{\rm g} = kl = 2\pi l/\lambda$, where we use the classical Fresnel approximation
\begin{equation}
l \approx  \left( \Delta + \frac{ r^2}{2\Delta} \right) + \frac{R^2 - 2Rr\cos \theta}{2\Delta}.
\end{equation}
Thus,
\begin{equation}
\varphi_{\rm g} (R,\theta) = \varphi_0 + \frac{\pi}{\lambda \Delta} (R^2 - 2Rr\cos \theta),
\label{eq_phase_g}
\end{equation}
where $\varphi_0 = (2\pi/\lambda)[\Delta + (r^2/2\Delta)]$. The complex normalized amplitude of the spherical wave sent to P by a surface element 
$R{\rm d}R{\rm d}\theta$ around S is, using Huygens' principle, 
\begin{equation}
\displaystyle
d^2 a(r) = (\cos \chi) \frac{\exp({\rm i}\varphi_0)}{{\rm i} \lambda \Delta} \exp[{\rm i}(\varphi_{\rm a} + \varphi_{\rm g})] R {\rm d}R {\rm d}\theta,
\label{eq_d2a}
\end{equation}
where ${\rm i}^2=-1$. The factor $(\cos \chi) [\exp({\rm i}\varphi_0)]/({\rm i} \lambda \Delta)$ ensures the conservation of energy of the wave and its correct phase at P. For $r$ fixed, the phase $\varphi_0$ is a constant that can be taken as equal to zero through an appropriate change of origin. Moreover, in the cases studied here (a remote object that subtends a small angle), the Rayleigh-Sommerfeld inclination factor $\cos \chi$ can be taken as equal to unity; see Appendix~\ref{app_Sommerfeld}. The wave amplitude received at P is then
\begin{equation}
\begin{array}{l}
\displaystyle
a(r) \approx  \frac{1}{2 {\rm i} \lambda_{\rm F}^2} 
\int_{R_0}^{+\infty} \int_0^{2\pi} 
\exp[{\rm i}(\varphi_{\rm a} + \varphi_{\rm g})] R \, {\rm d}R {\rm d}\theta = \\
 \\
\displaystyle
\frac{1}{2 {\rm i} \lambda_{\rm F}^2} \times \\
 \\
\displaystyle
\int_{R_0}^{+\infty} \int_0^{2\pi}
\exp \left\{ {\rm i} \left[ \varphi_{\rm a} + \left(\frac{\pi}{\lambda_{\rm F}^2}\right) 
\left( \frac{R^2}{2} - Rr \cos \theta \right) \right] \right\} R \, {\rm d}R {\rm d}\theta, 
\\
\end{array}
\label{eq_amplitude_ori}
\end{equation}
where
\begin{equation}
\lambda_{\rm F} := \sqrt{ \frac{\lambda \Delta}{2} }
\label{eq_Fresnel_scale}
\end{equation} 
denotes the Fresnel scale. To within about 10\%, $\lambda_{\rm F}$ represents the spacing between the first two Fresnel fringes seen near a shadow edge; see Fig.~\ref{fig_Fresnel_Poisson}. 

Using Eq.~\ref{eq_Bessel}, we can write Eq.~\ref{eq_amplitude_ori} as 
\begin{equation}
\begin{array}{l}
\displaystyle
a(r) = \frac{\pi}{ {\rm i} \lambda_{\rm F}^2}  \times \\
 \\
\displaystyle
\int_{R_0}^{+\infty} 
\exp \left\{ {\rm i} \left[ \varphi_{\rm a}(R) + \frac{\pi R^2}{2\lambda_{\rm F}^2}  \right]  \right\}
J_0 \left( \frac{\pi Rr}{\lambda_{\rm F}^2} \right) R \, {\rm d}R,
\end{array}
\label{eq_amplitude}
\end{equation} 
where we recall that $\varphi_{\rm a}$ only depends on $R$ due to the spherical symmetry of the atmosphere.
When we move away from centrality, and more precisely for 
\begin{equation}
u:= \frac{\pi Rr}{\lambda_{\rm F}^2} \gtrsim {\rm a~few~times~unity},
\label{eq_u}
\end{equation} 
the Bessel function $J_0$, and more generally the functions $J_n$, are well approximated by Hankel's asymptotic form 
\begin{equation}
J_n(u) \approx  \sqrt{\frac{2}{\pi u}} \cos \left( u - \frac{n\pi}{2} - \frac{\pi}{4} \right).
\label{eq_asymptotic}
\end{equation} 
For $n=0$, the expression above already describes to better than 1\% the first oscillation of the Poisson spot shown in Fig.~\ref{fig_Fresnel_Poisson}. Equation~\ref{eq_asymptotic} can be re-expressed as 
\begin{equation}
J_0(u) =  \frac{1}{\sqrt{4\pi u}} \left[ (1+{\rm i}) \exp(-{\rm i}u) + (1-{\rm i}) \exp({\rm i}u) \right].
\label{eq_asymptotic_bis}
\end{equation} 
This form has the advantage of splitting $J_0$ into the two terms $\exp(-{\rm i} u)$ and $\exp({\rm i} u)$, respectively associated with partial waves,
qualified here as primary and secondary waves, coming from points at $\theta=$ 0 and $\pi$ (Fig.~\ref{fig_geo_occ}). 
It also reveals the phase difference between these two waves.  
The prefactor $1-{\rm i}$ associated with $\exp(iu)$ (primary wave) in Eq.~\ref{eq_asymptotic_bis}  differs from the prefactor $1+{\rm i}$ of  $\exp(-{\rm i} u)$ (secondary wave) by a multiplicative factor of $-{\rm i}$, i.e. a phase shift of $\pi/2$ between the two waves. 
This comes from the term $-Rr\cos(\theta)$ in the phase $\varphi_{\rm g}(R,\theta)$ (Eq.~\ref{eq_phase_g}). Thus, $\varphi_{\rm g}$ reaches a local minimum at $\theta=0$ for $R$ fixed, which corresponds to the contribution of the primary wave to the global received wave. Conversely, $\varphi_{\rm g}$ reaches a local maximum at $\theta=\pi$, which corresponds to the contribution of the secondary wave. This difference is at the origin of the $\pi/2$ phase shift, as detailed in Section~\ref{sec_dense_atmo_extended_vicinity} and  Appendix~\ref{app_stationary_method}.

This phenomenon stems from the fact that the ray of the partial secondary wave arriving at P crosses the revolution axis $(OO')$ (Fig.~\ref{fig_geo_occ}). Along this axis, the wave is very different from a locally progressive plane wave, as
it presents a narrow irradiance peak (like the function $J_0^2$). 
Consequently, the spatial variation of its phase on a ray is different from the optical path multiplied by the wave number $k$ because the eikonal equation is not valid within $(OO')$.

Using Eq.~\ref{eq_asymptotic_bis}, Eq.~\ref{eq_amplitude} yields 
\begin{eqnarray}
a(r) = a_1(r) + a_2(r), {\rm~where}  \label{eq_amplitude_a1_plus_a2} \\ \nonumber \\
\displaystyle
a_1(r)= \left( \frac{1-{\rm i}}{2\lambda_{\rm F}} \right) \frac{1}{\sqrt{r}} \int_{R_0}^{+\infty} \exp[ {\rm i} \varphi_1 (R)] \sqrt{R} \, {\rm d}R,  \label{eq_amplitude_a1_ori} 
\\ \nonumber \\
\displaystyle
a_2(r)= -\left(  \frac{1+{\rm i}}{2\lambda_{\rm F}} \right) \frac{1}{\sqrt{r}} \int_{R_0}^{+\infty} \exp[ {\rm i} \varphi_2 (R)] \sqrt{R} \, {\rm d}R, \label{eq_amplitude_a2_ori} 
\end{eqnarray} 
and 
\begin{eqnarray}
 \displaystyle 
 \varphi_1 (R) := \varphi_{\rm a} (R) + \frac{\pi}{\lambda_{\rm F}^2} \left( \frac{R^2}{2}  - R r \right)  \label{eq_phase1} \\
 \nonumber \\
 \displaystyle 
 \varphi_2 (R) := \varphi_{\rm a} (R) + \frac{\pi}{\lambda_{\rm F}^2} \left( \frac{R^2}{2}  + R r \right). \label{eq_phase2}
\end{eqnarray}
Equations~\ref{eq_amplitude_a1_plus_a2}-\ref{eq_phase2} can also be retrieved by performing the integration with respect to $\theta$  in Eq.~\ref{eq_amplitude_ori} under the approximation of the stationary phase method (Appendix~\ref{app_stationary_method}), 
using neither $J_0$ nor the asymptotic forms given by Eqs.~\ref{eq_asymptotic}-\ref{eq_asymptotic_bis}. This said, Eqs.~\ref{eq_amplitude_a1_plus_a2}-\ref{eq_phase2} contain all the information necessary to calculate the amplitudes of the waves due to each image, their sum, and finally the flux received by the observer. 

We now construct a hierarchical sequence of shadow models, starting with the airless body, then examining the tenuous atmosphere regime, and ending with the dense atmosphere case. During this sequence, we can see how the central flash structure evolves.

\section{The airless body case}
\label{sec_airless}

For an airless body, we have $ \varphi_{\rm a}(R) \equiv 0$ in Eq.~\ref{eq_amplitude}. We consider two cases, one well inside the dark shadow (Fig.~\ref{fig_tenuous_vs_dense}), with the Poisson spot at the center, and one near the shadow edge, where the Fresnel fringes prevail.  

\subsection{Inside the shadow, far from the edge}

\begin{figure}[!t]
\centerline{\includegraphics[totalheight=55mm,trim=0 0 0 0]{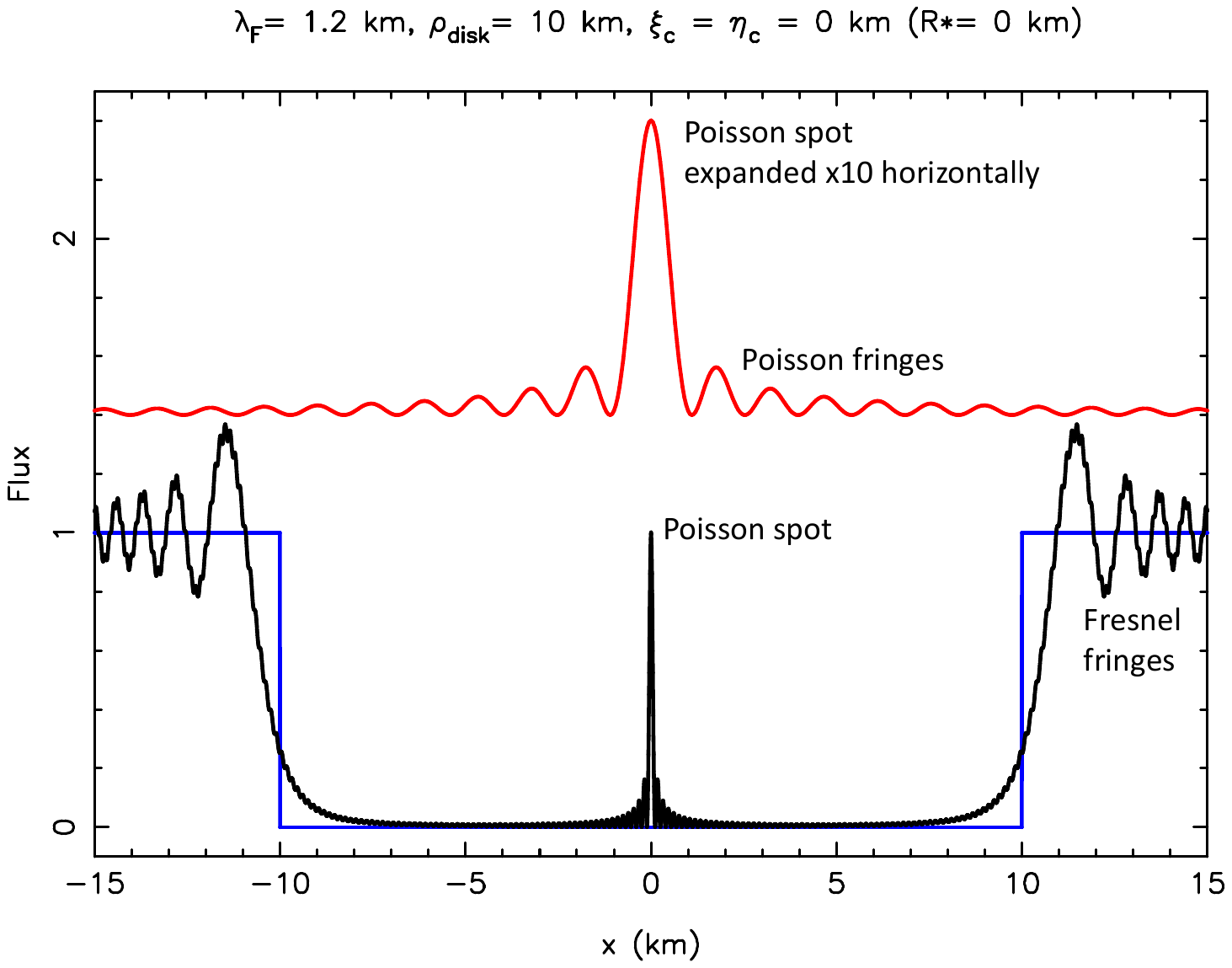}}
\caption{Diametric profile of the shadow cast by an opaque circular mask of radius $R_0=10$~km, illuminated by a monochromatic wave that provides a Fresnel scale of $\lambda_{\rm F}=1.2$~km. 
The profile has been obtained through the numerical integration of Eq.~\ref{eq_amplitude}, where $\varphi_{\rm a} (R) \equiv 0$.
The blue profile shows the shadow within the limit of geometrical optics, while the black profile accounts for diffraction effects. 
The edge of the shadow displays the Fresnel fringes with typical spacings of $\lambda_{\rm F}$ (Eq.~\ref{eq_Fresnel_scale}), while the shadow center hosts the Poisson spot that peaks at the value of the flux far away outside the shadow, normalized here to unity. 
The upper red curve shows a shifted and expanded  view of the Poisson spot. The fringes around the Poisson spot have a spacing of $\lambda_{\rm F}^2/R_0 = 0.144$~km (Eq.~\ref{eq_inter_fringe}).}
\label{fig_Fresnel_Poisson}
\end{figure}

For $r \lesssim \lambda^2_{\rm F}/(\pi R_0)$, we have
\begin{equation}
a(r) = 
\frac{\pi}{ {\rm i} \lambda_{\rm F}^2} 
\int_{R_0}^{+\infty} 
\exp  \left( \frac{ {\rm i} \pi R^2}{2 \lambda_{\rm F}^2}\right)
J_0 \left( \frac{\pi Rr}{\lambda_{\rm F}^2} \right) R \, {\rm d}R.
\label{eq_amplitude_airless}
\end{equation}
The Sommerfeld lemma (Appendix~\ref{app_Sommerfeld}) provides
\begin{equation}
a(r) \approx \exp \left( \frac{ {\rm i} \pi R_0^2}{2 \lambda_{\rm F}^2} \right) J_0 \left( \frac{\pi R_0 r}{\lambda_{\rm F}^2} \right),
\label{eq_amplitude_Poisson}
\end{equation} 
from which the Poisson spot profile is obtained, 
\begin{equation}
\phi_{\rm Pois}(r) = |a(r)|^2 =  J_0^2 \left( \frac{\pi R_0 r}{\lambda_{\rm F}^2} \right).
\label{eq_Poisson_spot}
\end{equation}
The peak value of the spot is $\phi_{\rm Pois} (0) = J_0^2(0) = 1$, which is the value of the flux due to the unocculted star; see Fig.~\ref{fig_Fresnel_Poisson}. Other examples of Poisson spots where the value of $\lambda_{\rm F}$ is varied can be found in \cite{Roques1987}. 
As $r$ increases, Eq.~\ref{eq_asymptotic} shows that $\phi(r) = J_0^2 ( \pi R_0 r/\lambda_{\rm F}^2 )$ oscillates 
with an amplitude that decreases as $1/r$, forming fringes radially spaced by  
\begin{equation}
\lambda_{\rm P}=  \frac{\lambda_{\rm F}^2}{R_0} = \frac{\lambda \Delta}{2R_0}.
\label{eq_inter_fringe}
\end{equation}

The diameter $D_{\rm Pois}$ of the Poisson spot can be defined as the diameter of the first dark fringe, which occurs at the first zero $u_0= 2.405...$ of  $J_0(u)$, yielding 
\begin{equation}
D_{\rm Pois} 
= \frac{2u_0 \lambda_{\rm F}^2}{\pi R_0} 
\approx 1.53 \frac{\lambda_{\rm F}^2}{R_0} 
\approx 0.766 \frac{\lambda \Delta}{R_0}.
\label{eq_diam_Poisson}
\end{equation}
We note that the Poisson spot disappears as $\lambda$ approaches zero, in the sense that its width (and also is total flux) tends to zero while its peak value remains equal to one.

\subsection{Near the shadow edge}

As one approaches the shadow edge, the phase $\varphi_1(R)$ admits a stationary value at $R=r \approx R_0$, so that the Sommerfeld lemma cannot be used anymore. The stationary condition (Appendix~\ref{app_stationary_method}) then provides an evaluation of the integral in Eq.~\ref{eq_amplitude_a2_ori} through Eq.~\ref{eq_Fresnel_int_edge}, where $a=R_0$, $b=+\infty$, $f(v)= \sqrt{v}$ , and $g(v)= \pi [(v^2/2) - vr]/\lambda_{\rm F}^2$.
Elementary calculations then provide 
\begin{equation}
\phi_{\rm Fr}(r)=  |a_1(r)|^2 = 
\left| \frac{1}{2} + \frac{1}{1+{\rm i}}  F_{\rm r} \left( \frac{r - R_0}{\lambda_{\rm F}} \right) \right|^2,
\label{eq_Fresnel_fringes}
\end{equation}
where $ F_{\rm r} $ is the Fresnel function, defined by Eq.~\ref{eq_Fresnel_func}. 
This describes the classical Fresnel fringes near the edge of a shadow cast by an opaque body; see Fig.~\ref{fig_Fresnel_Poisson}. The flux decays equivalently to  
$[\lambda_{\rm F}/(r-R_0)]^2/(2\pi^2)$ for $r \lesssim R_0$ and the oscillations of the Fresnel fringes damps equivalently to  
$(\sqrt{2}/\pi)[\lambda_{\rm F}/(r-R_0)]$ for $r \gtrsim R_0$; see Appendix~\ref{app_stationary_method}. These Fresnel fringes are the only diffraction effect considered in the early work of \cite{Fabry1929}.

In the presence of an atmosphere, we find that both the Poisson spot and the Fresnel fringes survive, but with a different peak value and a different spacing, respectively. In spite of these differences, we still use the nomenclature Poisson spot, Poisson fringes, or Fresnel diffraction to describe these features. 

\section{The tenuous atmosphere case}
\label{sec_tenuous_atmo}

\subsection{Inside the dark shadow, far from the edge}

The presence of an atmosphere is accounted for by the term $\varphi_{\rm a}(R)$ in Eq.~\ref{eq_amplitude}, where we define the phase 
\begin{equation}
\varphi (R) :=  \varphi_{\rm a}(R) + \frac{\pi R^2}{2\lambda_{\rm F}^2} {\rm~for~} R \geq R_0.
\label{eq_phase}
\end{equation}
If the observer is deep inside the dark shadow of the object (see Fig.~\ref{fig_tenuous_vs_dense}), then
no stellar image is seen by the observer. This implies that $\varphi (R)$ is nowhere stationary, i.e. that $\varphi' (R) \neq 0$ for any $R > R_0$. The Sommerfeld lemma (Eq.~\ref{eq_Sommerfeld}) can then be applied. As detailed in Appendix~\ref{app_Sommerfeld}, the flux observed near the shadow center is then 
\begin{equation}
\phi_{\rm Diff} (r) =  \left( \frac{R_0}{r_0} \right)^2 J_0^2 \left( \frac{\pi R_0 r}{\lambda_{\rm F}^2} \right).
\label{eq_Poisson_tenuous}
\end{equation} 
This is Eq.~\ref{eq_Poisson_spot}, except for the amplification factor $(R_0/r_0)^2$. 
For an airless body, we retrieve the classical Poisson spot with $R_0= r_0$. 
As the atmosphere increases in density while remaining tenuous, the Poisson spot is amplified by the ratio of the projected surface area of the body to the surface area of its shadow, a factor that can reach several times unity. Remarkably, this amplification occurs even though no rays are focused toward the shadow center in the geometrical optics regime. This is actually due to the fact that the atmosphere decreases the rate of change of $\varphi(R)$ near $R_0$, when compared to the airless case. This stems from the fact that $\varphi'_{\rm a}(R_0)<0$ (Eq.~\ref{eq_dphi}), which decreases the value of $g'(v)$ in Eq.~\ref{eq_Sommerfeld}, thus increasing the value of $\phi_{\rm Diff} (r)$.

\subsection{Outside the dark shadow, far from the edge}

We assume here that $r_0 \gg \lambda^2_{\rm F}/(\pi R_0)$, i.e. that the size of the dark shadow is much larger that the diameter of the Poisson spot.
We consider an observer well outside the dark shadow (in the sense that Fresnel diffraction is negligible).
Such an observer sees the primary stellar image, but not the secondary image. This means that the integral in Eq.~\ref{eq_amplitude_a2_ori} is negligible compared to the integral in Eq.~\ref{eq_amplitude_a1_ori}.
The contribution to the integral  in Eq.~\ref{eq_amplitude_a1_ori} mainly comes from the neighborhood of $R_1$ where $\varphi_1 (R)$ is stationary, i.e. where $\varphi'_1 (R)=0$. Using Eq.~\ref{eq_phase1}, this yields $\varphi_{\rm a}' (R_1)= -\pi (R_1 -r)/\lambda_{\rm F}^2$. As the deviation angle at $R_1$ is $\omega(R_1)= (R_1 - r)/\Delta$ (Fig.~\ref{fig_tenuous_vs_dense}), we have $\varphi_{\rm a}' (R_1)= -(\pi \Delta/\lambda_{\rm F}^2)\omega(R_1)$. We note that the stationary condition must be satisfied for any value of $r$, hence of $R_1(r)$, thus 
\begin{equation}
\varphi'_{\rm a} (R_1) =  -\frac{\pi \Delta}{\lambda_{\rm F}^2} \omega(R_1) = -k \omega(R_1).
\label{eq_dphi}
\end{equation}
This is an expression of the Malus-Dupin theorem, which states that the pencil of light emerging from the atmosphere is perpendicular to the local wave front.
From Eq.~\ref{eq_dphi}, we have $\varphi_1'' =  {\rm d} \varphi'_1/{\rm d}R =  (\pi /\lambda_{\rm F}^2)(1 - \Delta {\rm d}\omega/{\rm d}R)$. From Eq.~\ref{eq_flash_geo_optics_perp}, we obtain 
\begin{equation}
\varphi''_1 (R_1) = \frac{1}{\phi_{\perp 1}(r)} \left( \frac{\pi}{\lambda_{\rm F}^2} \right),
\label{eq_ddphi}
\end{equation} 
where we recall that $\phi_{\perp 1}(r)$ is the flux received at $r$ from the primary stellar image in the geometrical optics limit, without the focusing effect of the limb curvature.

Because $\varphi'_1=0$ at $R_1 > R_0$, the Sommerfeld lemma (Eq.~\ref{eq_Sommerfeld}) cannot be used anymore. In order to evaluate the integral in Eq.~\ref{eq_amplitude_a1_ori}, we have instead to use Eq.~\ref{eq_Fresnel_int}, where $f(R)= \sqrt{R}$ and $g(R)= \varphi_1(R)$.
Let us consider the width $A$ of the first Fresnel zone in $R - R_1$, $A= \sqrt{2\pi/\varphi_1''(R_1)}=\lambda_{\rm F} \sqrt{2\phi_{\perp 1}(r_0)}$.  Since $\phi_{\perp 1} < 1$, the effect of the refraction is to diminish $A$.  This is one of the aspects of the Fresnel zone flattening considered by \cite{Young1976}.
This said, for $R_1 \gtrsim R_0 +$ several times $A$, then the bounds of the integral ($R_0$ and $+\infty$) can be replaced by $-\infty$ and $+\infty$, providing 
\begin{equation}
a_1(r) \approx  \sqrt{\phi_1(r)} \exp[ {\rm i} \varphi_1(R_1)].
\label{eq_amplitude_a1}
\end{equation} 
Here we have used the fact that $\phi_1 (r) = (R_1/r) \phi_{\perp 1}(r)$ is the flux received at P from the primary stellar images accounting for the focusing effect of the limb curvature (Eq.~\ref{eq_phi_to_phi_perp}). Equation~\ref{eq_amplitude_a1} merely states that the observed flux, 
\begin{equation}
\phi_{\rm Diff} (r)= |a_1(r)|^2 = \phi_1(r),
\end{equation} 
is the flux calculated in the geometrical optics approximation (Eq.~\ref{eq_phi_to_phi_perp}), with no noticeable diffraction effects. 

\subsection{Near the dark shadow edge}

As $R_1 \approx R_0$, edge effects come into play and we meet again the Fresnel function $F_{\rm r}$.
Using Eq.~\ref{eq_Fresnel_int_edge}, we get 
\begin{equation}
a_1(r) \approx  \sqrt{\phi_1} \exp({\rm i} \varphi_1) 
\left[ \frac{1}{2} + \frac{1}{1+{\rm i}}  F_{\rm r} \left( \frac{R_1-R_0}{\sqrt{\phi_{\perp 1}(r)} \lambda_{\rm F}} \right) \right]
\label{eq_amplitude_a1_Fresnel}
\end{equation} 
and the corresponding flux 
\begin{equation}
\phi_{\rm Diff}(r)=  \phi_1(r) \left| \frac{1}{2} + \frac{1}{1+{\rm i}}  F_{\rm r} \left( \frac{R_1-R_0}{\sqrt{\phi_{\perp 1}(r)} \lambda_{\rm F}} \right) \right|^2.
\label{eq_flux_tenuous_atmosphere_R}
\end{equation}

Formally, $R_1$ no longer exists for $r < r_0$. However, $\phi_{\rm Diff}(r)$ can still be evaluated by noting that from Eq.~\ref{eq_flash_geo_optics_perp} we can replace $R_1 - R_0$ by $\approx \phi_{\perp 1}(r_0) (r-r_0)$. 
This is equivalent to extrapolating the atmospheric profile a little bit below the radius of the body, which provides 
\begin{equation}
\phi_{\rm Diff}(r) \approx  \phi_1(r_0) 
\left| \frac{1}{2} + \frac{1}{1+{\rm i}}  F_{\rm r} \left( \frac{\sqrt{\phi_{\perp 1}(r_0)}(r-r_0)}{\lambda_{\rm F}} \right) \right|^2.
\label{eq_flux_tenuous_atmosphere_r}
\end{equation} 
Hence for ($r>r_0$), and compared again to the airless case, the Fresnel fringes are locally stretched by the factor $1/\sqrt{\phi_{\perp 1}(r_0)}$ as seen in the observer plane, a result already obtained by \cite{French1976}. For $r < r_0$, the flux approaches zero, as predicted by Eq.~\ref{eq_damping_Fresnel_fringes}. For $r$ smaller than $r_0$ minus a few times $\lambda_{\rm F}/\sqrt{\phi_{\perp 1}(r_0)}$, this yields the flux near the edge of the shadow,
\begin{equation}
\phi_{\rm Edge}(r) \approx
\frac{R_0}{2\pi^2 r_0} \left( \frac{\lambda_{\rm F}}{r-r_0} \right)^2.
\label{eq_phi_edge}
\end{equation} 
This equation describes how the flux asymptotically decays to zero as the observer gets deeper into the dark shadow. The flux reaches half the value $\phi_1(r_0)$ for 
\begin{equation}
r - r_0 \approx -\frac{\lambda_{\rm F}}{\pi} \sqrt{\frac{R_0}{r_0}} \frac{1}{\sqrt{\phi_1(r_0)}}.
\label{eq_phi_edge_half_light}
\end{equation} 
This shows that if $\phi_1(r_0)$ is small, the damping distance of the flux in the observer plane may be significantly greater than $\lambda_{\rm F}$.

In principle, the measurement of the stretching factor $1/\sqrt{\phi_{\perp 1}(r_0)}$ can be used to detect tenuous atmospheres. In practice, however, this requires a high signal-to-noise ratio and a high cadence of acquisition, as the first Fresnel fringes cover a fraction of a second only for typical Earth-based occultations. Also, it requires a good knowledge of the topography of the object, as the fringe spacing depends on the angle between the apparent path of the star relative to the local limb.
Another way to detect atmospheres is to measure the shrinking $R_0 - r_0$ of the shadow radius with respect to the actual radius $R_0$ of the object. However, this requires the size of the object to be known using independent results, for instance from in situ space missions. This method was used for a Titania occultation observed in 2001, by comparing the radius derived from the occultation ($r_0= 788.4\pm$0.6~km) and the Voyager 1 result ($R_0=788.9 \pm 1.8$~km), which provided an upper limit of 70~nbar for the surface pressure of a CO$_2$ atmosphere \citep{Widemann2009}.

\section{The dense atmosphere case}
\label{sec_dense_atmo}

\subsection{Immediate vicinity of the center}

As the atmosphere becomes denser, the radius $r_0$ of the dark shadow decreases. At some point, $r_0$ approaches zero, which marks the transition between the tenuous and dense atmosphere cases (Fig.~\ref{fig_tenuous_vs_dense}). In principle, Eq.~\ref{eq_Poisson_tenuous} predicts that $\phi_{\rm Diff}(r)$ diverges. However, the Sommerfeld lemma can no longer be used since the phase $\varphi (R)$ in Eq.~\ref{eq_phase} becomes stationary ($\varphi' (R)=0$) at the radius $R_{\rm CF} > R_0$. Again using Eq.~\ref{eq_Fresnel_integral_g"_positive} and for $R_{\rm CF} \gtrsim R_0 +$ several times $\lambda_{\rm F} \sqrt{\phi_{\perp}(0)}$, we have, in the immediate vicinity of the center,
\begin{equation}
a(r) = 
(1-{\rm i})\pi \left( \frac{R_{\rm CF}}{\lambda_{\rm F}} \right) \sqrt{\phi_\perp(0)} \exp \left[ {\rm i} \varphi(R_{\rm CF}) \right] J_0 \left(\frac{\pi R_{\rm CF}r}{\lambda_{\rm F}^2} \right),
\label{eq_amplitude_J0}
\end{equation}
so that 
\begin{equation}
\phi_{\rm Diff} (r) = |a(r)|^2 =  2\pi^2 \left( \frac{R_{\rm CF}}{\lambda_{\rm F}} \right)^2  \phi_\perp(0) 
J_0^2 \left( \frac{\pi R_{\rm CF} r}{\lambda_{\rm F}^2} \right).
\label{eq_flash_J0}
\end{equation} 
This result is true for any spherically symmetric dense atmosphere, regardless of the details of the $\tau(R)$ expression in Eq.~\ref{eq_phase_a}. In the particular case of an isothermal atmosphere, $\phi_\perp (0) \approx H/R_{\rm CF}$ (Eq.~\ref{eq_phi_perp_0}), so that Eq.~\ref{eq_flash_J0} becomes 
\begin{equation}
\phi_{\rm Diff} (r) \approx 2\pi^2 \left( \frac{R_{\rm CF} H}{\lambda_{\rm F}^2} \right)
J_0^2 \left( \frac{\pi R_{\rm CF} r}{\lambda_{\rm F}^2} \right),
\label{eq_flash_J0_iso}
\end{equation}
which is \cite{Hubbard1977}'s result (Eq.~\ref{eq_CF_Hubbard}).

In the transition regime where  $R_{\rm CF} - R_0 \approx \lambda_{\rm F} \sqrt{\phi_{\perp}(0)}$, i.e. when the bright annulus causing the central flash is seen by the observer as lying just about the limb of the occulter, Fresnel diffraction occurs. This requires the corrective factor introduced in eq.~\ref{eq_Fresnel_int_edge}, that is, 
\begin{equation}
\begin{array}{l}
\displaystyle
\phi_{\rm Diff} (r) =  
2\pi^2 \left( \frac{R_{\rm CF}}{\lambda_{\rm F}} \right)^2 J_0^2 \left( \frac{\pi R_{\rm CF} r}{\lambda_{\rm F}^2} \right) \phi_\perp(0) \times \\ \\
\displaystyle
\left| \frac{1}{2} + \frac{1}{1+{\rm i}}  F_{\rm r} \left( \frac{R_{\rm CF} - R_0}{\sqrt{\phi_\perp (0)} \lambda_{\rm F}} \right) \right|^2.
\end{array}
\label{eq_flash_J0_Fresnel_R}
\end{equation}

The comparison of Eq.~\ref{eq_flash_J0} with Eq.~\ref{eq_Poisson_spot} shows that the central flash caused by a dense atmosphere has the same functional dependence on $r$ as the Poisson spot. However, instead of peaking at one, it peaks at 
\begin{equation}
\phi_{\rm Diff} (0)
= 2\pi^2 \left( \frac{R_{\rm CF}}{\lambda_{\rm F}} \right)^2 \phi_\perp(0)
\approx 2\pi^2 \left( \frac{R_{\rm CF} H}{\lambda_{\rm F}^2} \right),
\label{eq_max_CF_diffrac}
\end{equation} 
in the general case and in the isothermal case, respectively.

We note that the flash has a typical width of $\sim \lambda_{\rm F}^2/R_{\rm CF}$ (Eq.~\ref{eq_diam_Poisson}) and a height of $2\pi^2(R_{\rm CF}^2/\lambda_{\rm F}^2)  \phi_\perp(0)$. So contrarily to the classical Poisson spot (Eq.~\ref{eq_Poisson_spot}), it does not disappear when $\lambda$ approaches zero, as expected from geometrical optics. In Sect.~\ref{sec_Pluto_Triton} we see that for objects like Pluto or Triton, $\phi_{\rm Diff} (0)$ can be very large (some $10^4$-$10^5$) while $\lambda_{\rm F}^2/R_{\rm CF}$ is very small (a few meters).

The thickness $\Delta R_{\rm Diff}$ of the central flash layer can be defined as the interval of $R$ which most contributes to the integral in Eq.~\ref{eq_Fresnel_integral_g"_positive}, i.e. a few times $1/\sqrt{\varphi''(R_{\rm CF})}$.
Based on Eq.~\ref{eq_ddphi}, this provides a thickness of the order of 
\begin{equation}
\Delta R_{\rm Diff} \sim \lambda_{\rm F} \sqrt{\phi_\perp(0)} \approx \lambda_{\rm F} \sqrt{\frac{H}{R_{\rm CF}}},
\end{equation} 
in the general and isothermal cases, respectively.

The flash layer thus acts as an annular lens with radius $R_{\rm CF}$, width $\Delta R_{\rm Diff}$, and  a focal length of $\Delta$. The image of the star produced by this lens  in the shadow plane (with radial profile $\propto J_0^2(u)$) resembles, but is not identical to, the Airy disk (radial profile $\propto [J_1(u)/u]^2$) produced by a lens of radius $R_{\rm CF}$, where $u$ is defined in Eq.~\ref{eq_u}. A comparison between the Poisson spot and the Airy disk in given in Appendix~\ref{app_Poisson_vs_Airy}.

\subsection{Extended vicinity of the center}
\label{sec_dense_atmo_extended_vicinity}

For $r \gtrsim \lambda^2_{\rm F}/(\pi R_{\rm CF})$, and in the gray shaded region shown in the lower panel of Fig.~\ref{fig_tenuous_vs_dense}, the two stellar images are seen by the observer. The amplitude caused by the primary image (Eq.~\ref{eq_amplitude_a1}) is unchanged, while the stationary phase method can be used to calculate $a_2(r)$, which yields 
\begin{eqnarray}
 \displaystyle a_1(r) \approx  \sqrt{\phi_1 (r)} \exp[ {\rm i} \varphi_1(R_1)],  \label{eq_amplitude_a1_bis}\\
 \nonumber \\
 \displaystyle a_2(r) \approx   -{\rm i} \sqrt{\phi_2 (r)} \exp[ {\rm i} \varphi_2(R_2))]. \label{eq_amplitude_a2}
\end{eqnarray} 
This provides the total normalized flux, 
\begin{equation}
\phi_{\rm Diff} (r)= 
|a_1 + a_2|^2 =
 \phi_1 + \phi_2 + 2\sqrt{\phi_1 \phi_2} \sin \left( \varphi_2 - \varphi_1 \right),
\label{eq_phi_phi1_phi2}
\end{equation} 
where $\varphi_1$ and $\varphi_2$ are short-hand notations for  $\varphi_1(R_1)$ and $\varphi_2(R_2)$, respectively.

We retrieve the $-{\rm i}$ prefactor in the expression of $a_2$ (hence its quadrature advance), which was already discussed in Sect.~\ref{sec_wave_optics}. This explains the presence of the term $\sin (\varphi_2 - \varphi_1)$ in Eq.~\ref{eq_phi_phi1_phi2}, instead of the usual factor $\cos (\varphi_2 - \varphi_1)$ encountered in the two-slit Young's experiment. 

Apart from this difference, we obtain the simple result that for $r \gtrsim \lambda_{\rm F}^2/R_{\rm CF}$, the Poisson fringes merely result from the interferences of two sources with respective fluxes $\phi_1$ and $\phi_2$. 
If one of the two images is blocked (e.g., if it is absorbed by an atmospheric haze layer), or if the secondary image does not exist
(as in Sect.~\ref{sec_tenuous_atmo}), then the received flux reduces to $\phi_1$ or $\phi_2$, with no more interference pattern.

For $\lambda_{\rm P} \ll r \ll R_1, R_2$, we have $R_1 \approx R_2 \approx R_{\rm CF}$ and $\varphi_2(R_2) - \varphi_1(R_1) \approx 2\pi R_{\rm CF} r/\lambda_{\rm F}^2$. Moreover, for an isothermal atmosphere, we have $\phi_1 \approx \phi_2 \approx H/r$ near centrality, so that 
\begin{equation}
\phi_{\rm Diff} (r) =  \frac{2H}{r} + \frac{2H}{r} \sin \left( \frac{2\pi R_{\rm CF}r}{\lambda_{\rm F}^2} \right).
\label{eq_phi_near_center}
\end{equation} 
The flux oscillates around the average value $2H/r$ (consistent with Eq.~\ref{eq_flash_geo_optics_iso}) with a fringe spacing 
\begin{equation}
\lambda_{\rm P}=  \frac{\lambda_{\rm F}^2}{R_{\rm CF}} = \frac{\lambda \Delta}{2R_{\rm CF}},
\label{eq_inter_fringe_atmo}
\end{equation} 
almost identical to Eq.~\ref{eq_inter_fringe} and to the fringe separation obtained in the Young's experiment with two slits separated by $2R_{\rm CF}$.

As $r$ increases, the secondary image gets deeper in the atmosphere while the primary image gets higher, so that $R_2 < R_1$ and $\varphi''_2(R_2) > \varphi''_1(R_1)$, hence $\phi_1 > \phi_2$. Consequently the visibility of the fringes,
\begin{equation}
V = \frac{2\sqrt{\phi_1 \phi_2}}{\phi_1 + \phi_2},
\label{eq_visibility}
\end{equation}
decreases from unity in the vicinity of the shadow center, where $\phi_1 \approx \phi_2$, and becomes weak when $\phi_1 \gg \phi_2$. 

\subsection{Outer domain}

As $r$ becomes larger than $r'_0$ (see the lower panel of Fig.~\ref{fig_tenuous_vs_dense}), the secondary image disappears behind the limb, which produces Fresnel diffraction fringes just before its disappearance. As $r$ continues to increase, the observer enters the outer domain of the shadow where the secondary wave cannot be approximated by geometrical optics (modified by the $-i$ prefactor as in Eq.~\ref{eq_amplitude_a2}), but is rather a weak, purely diffracted wave. As $\varphi_2(R)$ no longer admits a stationary value for $R > R_0$, the integral in Eq.~\ref{eq_amplitude_a2_ori} must be evaluated using Sommerfeld's lemma. From Eq.~\ref{eq_Fr_asymptot_minus_infty}, we obtain 
\begin{equation}
a_2(r) \approx \frac{1}{\pi} \sqrt{\frac{R_0}{2r}} \left( \frac{\lambda_{\rm F}}{r-r'_0} \right) \exp \left\{ {\rm i} \left[ \varphi_2(R_0) - \frac{\pi}{4} \right] \right\}.
\end{equation} 
Using the value of $a_1(r)$ given by Eq.~\ref{eq_amplitude_a1_bis}, we obtain the flux observed in the outer domain, 
\begin{equation}
\begin{array}{l}
\displaystyle
\phi_{\rm out} (r)= |a_1(r) + a_2(r)|^2 = \phi_1 + \frac{R_0}{2\pi^2 r} \left( \frac{\lambda_{\rm F}}{r-r'_0} \right)^2 + \\ \\ 
\displaystyle
\sqrt{\phi_1} \sqrt{\frac{2 R_0}{\pi^2 r}} \left( \frac{\lambda_{\rm F}}{r-r'_0} \right) \cos \left[ \varphi_2(R_0) - \varphi_1(R_1) - \frac{\pi}{4} \right]. \\
\end{array}
\end{equation} 
The term $1/(r-r'_0)$ describes the evanescent side of the Fresnel fringes associated with the disappearance of the secondary image, over a scale in $r$ of $\sim \lambda_{\rm F} \sqrt{R_0/r'_0}/[\pi \sqrt{\phi_2(r'_0})]$, by similarity with Eq.~\ref{eq_phi_edge_half_light}.

\section{Effect of stellar diameter in geometrical optics}
\label{sec_stellar_diam}

Equation~\ref{eq_flash_geo_optics_iso} assumes geometrical optics with a point-like star and predicts an infinite flux at the exact alignment of the star, the occulter, and the observer ($r=0$). In practice, however, the star appears as a disk with a finite radius $r_*$ when projected at the occulter distance.

As discussed later (Sect.~\ref{sec_discussion}), during ground-based stellar occultations by remote objects observed in the visible, diffraction and interference effects are usually averaged out by the stellar diameter. The flash profile can then be calculated using geometrical optics in the case of a dense atmosphere.
In this context, we note
that the atmosphere forms two distorted images of this disk. Clausius' theorem states that for a transparent atmosphere, the radiance (also called  specific intensity or brightness) of these images is conserved \citep{Fabry1929}. Consequently, if the radiance of the stellar photosphere is uniform, the received flux is proportional to the sum of the surface areas of these two images projected at distance $\Delta$.

This means in particular that near centrality the two images are flattened perpendicularly to the limb by a factor $\phi_\perp (0)$ (Eq.~\ref{eq_flash_geo_optics_perp}). At the same time, they are stretched by a factor $\approx R_{\rm CF}/r$ along the limb as long as $r > r_*$; see Eq.~\ref{eq_flash_geo_optics}.

\begin{figure}[!t]
\centerline{\includegraphics[totalheight=55mm,trim=0 0 0 0]{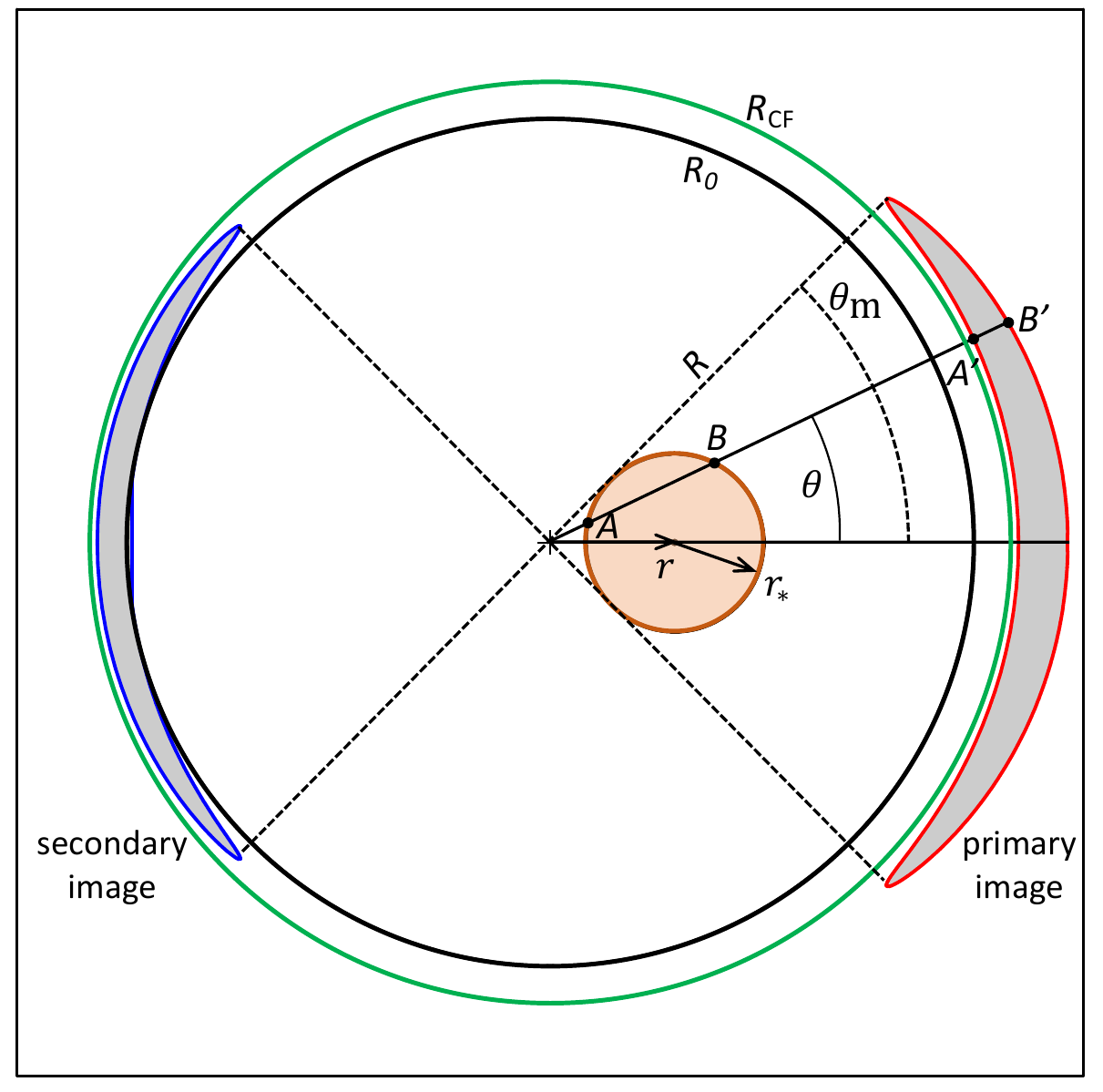}}
\centerline{\includegraphics[totalheight=55mm,trim=0 0 0 0]{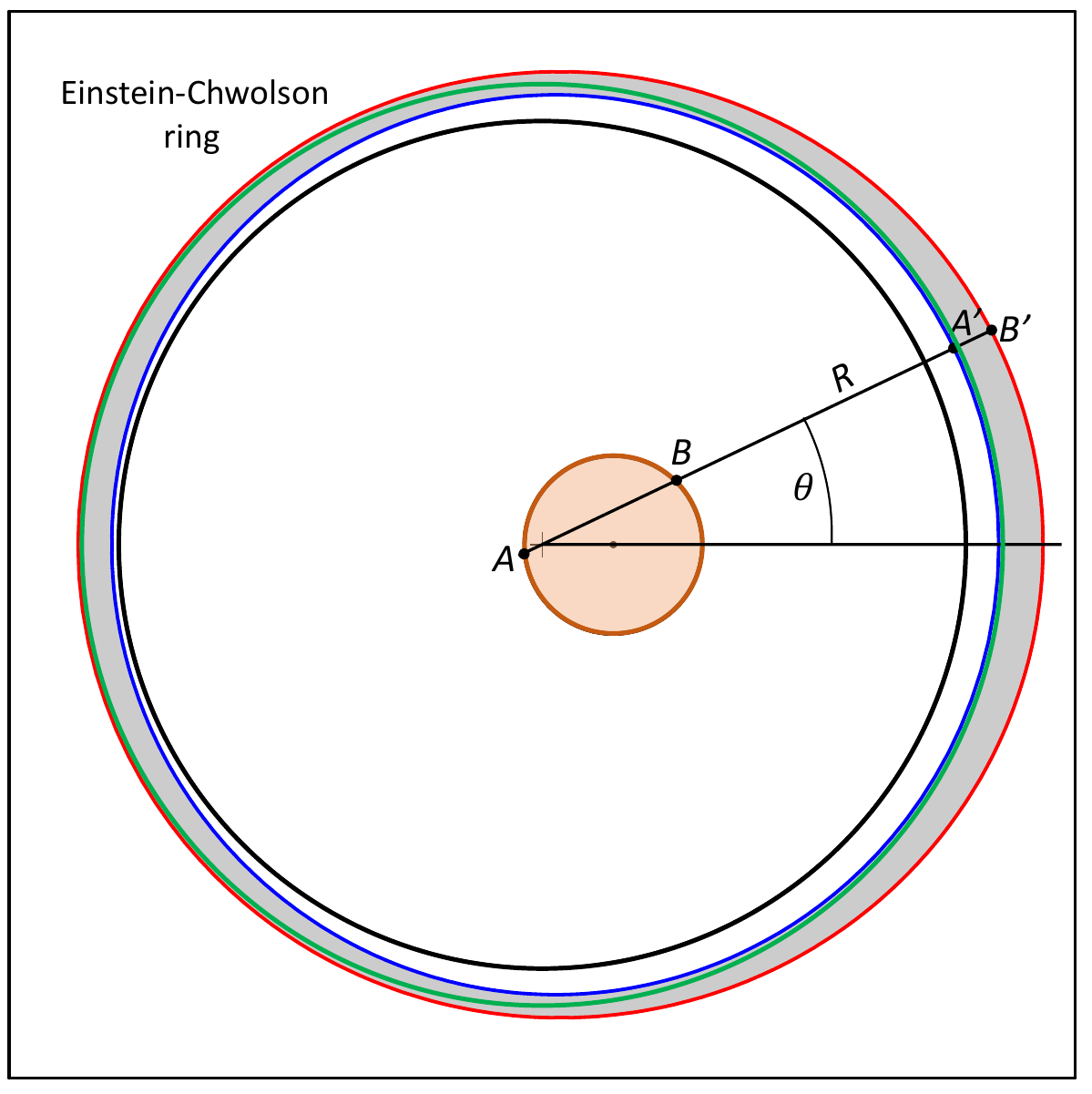}}
\caption{Star (orange disk) located behind an object of radius $R_0$, as seen by an observer at P (Fig.~\ref{fig_geo_occ}).
The central flash layer of radius $R_{\rm CF}$ is shown in green.
The star appears as a disk with radius $r_*$ projected at the body distance, its center being at distance $r$ from the projected body center. 
\it Upper panel\rm:  Case $r > r_*$, the atmosphere produces two images, a primary (secondary) image sketched as shaded the region delimited by the red (blue) line. 
In a given direction defined by the angle $\theta$, the two points A and B along the stellar limb have points A' and B' as images, respectively. 
\it Lower panel\rm: Case $r < r_*$, the two stellar images merge and form an Einstein-Chwolson ring. 
The apparent stellar radius $r_*$ and the atmospheric scale height $H$ have been greatly exaggerated for better visibility. 
In real cases, the stellar images are much more compressed.}
\label{fig_prim_second}
\end{figure}

\begin{figure}[!t]
\centerline{\includegraphics[totalheight=55mm,trim=0 0 0 0]{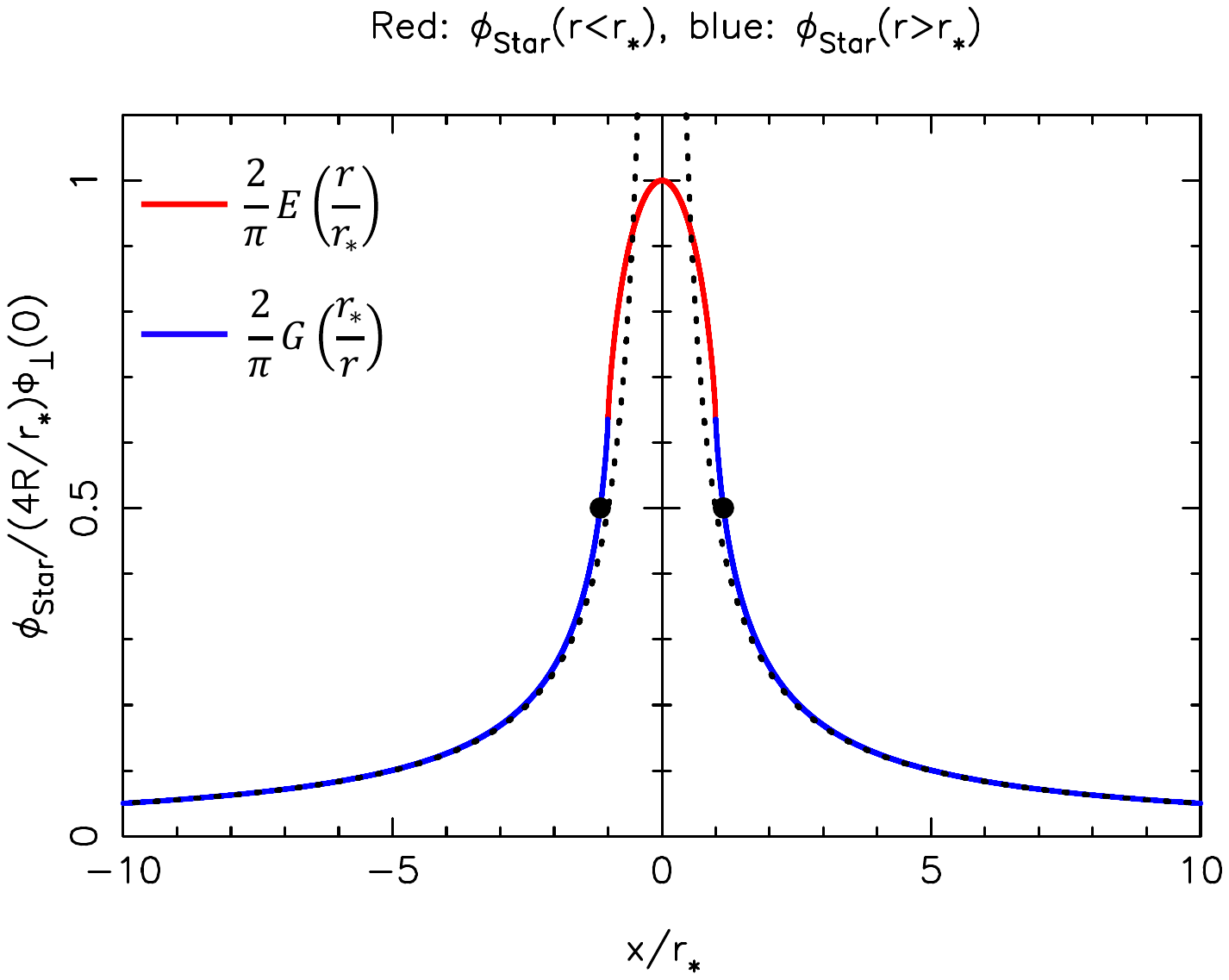}}
\caption{Central flash resulting from a finite stellar radius $r_*$ projected at the body distance, in the geometrical optics regime. 
The plot has been generated by used Eq.~\ref{eq_phi_r_gt_r*} (blue part) and Eq.~\ref{eq_phi_r_lt_r*} (red part). The functions $E$ and $G$ are given in Eqs.~\ref{eq_G} and Table~\ref{tab_CF_formulae}. 
The values along the horizontal axis have been normalized to the stellar radius $r_*$, while the flux along the vertical axis has been normalized to its peak value $4(R/r_*) \phi_\perp(0)$ (Eq.~\ref{eq_max_CF_stellar_diam}). 
The two bullets show the values $x/r_*= \pm 1.14$ where the flash reaches half of its maximum value. 
The dotted line shows the diverging flash profile produced by a point-like star in the geometrical optics approximation (Eq.~\ref{eq_flash_geo_optics_iso}).} 
\label{fig_CF_stellar_diameter}
\end{figure}

We first consider the case $r \geq r_*$ and a segment $AB$ defined by two points $A$ and $B$ along the stellar limb and inclined by an angle $\theta$ with respect to the reference axis (Fig.~\ref{fig_prim_second}). This segment has an image $A'B'$ with a width $\Delta R = \phi_\perp(0) AB$ across the stellar image. Elementary calculations show that $AB= 2\sqrt{r_*^2 - r^2 \sin^2 \theta}$, so that the surface area of any one of the stellar images is 
\begin{equation}
S \approx  \phi_\perp (0) R_{\rm CF} \int_{-\theta_{\rm m}}^{+\theta_{\rm m}} 2 \sqrt{r_*^2 - r^2 \sin^2 \theta} \, {\rm d}\theta,
\label{eq_S}
\end{equation} 
where $\theta_{\rm m}= \arcsin(r_*/r)$. 
Near the shadow center, the two stellar images have similar surface areas, so the quantity above must be multiplied by two. Considering the parity of the function $\sin^2 \theta$, and recalling that here we assume that the disk has a uniform radiance and that the atmosphere is transparent, the normalized central flash profile $\phi_{\rm Star}(r) = 2S/(\pi r_*^2)$ is eventually given by 
\begin{equation}
 \begin{array}{l}
 \phi_{\rm Star} (r \geq r_*)  \displaystyle =  \left( \frac{8R_{\rm CF}}{\pi r_*}\right) \phi_\perp(0) \, G \left( \frac{r_*}{r} \right), {\rm where} \\ 
  \\
 \displaystyle G(w) := \frac{1}{w} \int_0^{\arcsin(w)} \sqrt{w^2 - \sin^2 \theta} \, {\rm d}\theta.
 \\
  \end{array}
 \label{eq_phi_r_gt_r*}
\end{equation} 
Using the identity (\citealt{Gradshteyn1980}) 
\begin{equation}
 \begin{array}{l}
 \displaystyle
 \int_0^{\theta_0} \!\! \sqrt{1-p^2 \sin^2 \theta} \, {\rm d}\theta = 
 p \!\! \int_0^{\arcsin(p \sin \theta_0)} \!\! \sqrt{1 - (\sin^2\theta/p^2)} \, {\rm d}\theta \\
 \\
 \displaystyle 
 + \frac{1-p^2}{p} \int_0^{\arcsin(p \sin \theta_0)} \frac{{\rm d}\theta}{\sqrt{ 1- (\sin^2\theta/p^2)}} {\rm~(for~} p^2 > 1), \\
 \end{array}
\end{equation} 
posing $p= 1/w$ and taking $\theta_0= \arcsin{w}$, we obtain 
\begin{equation}
\begin{array}{ll}
G(w)= &
\displaystyle
\frac{1}{w} \left[  E(w) - (1 - w^2) F(w) \right], \\
 & \\
{\rm where} & 
\displaystyle E(w) := \int_0^{\pi/2} \sqrt{1 - w^2 \sin^2 \theta} \, {\rm d}\theta \\
 & \\
{\rm and} & 
\displaystyle F(w) := \int_0^{\pi/2} \frac{{\rm d}\theta}{\sqrt{1 - w^2 \sin^2 \theta}} \\
\end{array}
\label{eq_G}
\end{equation} 
are the complete elliptic integrals of the second and first kinds, respectively. 
It can be verified that as $w$ approaches zero, $G(w) \approx w\pi/4$. Thus, $\phi_{\rm Star} (r \gg r_*) \approx 2(R_{\rm CF}/r) \phi_\perp(0)$. For an isothermal atmosphere, $\phi_\perp(0) \approx H/R_{\rm CF}$, so that $\phi_{\rm Star} (r \gg r_*) \approx 2H/r$, as predicted by Eq.~\ref{eq_flash_geo_optics_iso}. 
For $ r= r_*$, we have $w=1$, so we obtain from $G(1)=1$,
\begin{equation}
\phi_{\rm Star} (r = r_*) = \left( \frac{8R_{\rm CF}}{\pi r_*} \right) \phi_\perp(0) =  \frac{8H}{\pi r_*},
\end{equation}
in the general and isothermal cases, respectively.

When $r \leq r_*$, the two stellar images merge and form an Einstein-Chwolson ring (Fig.~\ref{fig_prim_second}), similar to the gravitational lensing of distant galaxies by a foreground mass. The integral in Eq.~\ref{eq_S} must then be carried out from 0 to $2\pi$ (instead of from $-\theta_{\rm m}$ to $+\theta_{\rm m}$), providing 
\begin{equation}
\phi_{\rm Star} (r \leq r_*) = \left(  \frac{8R_{\rm CF}}{\pi r_*} \right) \phi_\perp(0) E \left( \frac{r}{r_*} \right).
\label{eq_phi_r_lt_r*}
\end{equation} 
From $E(0)=\pi/2$, we obtain a finite value for the flux at $r=0$, 
\begin{equation}
\phi_{\rm Star} (0) = \left( \frac{4R_{\rm CF}}{r_*} \right) \phi_\perp(0)
=  \frac{4H}{r_*}  = \frac{8H}{D_*},
\label{eq_max_CF_stellar_diam}
\end{equation} 
in the general and isothermal cases, respectively, where $D_* := 2 r_*$ is the apparent stellar diameter (a result already mentioned by \citealt{Elliot1977}
for an isothermal atmosphere). We note that we retrieve the divergence of the flux at $r=0$ (Eq.~\ref{eq_flash_geo_optics}) as $D_*$ approaches zero.

The numerical integration of Eq.~\ref{eq_phi_r_gt_r*} shows that the flash reaches half of its peak flux at $r \approx 1.14 r_*$, which means that the full width at half-maximum (FWHM) of the flash is 
\begin{equation}
{\rm FWHM} (\phi_{\rm Star})= 1.14 D_*.
\label{eq_FWHM_stellar_diam}
\end{equation} 
The half-maximum value is not reached at the junction of the $E$ and $G$ functions,
but on the branch describing the $G$ function (Fig.~\ref{fig_CF_stellar_diameter}).
At that point, the two stellar images are separated, as in the upper panel of Fig.~\ref{fig_prim_second}. 
The various properties of the flash in the presence of a finite stellar radius $\lambda_{\rm P} \ll r_* \ll R_0$ are displayed in Fig.~\ref{fig_CF_stellar_diameter}, and the comparison with the case  $r_* \ll \lambda_{\rm P}$ is presented in Appendix~\ref{app_point_vs_finite_size}.

We recall that these results assume a stellar disk with uniform radiance.
The effects of limb darkening are examined in Appendix~\ref{app_limb_dark}. In particular, the height of the flash is changed according to Eq.~\ref{eq_phi_max_with_LD}. For practical cases (Eq.~\ref{eq_limb_darkening}), the numerical schemes that permit the evaluation of the flash profile (Eq.~\ref{eq_phi_LD_reduced}) are based on Eqs.~\ref{eq_int_I_du}-\ref{eq_int_f_I'_u_du}. 
Ignoring limb darkening, a summary of the flash properties for both a monochromatic point-like source with diffraction and a finite-size star in geometrical optics is given in Appendix~\ref{app_table_general}.

\section{Applications to Pluto and Triton}
\label{sec_Pluto_Triton}

\begin{table}[!t]
\caption{Parameters of central flashes caused by the atmospheres of Triton and Pluto; also see Table~\ref{tab_CF_formulae}.}
\label{tab_CF_Triton_Pluto}
\renewcommand{\arraystretch}{1.2} 
\begin{tabular}{lll}
\hline
\hline
 &Triton\tablefootmark{1} & Pluto\tablefootmark{2} \\
\hline
$\phi_\perp(0)$ & $1.1 \times 10^{-2}$ & $2.6 \times 10^{-3}$ \\
Mean limb radius                         & 1353 km & 1190 km \\
Flash layer radius $R_{\rm CF}$ & 1361 km & 1191 km \\
\hline
\multicolumn{3}{c}{Diffraction, point-like star, and monochromatic wave} \\
\hline
Thickness of  flash layer & 130 m & 60 m \\
Peak value\tablefootmark{3}  & $3 \times10^5$ &  $5 \times10^4$ \\
Diameter first dark fringe & 1.6 m & 1.8 m \\
Poisson fringe spacing $\lambda_{\rm P}$ & 1.1 m & 1.2 m \\
\hline
\multicolumn{3}{c}{Finite stellar diameter, geometrical optics} \\
\hline
Peak value\tablefootmark{4}     & 240 & 50  \\
Flash FWHM & 570 m & 570 m \\
\hline
\end{tabular}
\tablefoot{
The quantities listed here are calculated by assuming 
a projected stellar diameter $D_*=0.5$~km and
a Fresnel scale $\lambda_{\rm F} = 1.2$~km, typical of Pluto and Triton occultations.
\tablefoottext{1}{\cite{McKinnon1995} and \cite{Marques2022}}
\tablefoottext{2}{\cite{Hinson2017},  \cite{Dias2015} and \cite{Sicardy2021}.}
\tablefoottext{3}{From Eq.~\ref{eq_max_CF_diffrac}.}
\tablefoottext{4}{From Eq.~\ref{eq_max_CF_stellar_diam}.}
}
\end{table}

We now apply our results to the cases of Pluto's and Triton's atmospheres. We consider these two objects because their atmospheres are close to spherical and are essentially transparent, with modest local density fluctuations.
In contrast, other bodies like Titan or the giant planets have atmospheres that significantly depart from sphericity, and/or contain absorbing hazes, and/or harbor vigorous internal gravity waves that create a strong stellar scintillation that blurs the central flash profile (Sect.~\ref{sec_discussion}). 

Table~\ref{tab_CF_Triton_Pluto} lists the physical parameters of flashes caused by Pluto's and Triton's dense atmospheres, splitting the case of a flash dominated by diffraction with a point-like star and a monochromatic wave, and a flash dominated by the stellar diameter $D_*$ in the geometrical optics regime. 
Considering the typical geocentric distances $\Delta \approx 30$~au of Pluto and Triton, and assuming that observations are made in the visible ($\lambda \approx 0.6$~$\upmu$m), here we use a representative value $\lambda_{\rm F} = 1.2$~km for the Fresnel scale. The projected stellar dia\-meter $D_*$ depends on $\Delta$ and on the magnitude and spectral type of the occulted star. Here we use $D_* = 0.5$~km, which is typical of stars with magnitudes $\approx$13 for $\Delta \sim30$~au \citep{Sicardy2024}.

\begin{figure}[!t]
\centerline{\includegraphics[totalheight=45mm,trim=00  00 00 00]{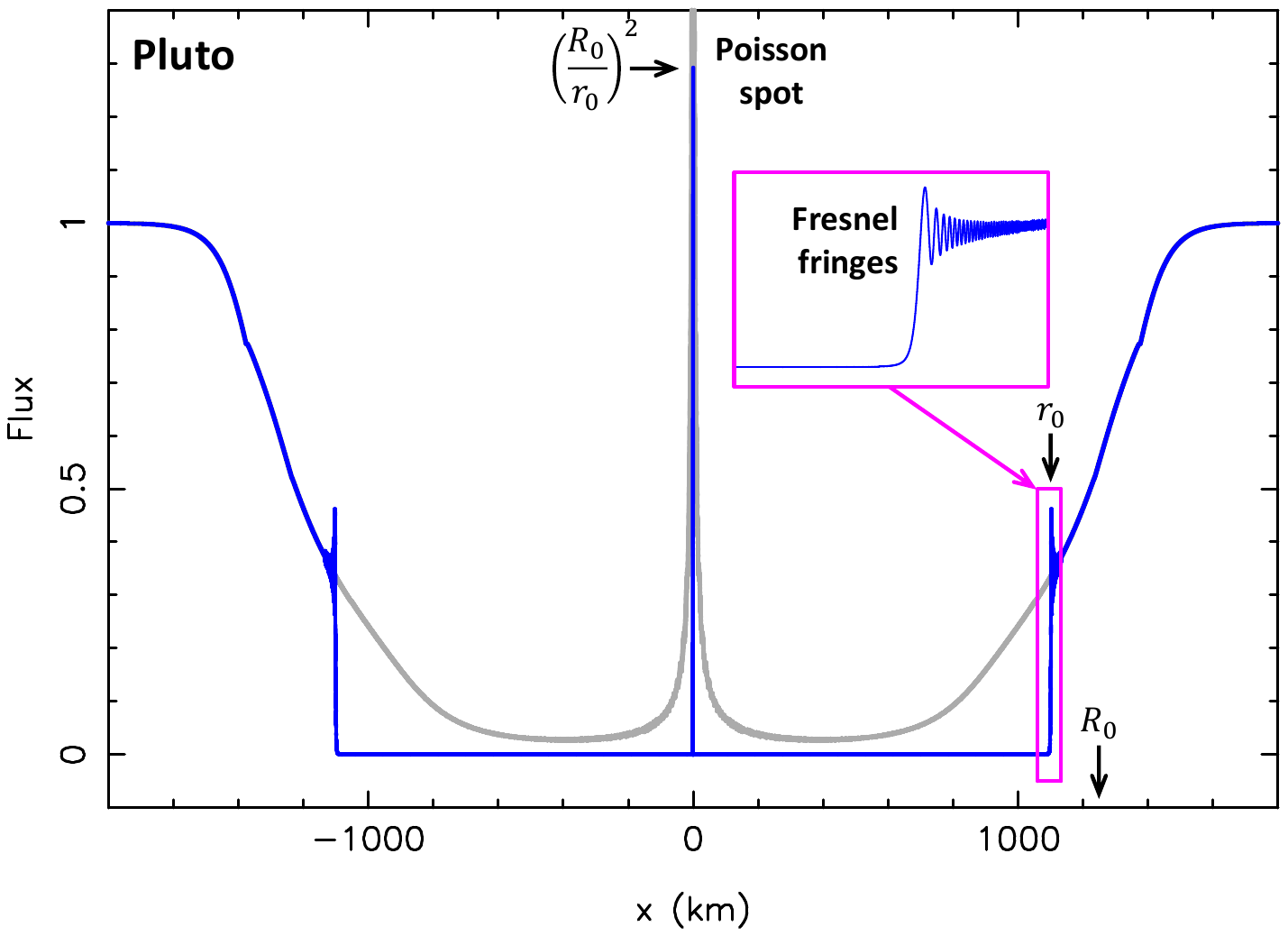}}
\centerline{\includegraphics[totalheight=45mm,trim=00  00 00 00]{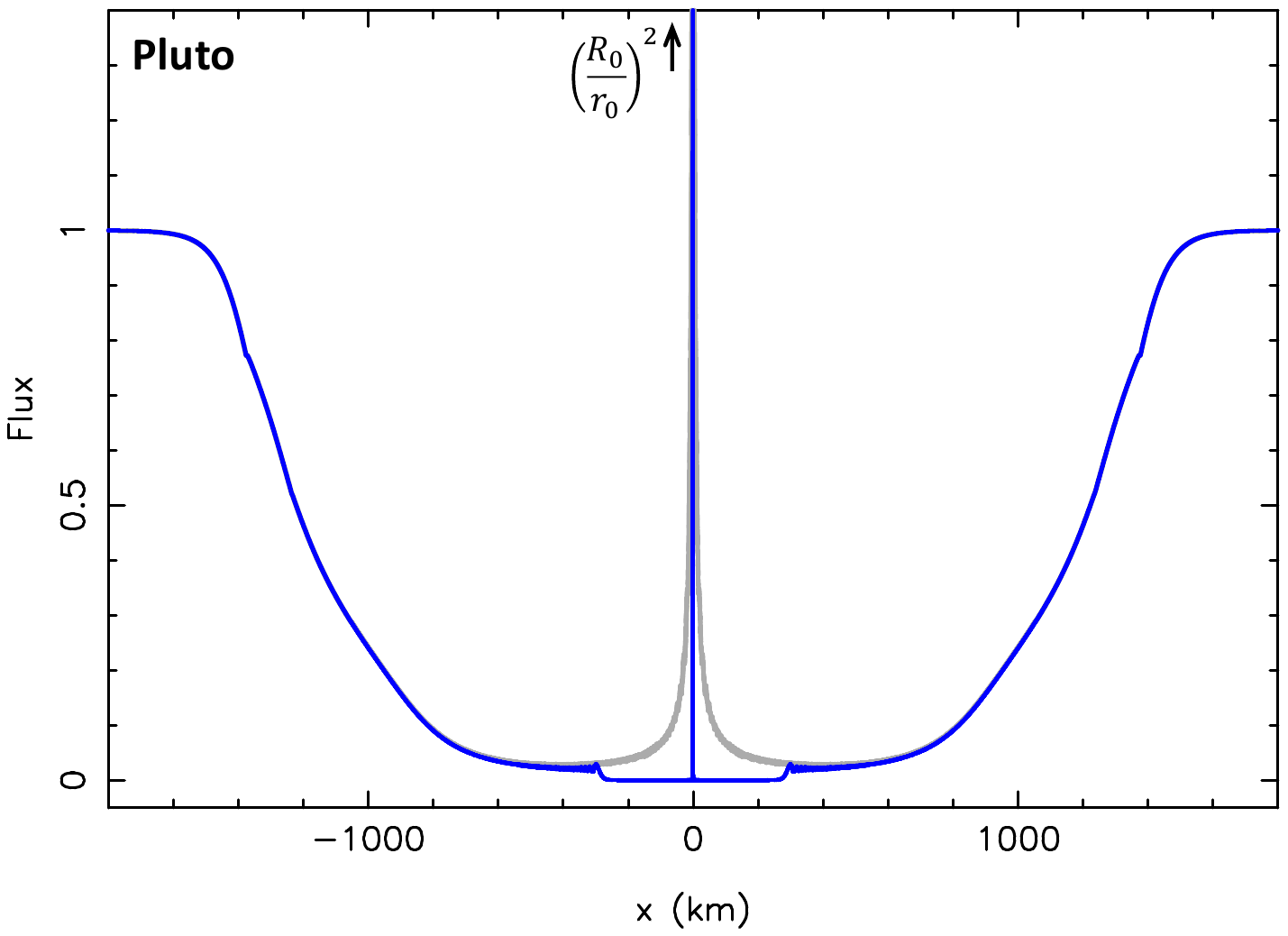}}
\centerline{\includegraphics[totalheight=45mm,trim=00  00 00 00]{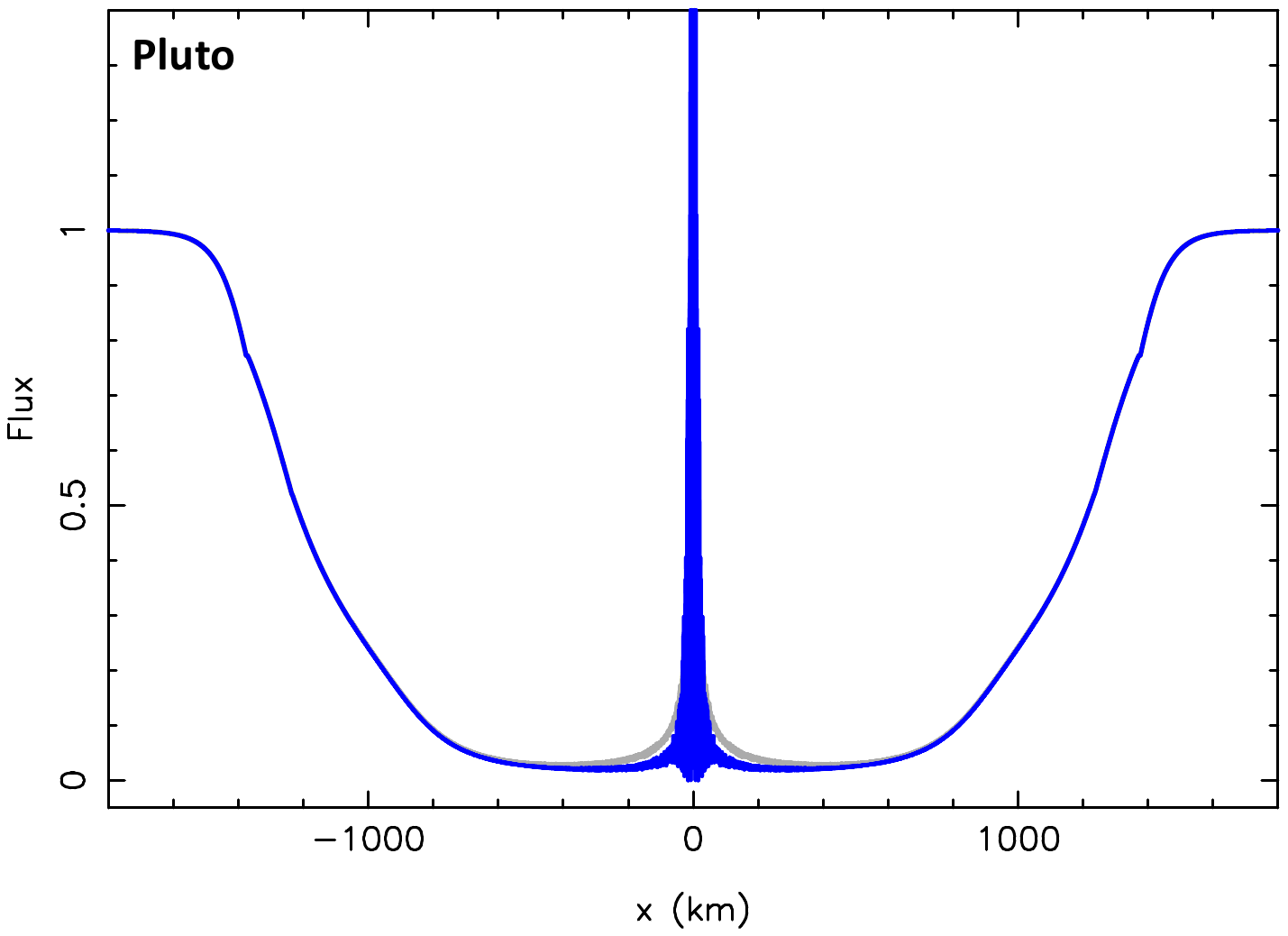}}
\caption{Synthetic light curves showing diametral occultations of a point-like star 
by Pluto, taking diffraction into account.The gray curve is the synthetic light curve generated by a ray-tracing code \citep{Dias2015} based on geometrical optics, with a Pluto geocentric distance $\Delta=33$~au and a wavelength $\lambda=0.6$~$\upmu$m, which yield a Fresnel scale $\lambda_{\rm F}=1.2$~km. The code uses the state of Pluto's atmosphere as of 6 June 2020, assuming pure nitrogen $N_2$ with a surface pressure of 12.2~$\upmu$bar and a Pluto radius of 1187~km \citep{Sicardy2021}.
\it Upper panel\rm:
Case of mask with radius $R_0=1250$~km placed in front of Pluto, creating a dark shadow of radius $r_0=1100$~km. The blue curve was generated using Eqs.~\ref{eq_Poisson_tenuous} and \ref{eq_flux_tenuous_atmosphere_r}. It shows the Fresnel fringes caused by diffraction as the primary image approaches the edge of the mask. The 50-km-wide magenta box shows an expanded view of the Fresnel diffraction, with the first fringes separated by $\lambda_{\rm F}/\sqrt{\phi_{\perp 1}(r_0)} \approx 6$~km (Eq.~\ref{eq_flux_tenuous_atmosphere_r}).
The Poisson spot is amplified by the factor $(R_0/r_0)^2$ compared to the classical Poisson spot created by an airless object (Fig.~\ref{fig_Fresnel_Poisson}). 
\it Middle panel\rm: 
Case $R_0= 1192$~km, the dark shadow is about to disappear. 
The central flash predicted by Eq.~\ref{eq_Poisson_tenuous} is now too high to fit in the figure. 
The Fresnel fringes are barely visible, as their signal is weak due to the smallness of the factor $\phi_1(r_0)$ in Eq.~\ref{eq_flux_tenuous_atmosphere_r}.
\it Lower panel\rm: 
Case $R_0= 1191$~km marks the transition between the tenuous and dense atmospheres. The dark shadow has disappeared and the highly amplified flash is now described by Eq.~\ref{eq_flash_J0_Fresnel_R}.}
\label{fig_z_flux_Pluto_06jun20_R0_1250_1192_1191_km}
\end{figure}

We first consider occultations by Pluto. In order to illustrate the various cases considered in Sect.~\ref{sec_tenuous_atmo}, 
we made thought experiments in which we changed the parameter $R_0$, and kept everything else equal. 
This means that we placed a circular opaque mask of radius $R_0$ in the plane of the sky in front of Pluto. This approach is admittedly artificial because $R_0$ is no longer the physical radius of the body, but it has the advantage of comparing the various cases using a unique atmospheric structure (partially hidden when $R_0 > 1190$~km).

The upper panel of Fig.~\ref{fig_z_flux_Pluto_06jun20_R0_1250_1192_1191_km} shows the effect of a mask with radius $R_0=1250$~km, and illustrates
the tenuous atmosphere case. This atmosphere creates a dark shadow of radius $r_0=1100$~km, with a Poisson spot amplified by the factor $(R_0/r_0)^2 \approx 1.3$ (Eq.~\ref{eq_Poisson_tenuous}). 
The middle and lower panels of the figure show the transition between the tenuous and the dense atmosphere cases, which occurs between $R_0=1192$~km and $R_0=1191$~km. 

\begin{figure}[!t]
\centerline{\includegraphics[totalheight=45mm,trim=00  00 00 00]{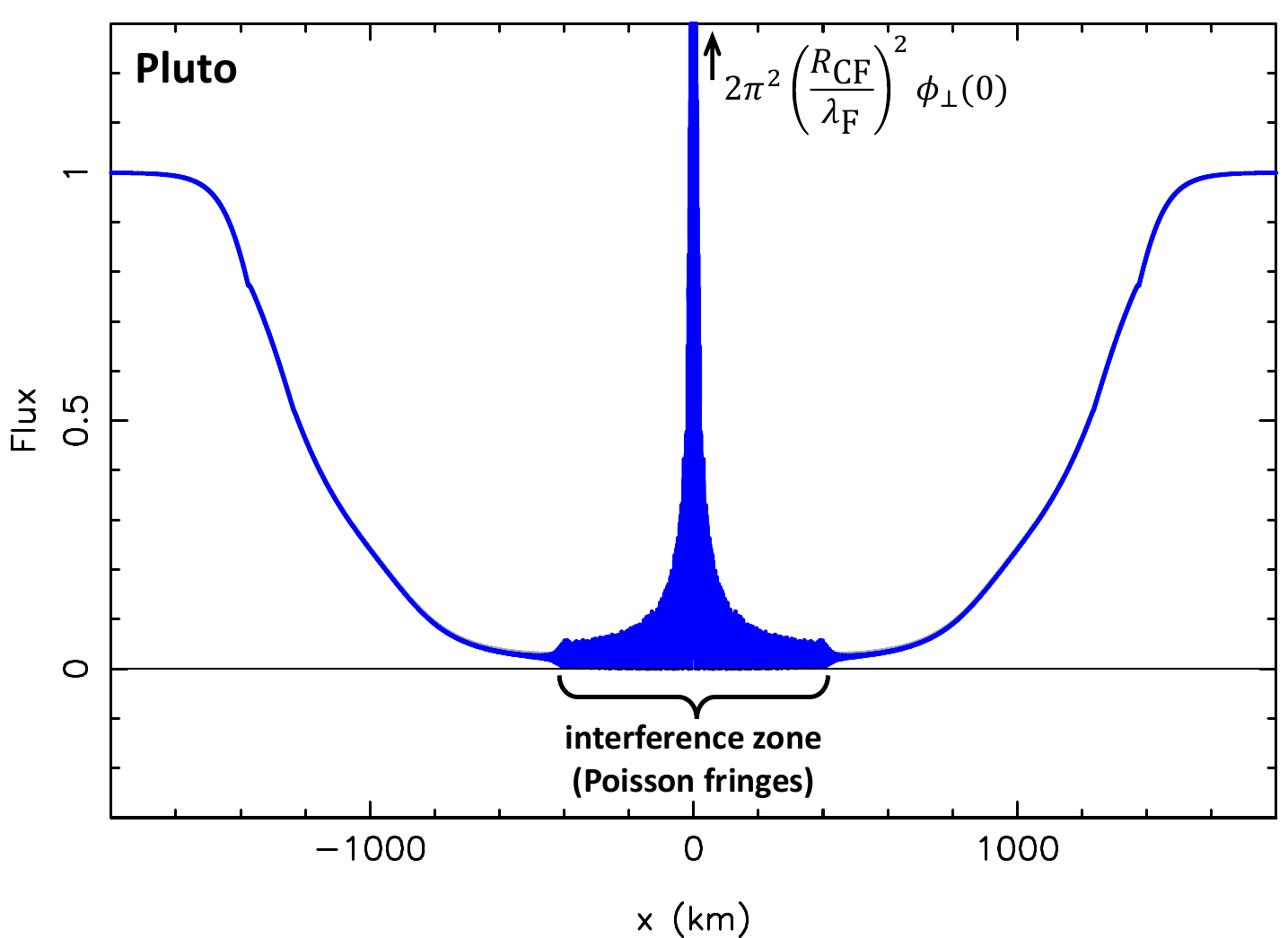}}
\centerline{\includegraphics[totalheight=47mm,trim=00  00 00 00]{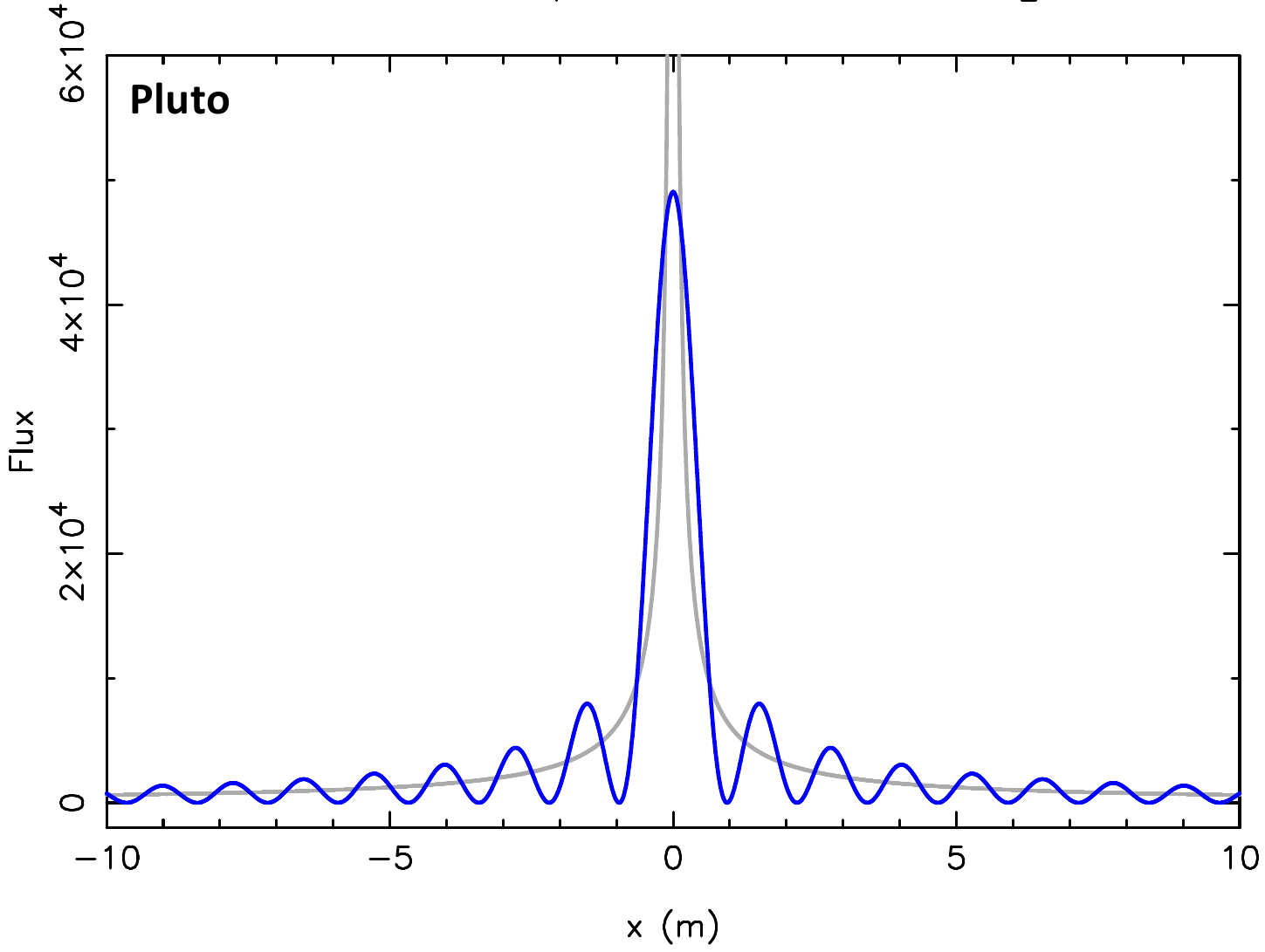}}
\caption{\it Upper panel\rm:
Same as Fig.~\ref{fig_z_flux_Pluto_06jun20_R0_1250_1192_1191_km} but for $R_0=1190$~km, which corresponds
to the dense atmosphere case. 
The flash now reaches its full height, given by Eq.~\ref{eq_flash_J0}. 
It is too high to fit in the figure, as it reaches a peak value of about $5 \times 10^4$; see Table~\ref{tab_CF_Triton_Pluto} and lower panel.
In the region with $|x| \lesssim 400$~km, the two stellar images interfere, causing Poisson fringes in the wings of the central flash. 
\it Lower panel\rm:
Expanded view of the central flash. Note the very large differences in both the horizontal and vertical scales compared to the upper panel. In particular position $x$ is counted in kilometers in the upper panel, and in meters in the lower panel.
At that scale, the Poisson fringes are resolved, with a separation of $\lambda_{\rm P}= \lambda_{\rm F}^2/R_0 \approx 1.2$~m (Eq.~\ref{eq_inter_fringe}).}
\label{fig_z_flux_Pluto_06jun20_R0_1190_km}
\end{figure}

\begin{figure}[!t]
\centerline{\includegraphics[totalheight=45mm,trim=00  00 00 00]{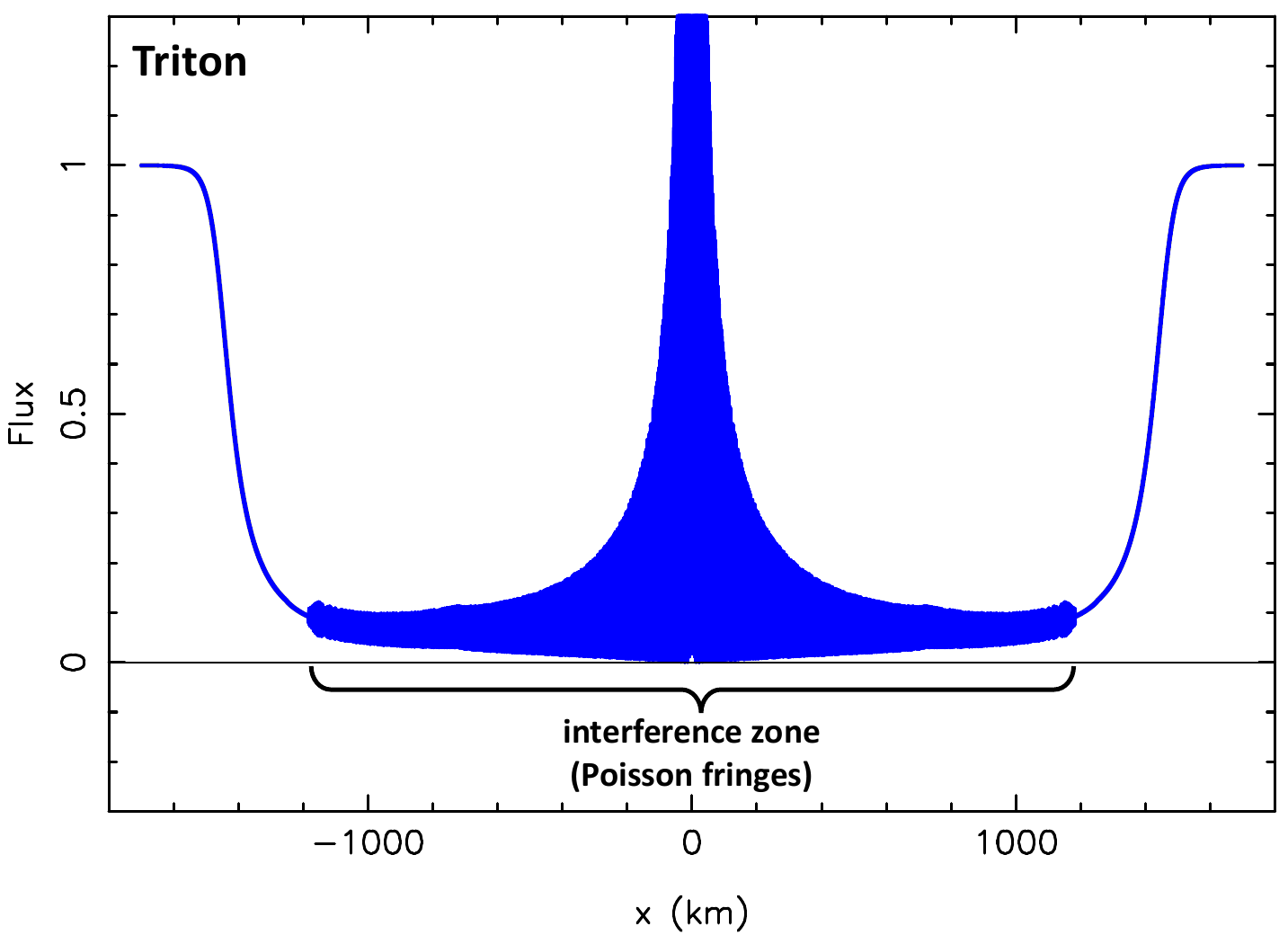}}
\caption{Same as Fig.~\ref{fig_z_flux_Pluto_06jun20_R0_1190_km} but for a Triton occultation observed at a geocentric distance $\Delta=29$~au. The atmospheric thermal structure used to generate this light curve is based on the 5 October 2017 occultation, assuming pure N$_2$, a surface pressure of 13.8~$\upmu$bar, and a Triton radius of $R_0=1353$~km \citep{Marques2022}.}
\label{fig_z_flux_Triton_05oct17_R0_1355_km}
\end{figure}

Figure~\ref{fig_z_flux_Pluto_06jun20_R0_1190_km} shows a synthetic profile using the actual average Pluto limb radius, $R_0=1190$~km \citep{Hinson2017}. A zone of full width $\sim$800~km in $x$ is now affected by the interferences given by  
the two stellar images. The same exercise for Triton provides a wide region of width $\sim$2400~km in $x$ where the two stellar images give interferences; see Fig.~\ref{fig_z_flux_Triton_05oct17_R0_1355_km}.

Concerning Pluto, Table~\ref{tab_CF_Triton_Pluto} shows that the central flash layer stays very close ($\sim$1~km) above the average limb of the dwarf planet. More precisely, it lies $\sim$4~km above the surface of the depression Sputnik Planitia (\citealt{Hinson2017}), which controls Pluto's general atmospheric pressure. Ray-tracing calculations show that the transition to the tenuous atmosphere regime will occur for a surface pressure of about 6~$\upmu$bar. According to seasonal volatile transport models, this should happen in the decade 2050-2060 \citep{Sicardy2021}. Occultations should then show a small drop of signal at mid-occultation, as illustrated in the middle panel of Fig.~\ref{fig_z_flux_Pluto_06jun20_R0_1250_1192_1191_km}. However, as significant topographic features of height $\sim$5~km with respect to the average radius are present on Pluto's surface \citep{Nimmo2017}, short drops in signal might be observed in the future at the bottom of Pluto occultation light curves as the primary and/or secondary images hit the top of these features.

\section{Discussion}
\label{sec_discussion}

\begin{figure}[!t]
\centerline{\includegraphics[totalheight=45mm,trim=00  00 00 00]{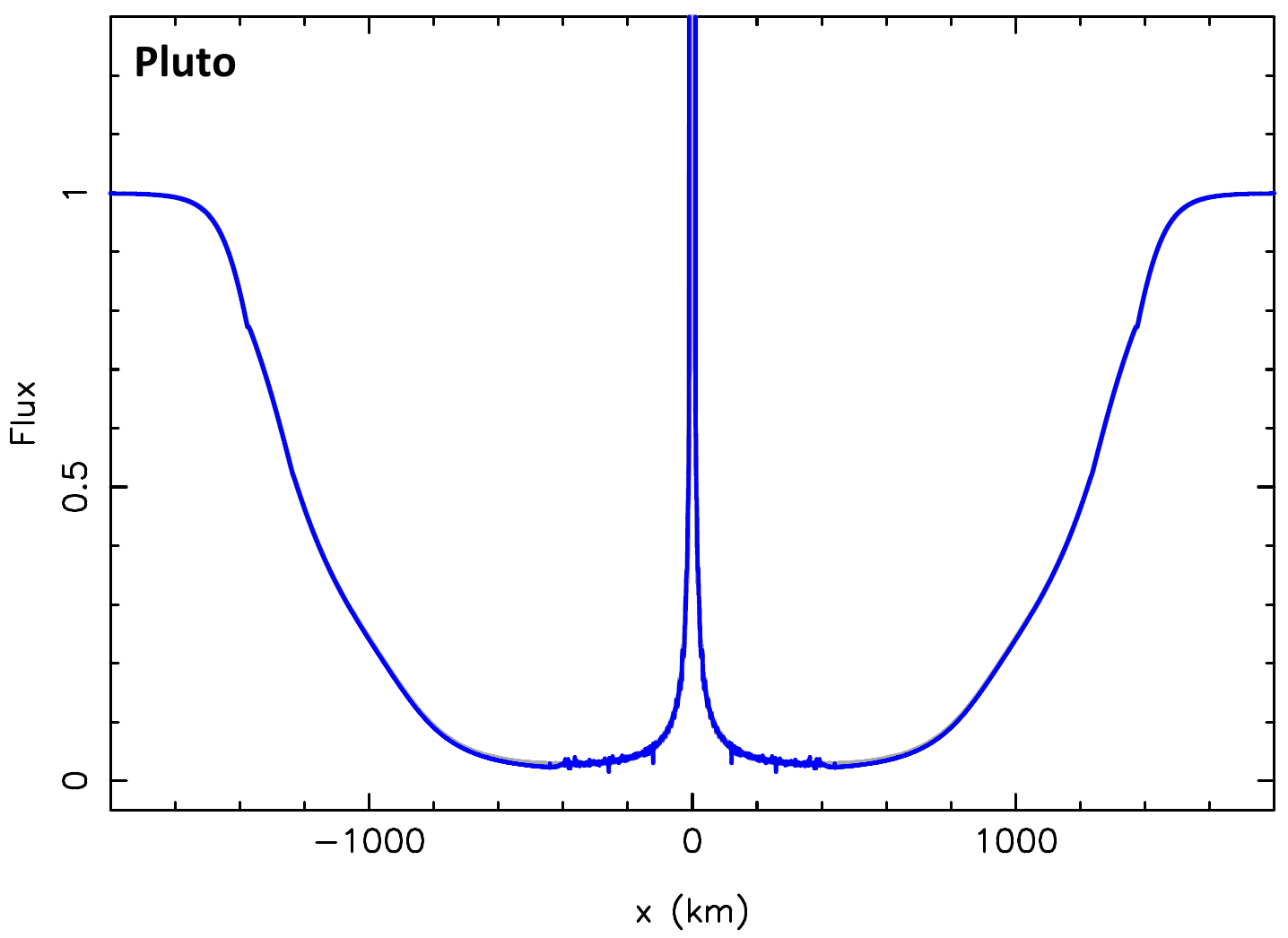}}
\caption{Same as the upper panel of Fig.~\ref{fig_z_flux_Pluto_06jun20_R0_1190_km}, except that the flux has been binned over radial intervals of 1~km in $x$, representative of the best time resolution currently obtained during Earth-based occultations by Pluto. The Poisson fringes are averaged out and are not visible anymore. The light curve is now undistinguishable from the geometrical optics model displayed in gray in the upper panel of Fig.~\ref{fig_z_flux_Pluto_06jun20_R0_1250_1192_1191_km}, except that in this example the flux remains finite with a value of about 990 at the center.}
\label{fig_z_flux_Pluto_06jun20_R0_1190_km_smoothed}
\end{figure}

The examples of Pluto and Triton occultations considered in the previous section concern Earth-based events observed in the visible, so that $\lambda_{\rm F} \sim 1$~km and $R_0 \sim 1000$~km. Consequently, $\lambda_{\rm F}/\sqrt{\phi_\perp}$ extends over several kilometers, while  $\lambda_{\rm F}^2/R_0$ represents about a meter. 
This means that Fresnel fringes can be resolved during occultations by outer Solar System objects (e.g., \citealt{Pereira2023}), while resolving Poisson fringes is much more challenging. 
A first reason for this is that the shadows of remote bodies typically move on Earth's surface at some 20~km~s$^{-1}$, so that resolving the Poisson fringes requires an acquisition rate of tens of thousands of images per second. With stars fainter than magnitude 13, this means an extremely small signal per exposure. 
The best resolutions currently attained in the shadow for ground-based occultation is about 1~km. At that scale, the Poisson fringes are averaged out and cannot be seen in occultation light curves; see Fig.~\ref{fig_z_flux_Pluto_06jun20_R0_1190_km_smoothed}.

\begin{figure}[!h]
\centerline{\includegraphics[totalheight=0045mm,trim=0 0 0 0]{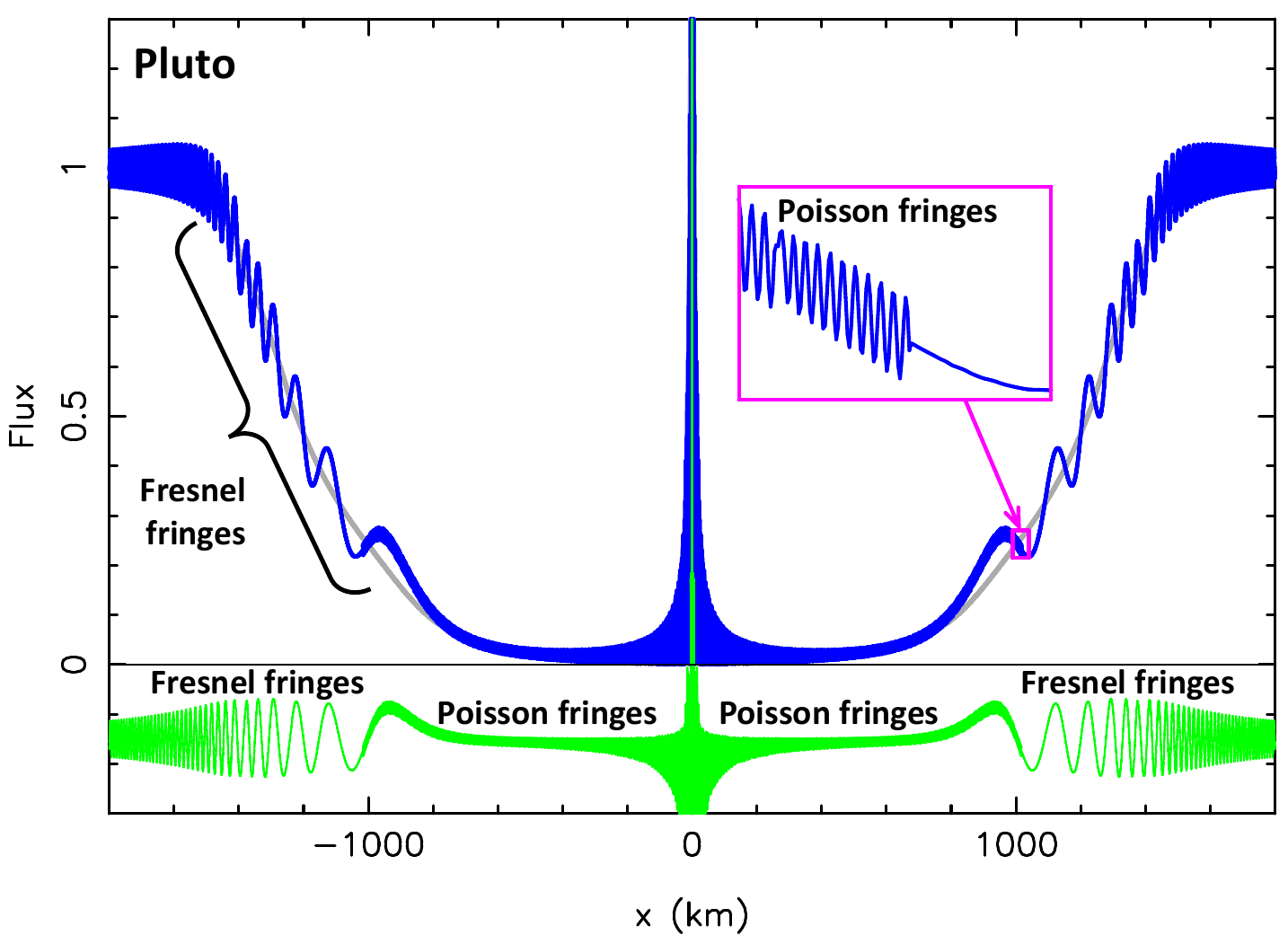}}
\centerline{\includegraphics[totalheight=0058mm,trim=0 0 0 0]{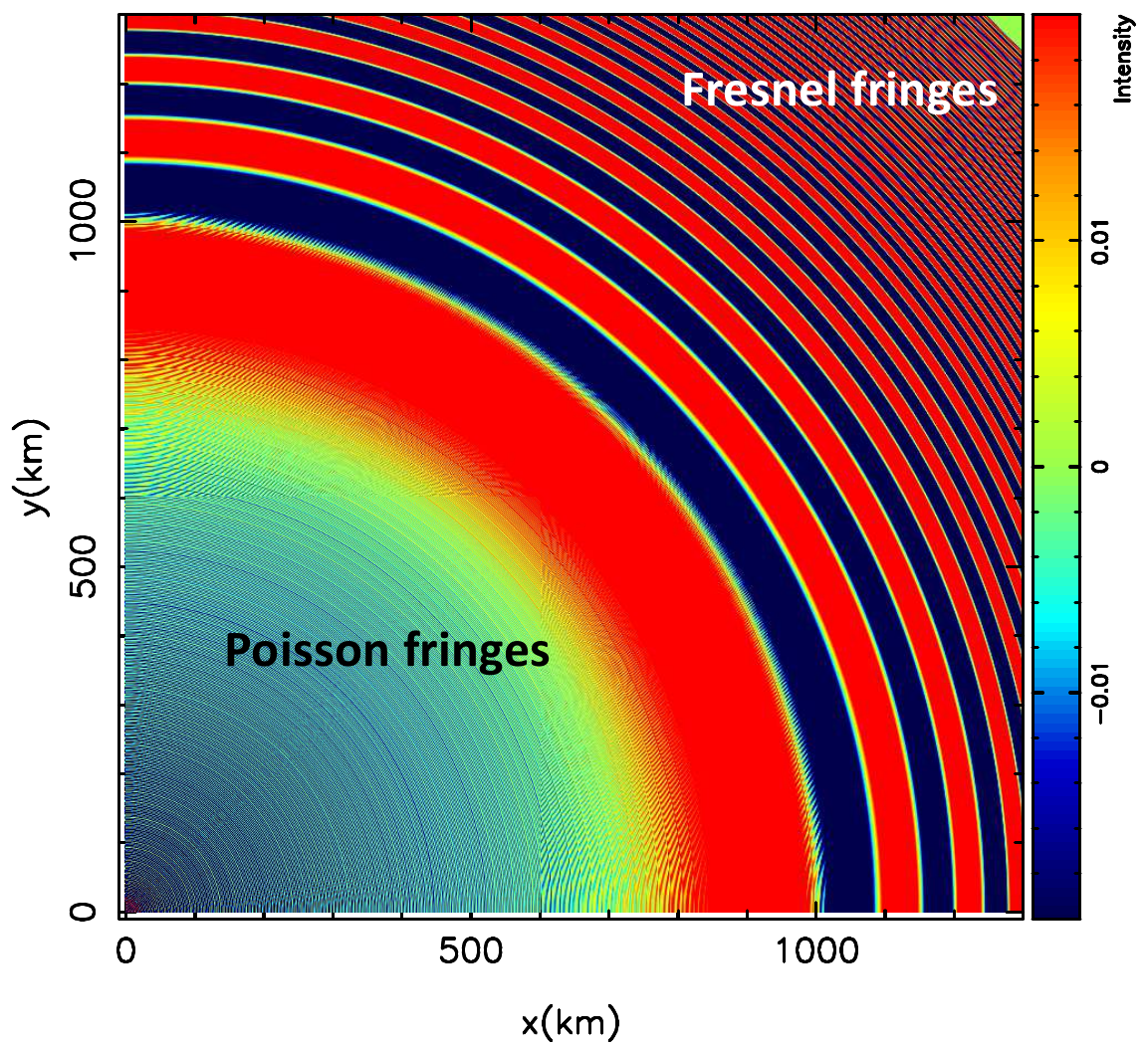}}
\centerline{\includegraphics[totalheight=0058mm,trim=0 0 0 0]{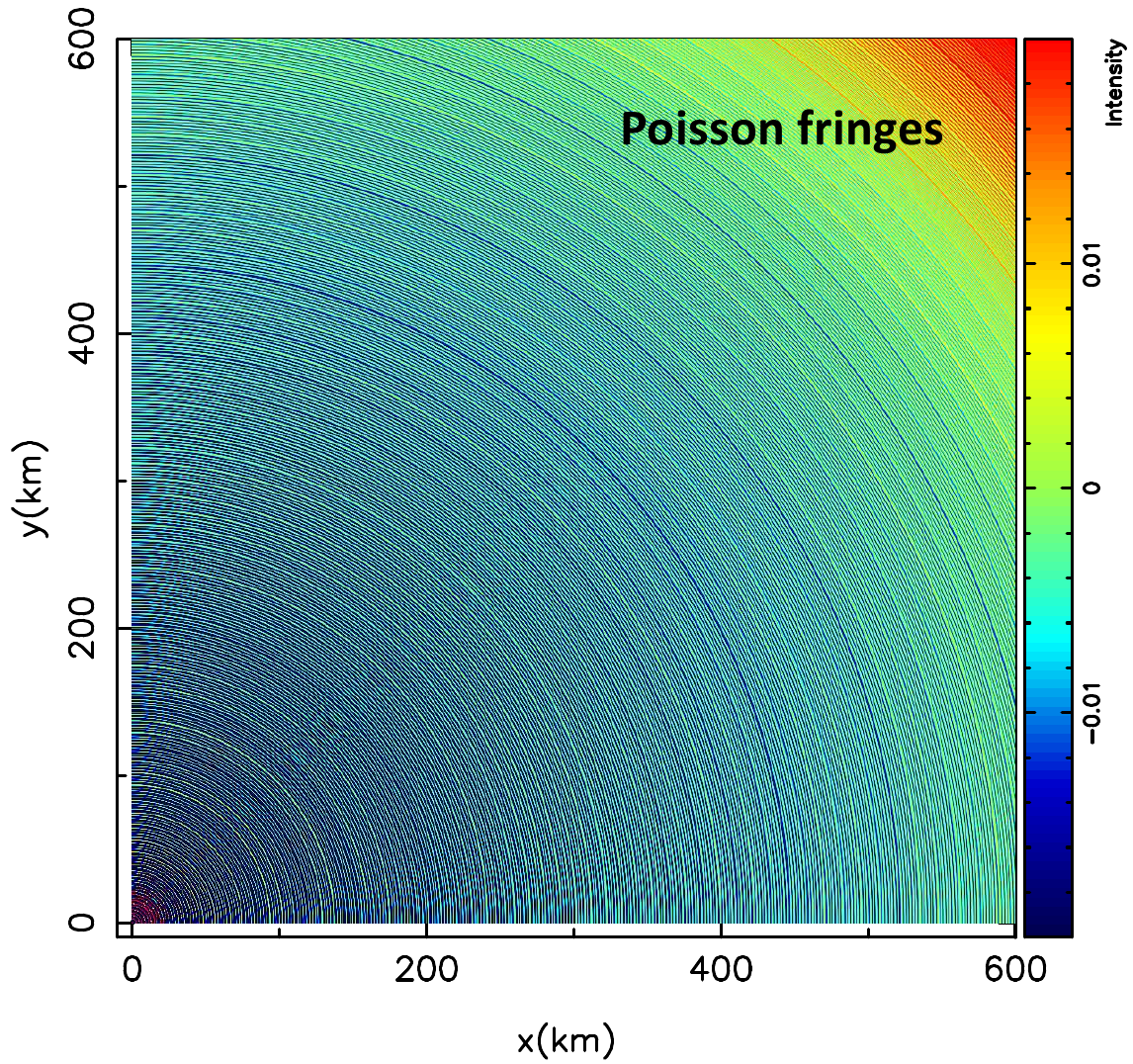}}
\caption{\it Upper panel\rm:
Same as Fig.~\ref{fig_z_flux_Pluto_06jun20_R0_1190_km} but with a larger wavelength $\lambda=1$~mm, hence a Fresnel scale $\lambda_{\rm F}=50$~km. The Fresnel fringes seen at the outer part of the shadow ($|x| \gtrsim 1020$~km) are separated by several tens of kilometers, while the Poisson fringes appearing at $|x| \lesssim 1020$~km, have separations of about 2~km and are not resolved in the panel. 
The 50-km-wide box (magenta) shows the resolved Poisson fringes vanishing as the secondary image disappears behind the limb. 
The green curve is the difference between the blue curve (wave optics) and the gray curve (geometrical optics), displaced vertically by -0.15 for better viewing. 
\it Middle panel\rm:
2D map based on the green curve of the upper panel, enhancing the difference between the Fresnel and Poisson fringes.
\it Lower panel\rm:
Enlargement of the middle panel, which reveals the fine striations caused by the Poisson fringes.}
\label{fig_z_flux_Pluto_06jun20_1mm}
\end{figure}

\begin{figure}[!t]
\centerline{\includegraphics[totalheight=90mm,trim=0 0 0 0]{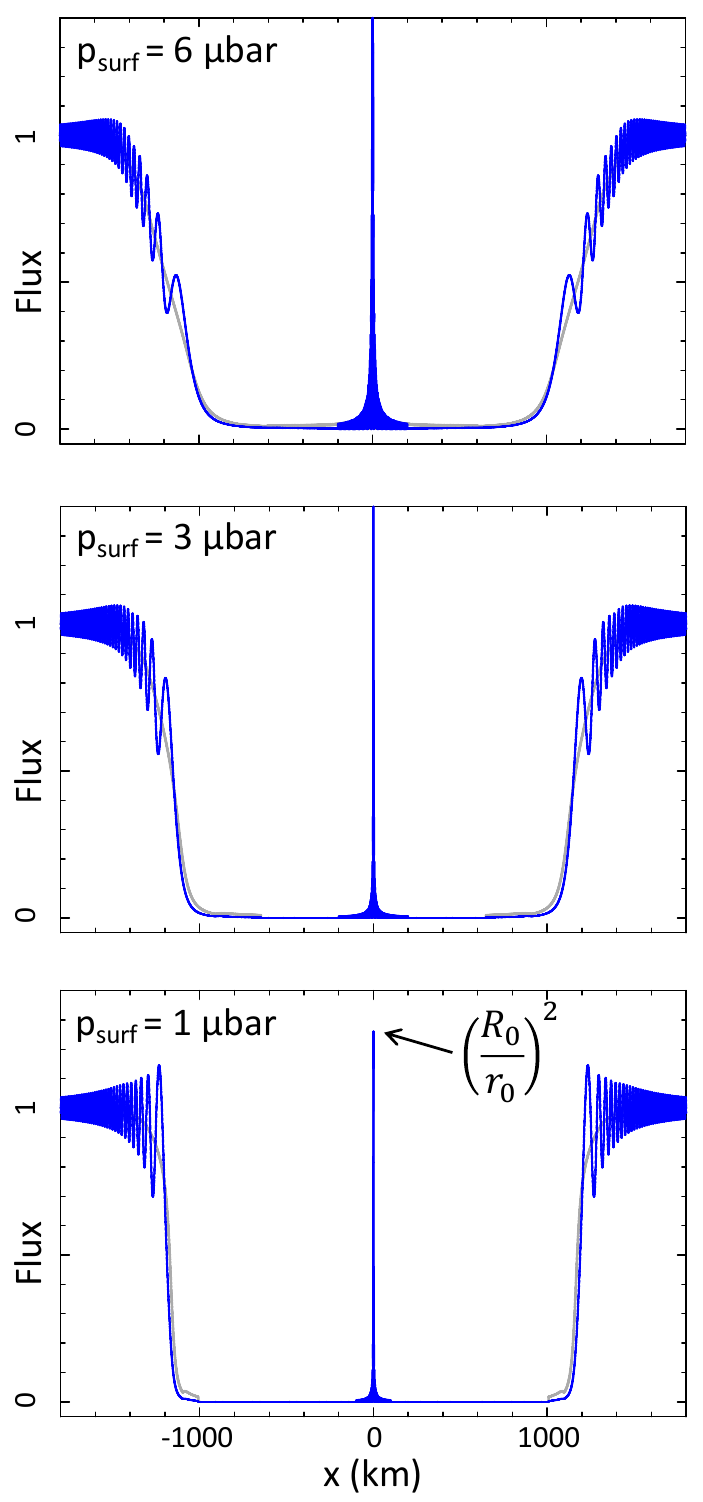}}
\caption{Same as Fig.~\ref{fig_z_flux_Pluto_06jun20_1mm} but reducing the nominal Pluto surface pressure ($p_{\rm surf} = 12.2$~$\upmu$bar) to 6~$\upmu$bar (upper panel), 3~$\upmu$bar (middle panel), and 1~$\upmu$bar (lower panel).
All the cases displayed here correspond to the tenuous atmosphere regime. The height of the central flash, $(R_0/r_0)^2$ (Eq.~\ref{eq_Poisson_tenuous}), exceeds the limits of the frame in the first two panels and is visible in the lower panel.  As $p_{\rm surf}$ approaches zero, the flash tends to the Poisson spot with a height equal to unity, as illustrated in Fig.~\ref{fig_Fresnel_Poisson}.}
\label{fig_z_flux_Pluto_06jun20_1mm_1_3_6_mubar}
\end{figure}

The Poisson fringes could be resolved by observing at longer wavelengths; see Fig.~\ref{fig_z_flux_Pluto_06jun20_1mm}, where a synthetic light curve of a Pluto occultation using  $\lambda=1$~mm is displayed\footnote{The molecular refractivity of $N_2$ decreases by only 1.6\% from the visible to 1~mm, so that Pluto's atmosphere remains in the dense atmosphere case.}. In this case, we have $\lambda_{\rm F} \approx 50$~km, so that the Fresnel fringes are clearly resolved. Meanwhile, the Poisson fringe separation is $\lambda_{\rm F}^2/R \approx 2$~km, which can also be resolved using cadences of a bit more than ten images per second. 

As the Pluto surface pressure $p_{\rm surf}$ decreases, the atmosphere enters  the tenuous regime. This is illustrated in Fig.~\ref{fig_z_flux_Pluto_06jun20_1mm_1_3_6_mubar}, where we consider the cases $p_{\rm surf}=$ 6, 3, and 1~$\upmu$bar. The Fresnel fringes then approach their classical shapes, while the central flash tends to the classical Poisson spot shown in  Fig.~\ref{fig_Fresnel_Poisson}.

The separation between the Poisson fringes may be comparable to that of the Fresnel fringes by making $\lambda_{\rm F}^2/R_0 \approx \lambda_{\rm F}$, i.e., $R_0 \approx \lambda_{\rm F}$. This can be achieved with occultations by small objects observed at large wavelengths. An example is given by \cite{Harju2018}, who describe an occultation by the Main-Belt Asteroid (372) Palma of radius $R_0 \sim 100$~km observed at $\lambda=4.2$~cm, so that $\lambda_{\rm F} \sim \lambda_{\rm F}^2/R_0 \sim 100$~km. In this case, the Poisson spot is actually resolved (Ibid.). 

Another way to resolve the Poisson fringes would be to observe very slow events. However, a limiting factor is that stars with magnitudes $\sim$13 have apparent diameters $D_*$  of hundreds of meters, so that $D_* \gg \lambda_{\rm P}$. 
This averages out the Poisson fringes, in the same way the exposure time does in Fig.~\ref{fig_z_flux_Pluto_06jun20_R0_1190_km_smoothed}.
Figure~\ref{fig_log_log_diffra_stellar_diam} compares a central flash with a point-like star observed in the visible with the flash produced by a star with $D_*=0.5$~km. It shows that the stellar diameter strongly reduces the Poisson spot and erases the Poisson fringes, turning the problem into a purely geometrical optics case.

Once smoothed by the stellar diameter, the maximum value of the flash (Eq.~\ref{eq_max_CF_stellar_diam}) still remains high, reaching typical values of 50 and 240 for Pluto and Triton (Table~\ref{tab_CF_Triton_Pluto}), while its FWHM ($1.14 D_*$, Eq.~\ref{eq_FWHM_stellar_diam}) is typically a fraction of a kilometer. 
In the post-Gaia era, the predictions of stellar occultations by Pluto reach accuracies of some $\pm$20~km \citep{Desmars2019} and about $\pm$60~km for Triton \citep{Marques2022}. Thus, observing the very core of the central flash to within a distance of $\sim D_*$ from centrality requires a demanding ``picket fence" approach, with tens of observers distributed across the band where the central flash is expected to move on the Earth surface.

A less demanding approach is to try to detect a significant increase in flux near the centrality, for instance within the distance $r$ where the flux becomes greater than the unocculted stellar flux, i.e., $\phi(r) > 1$. Inverting Eq.~\ref{eq_flash_geo_optics} and using the values in Table~\ref{tab_CF_Triton_Pluto}, this provides $r < 2R_{\rm CF} \phi_\perp(0)$, that is $r < 6$~km and $r < 30$~km for Pluto and Triton, respectively. Considering the current uncertainties on the predictions quoted above, this means that a picket fence strategy may efficiently probe the region where the surge due to the central flash reaches values $\phi(r) > 1$.

\begin{figure}[!t]
\centerline{\includegraphics[totalheight=90mm,trim=0 0 0 0]{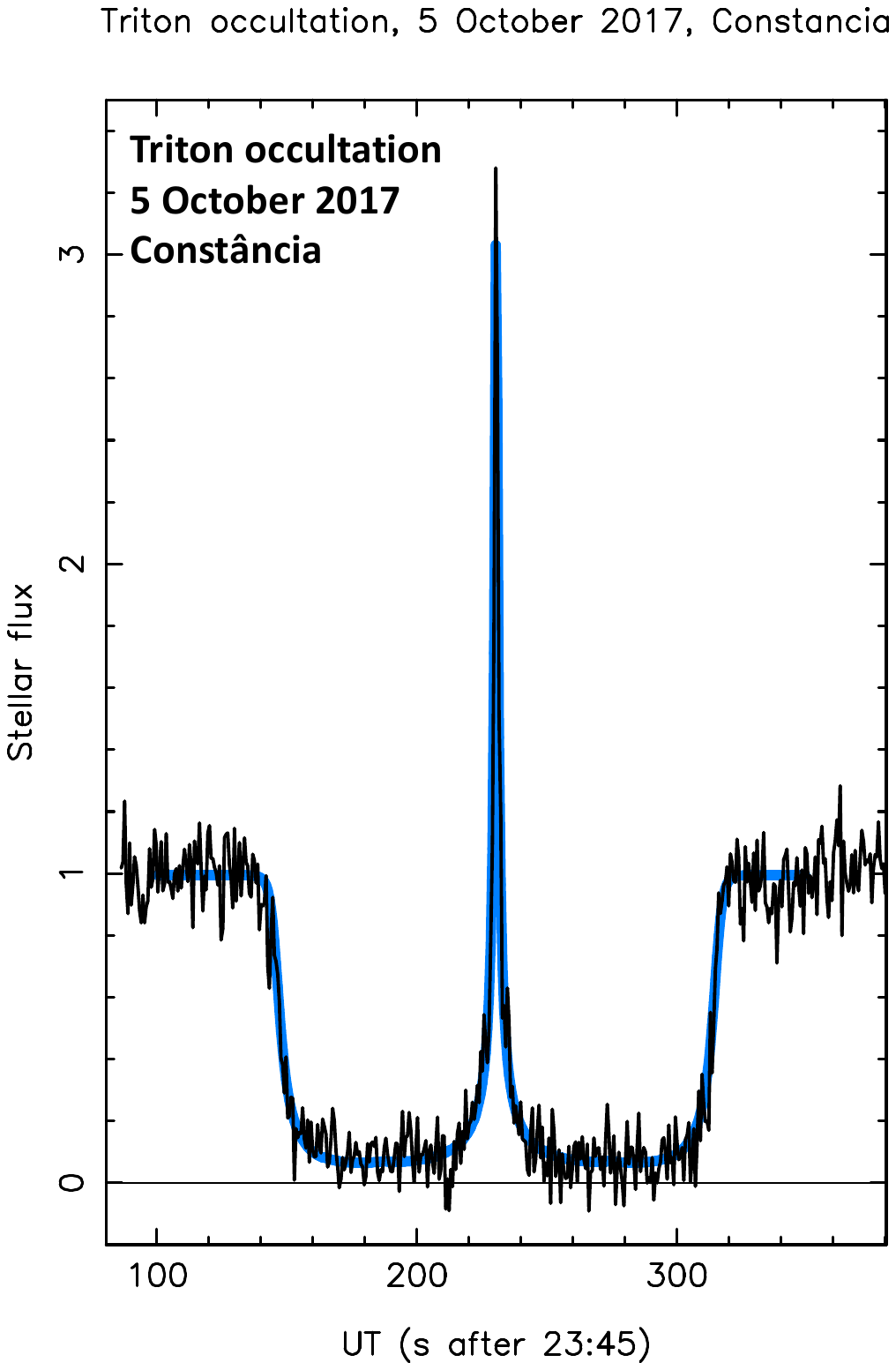}}
\caption{Occultation light curve observed from the Const\^ancia station during the stellar occultation by Triton on 5 October 2017.
The blue curve is a fit to the data using a ray tracing code that assumes a point-like star, a spherical atmosphere, and geometrical optics. The observed peak value of the flash is about 3.3 times the flux of the unocculted star.}
\label{fig_t_flux_Triton_05oct17_Constancia}
\end{figure}

In the post-Gaia era, the picket fence approach was successfully used for the 1 June 2022 Pluto stellar occultation, with a flash peak reaching about 2.5 times the unocculted stellar level \citep{Young2022}. 
In the same vein, a Triton occultation observed on 5 October 2017 resulted in 23 cuts through the central flash region, an observation that constrained the shape of the atmosphere, with an upper limit of 0.0019 for the oblateness of the central flash layer, i.e., a difference $\delta R < 2.5$~km between its polar and equatorial radii \citep{Marques2022}.
Figure~\ref{fig_t_flux_Triton_05oct17_Constancia} shows the light curve obtained from the Const\^ancia station, which had a closest approach distance $r_{\rm CA} \approx 8.4$~km  to the shadow center. Considering the stellar diameter $D_* = 0.65$~km projected at Triton for this event, and thus $D_* \gg \lambda_{\rm P}$ and $r_{\rm CA}/D_* \approx 13$, this means that the flash structure is undistinguishable from that caused by a point-like star in the geometrical optics regime; see Fig.~\ref{fig_CF_stellar_diameter}.

Putting aside the current technological limits on acquisition rates, one might consider observing very faint stars with apparent diameters $D_* \lesssim 1$~meter, so as to resolve the Poisson spot. However, this would require stars fainter than magnitude~20, and in this case their fluxes would be overwhelmed by those of Pluto or Triton (with magnitudes of about 14), which would make the observation unusable. 

Besides the practical difficulties in resolving the Poisson spot in the visible, physical effects in the occulter atmosphere can destroy the spot. 
One is the departure of the atmosphere from sphericity. A flattened layer with a difference of $\delta R$ between its equatorial and polar radii creates a diamond-shaped caustic curve in the shadow plane with sides of about $4\delta R$ \citep{Elliot1977}. 

The Poisson spot disappears when the diamond-shaped caustic is larger than the Poisson spot diameter, $4\delta R \gtrsim \lambda_{\rm F}^2/R$. 
For Pluto and Triton events observed in the visible, this limits  $\delta R$ to very small values of $\delta R \lesssim 1$~m. 
Pluto and Triton are slow rotators with periods of 6.4 and 5.9~days, respectively, which imply equatorial velocities $v_{\rm eq} \sim$15~m~s$^{-1}$ for both objects. Ignoring  possible zonal winds, these velocities cause a minimum atmospheric flattening $\delta R = R^2 v_{\rm eq}^2/(2GM)$, where $G$ is the gravitational constant and $M$ is the mass of the body \citep{Marques2022}. This yields $\delta R \sim 150$~m, which is much larger than the condition $\delta R \lesssim 1$~m derived above in the visible. 
However, observing at 1~mm would result in a value of $\lambda_{\rm F}^2/R$ one thousand times larger, thus preserving the Poisson spot  even with a flattening of $\delta R \sim 150$~m.

This said, we note that the Poisson fringes are more robust against departure from sphericity than the Poisson spot. This is due to the fact that at any time, the two stellar images can interfere even if the atmosphere is distorted.

\begin{figure*}[!t]
\centerline{\includegraphics[width=\textwidth,trim=0 0 0 0]{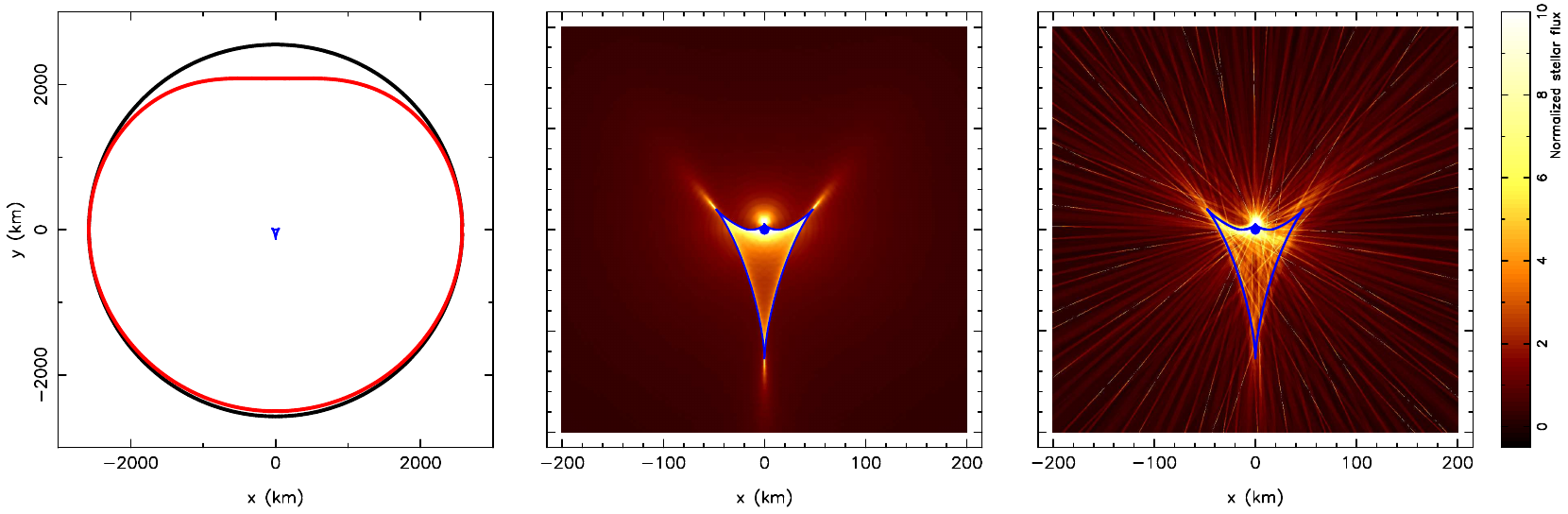}}
\caption{\it Left panel\rm: Titan's central flash layer as determined from a stellar occultation observed in 2003 (black line). The red line enhances the distortion of the layer by expanding by a factor of twenty its departure from sphericity. The small blue feature near the shadow center shows the caustic created by this model.
\it Middle panel\rm: Close-up view of the caustic and the distribution of light around it in the case of a smooth flash layer.
\it Right panel\rm: Same as middle panel in the presence of small corrugations of the flash layer caused by gravity waves. 
Adapted from \cite{Sicardy2022}.}
\label{fig_Titan_flash}
\end{figure*}

Another cause of destruction of the Poisson spot is that the atmospheres of Pluto and Triton are not perfect lenses. In particular, internal gravity waves cause local density fluctuations that induce stellar scintillation that eventually blurs the central flash (see for instance \cite{Hubbard1988,Sicardy2006,Sicardy2022} for the cases of Neptune's and Titan's central flashes).
This is illustrated in Fig.~\ref{fig_Titan_flash} for Titan. The non-centrosymmetric distortion of the atmosphere causes a more complex caustic than the diamond-shaped figure mentioned above, while gravity waves add streaks to the flash map.
These disruptive effects constitute a subject in their own right, and remain out of the scope of the present paper. 

Ignoring these difficulties, we consider a thought experiment in which a telescope with aperture $D_{\rm T}$ observes the occulter during an occultation. A natural question is then: what would an observer see through this telescope as the two stellar images move around the limb of the occulter, and in particular during the rapid flux variations caused by the interferences between the two stellar images? 

Such a telescope produces two pseudo-images, i.e., two Airy disks with angular sizes $\sim \lambda/D_{\rm T}$, i.e., with linear sizes $\sim 2 \lambda_{\rm F}^2/D_{\rm T}$ when projected at the body. These images are separated by $2R_{\rm CF}$, so that if $D_{\rm T} \lesssim \lambda_{\rm P}$, then $\lambda_{\rm F}^2/D_{\rm T} \gtrsim R_{\rm CF}$. This means that the observer sees two Airy disks that merge into a single Airy disk. As the two stellar images interfere constructively or destructively, the observer sees this single Airy disk switching on and off as $r$ varies.

Conversely, using a larger telescope with $D_{\rm T} \gtrsim \lambda_{\rm P}$ means that $\lambda_{\rm F}^2/D_{\rm T} \lesssim R_{\rm CF}$, which means that the two stellar images are separated. In this case the Poisson fringes are averaged out over the telescope aperture, and no more flux variations are observed. What the observer sees are now two distinct Airy disks, each with the flux predicted by geometrical optics.
A quantitative theory of a comparable situation is presented by \cite{Dettwiller2012} for the case where the fringes are due to diffraction by a single slit, with a discussion of the separation of the images of the two edges of the slit.

The separation of the stellar images around the limb was achieved during occultations by Titan observed in 2001 \citep{Bouchez2003} and 2022 \citep{Marlin2025}. The observers used adaptive optics systems with telescopes of diameters $D_{\rm T}=5$~m and $D_{\rm T}=10$~m, respectively, both using the  K-band ($\lambda=2.12$~$\upmu$m). Titan was  at geocentric distances of $\Delta \approx 1.25 \times 10^9$~km, so that $\lambda_{\rm P} \approx 0.5$~m. 
This means that the condition $D_{\rm T} \gg \lambda_{\rm P}$ was safely met for both observations, which allowed the separation of the primary and secondary 
images\footnote{Due to the distortion of Titan's atmosphere, more images were actually seen near the centrality during the 2022 observation.}.

\section{Conclusion}
\label{sec_conclu}

In this paper we studied the structure of the shadow cast by an opaque spherical body surrounded by a transparent and spherical atmosphere. In particular, we examined the effects of diffraction and finite stellar diameter in the structure of the central flash. The analytical expressions given in the main text are summarized in Table~\ref{tab_CF_formulae}, while numerical applications to ground-based occultations by Pluto and Triton are provided in Table~\ref{tab_CF_Triton_Pluto}.

Our starting point was an airless and opaque circular body of radius $R_0$ illuminated by a monochromatic wave of wavelength $\lambda$ coming from a star at infinity and of negligible  angular diameter. At distance $\Delta$, the body casts a shadow with Fresnel fringes at its edge and a Poisson spot at its center; see Eq.~\ref{eq_Poisson_spot} and Fig.~\ref{fig_Fresnel_Poisson}. The Poisson spot has a full width of $1.53 \lambda_{\rm F}^2/R_0$, where $\lambda_{\rm F}= \sqrt{\lambda \Delta/2}$ is the Fresnel scale, and a peak value of unity, when normalized to the unocculted stellar flux.

As a tenuous atmosphere is introduced, the ensuing deviation of the stellar rays creates a dark shadow of radius $r_0 < R_0$ (Fig.~\ref{fig_tenuous_vs_dense}). The Poisson spot remains unchanged in terms of radial profile, but it is amplified by a factor $(R_0/r_0)^2$ (Eq.~\ref{eq_Poisson_tenuous}). 
For a dense enough atmosphere, defined as an atmosphere that can focus the stellar rays to the shadow center through the existence of a central flash layer, the central flash still has the radial structure of the Poisson spot, but with a peak value of $2\pi^2 R_{\rm CF}H/\lambda_{\rm F}^2$ for an isothermal atmosphere with scale height $H$, where $R_{\rm CF}$ is the radius of the central flash layer. 
 
The central flash is surrounded by circular Poisson fringes whose oscillations decrease like $1/r$ and with separation $\lambda_{\rm P}= \lambda_{\rm F}^2/R_{\rm CF}$. This separation is the same as the one obtained during a Young's experiment with two sources separated by $2R_{\rm CF}$ interfering in the shadow plane.

For ground-based occultations by Pluto or Triton observed in the visible, the expected peak value of the flash is very large, some $10^4$-$10^5$ , and the Poisson fringe spacing $\lambda_{\rm P}$ 
is very small, about one meter (Table~\ref{tab_CF_Triton_Pluto}). In practice, the Poisson fringes are unobservable using current technology. However, observations conducted in radio wavelengths (Fig.~\ref{fig_z_flux_Pluto_06jun20_1mm}) may resolve both the central flash and its associated fringe pattern during Pluto or Triton occultations. 

In the visible and near IR domains where the Poisson fringes are usually blurred by the projected stellar diameter $D_*$, the flash profile is a combination of complete elliptic integrals of the first and second kinds (Eqs.~\ref{eq_phi_r_gt_r*} and \ref{eq_phi_r_lt_r*}). The FWHM of the flash is $1.14 D_*$ and its peak value is $8H/D_*$. For Pluto or Triton occultations with typical $D_* \sim 0.5$~km, $8H/D_*$ is of the order of 50 and 240, respectively. This implies that, in this case, the flash structure is largely dominated by $D_*$ and by geometrical optics. However, for observations conducted in millimetric wavelengths with $D_* \sim 0.5$~km, diffraction starts to dominate the smoothing effect of the stellar diameter.

Our calculations remain limited in terms of applications because, as mentioned in the Introduction, we assume a perfectly spherical atmosphere. In practice, even slight deviations from sphericity usually dominate the shape of the flash. However, our analysis serves as a firm basis for future calculations that we plan to do, in particular for describing the role of diffraction near the caustic created by a distorted atmosphere, such as the one displayed in Fig.~\ref{fig_Titan_flash}.

\begin{acknowledgements}
We thank the reviewers for comments that improved the presentation of this paper.
\end{acknowledgements}

\bibliographystyle{aa}
\bibliography{references}

\begin{appendix}

\section{The Sommerfeld lemma}
\label{app_Sommerfeld}

The Sommerfeld lemma (also known as the Sommerfeld expansion, \citealt{Sommerfeld1954}) states that 
if $f(v)$ and $g'(v)$ vary much more slowly than $\exp[ {\rm i} g(v)]$ and
if $g'(v) \neq 0$ everywhere in the interval $[a,b]$, then 
\begin{equation}
 \int_a^b f(v) \exp[ {\rm i} g(v)] \, {\rm d}v \approx 
 \left| \displaystyle f(v) \frac{\exp[ {\rm i} g(v)]}{ {\rm i} g'(v)} \right|_a^b
\label{eq_Sommerfeld}
\end{equation}
We use this lemma to evaluate Eq.~\ref{eq_amplitude_ori}, which should strictly speaking include the Rayleigh-Sommerfeld inclination factor $\cos \chi(R,\theta)$ (Fig.~\ref{fig_geo_occ}), 
\begin{equation}
\begin{array}{l}
\displaystyle 
a(r) =  \frac{1}{2 {\rm i} \lambda_{\rm F}^2} \times \\ \\
\displaystyle 
\int_{R_0}^{+\infty} \int_0^{2\pi}
[\cos \chi(R,\theta)]
\exp\{ {\rm i} [\varphi_{\rm a}(R) + \varphi_{\rm g}(R) ] \} R \, {\rm d}R {\rm d}\theta.
\end{array}
\end{equation} 
For values of  $r \ll R_0$, the dependence of $\chi(R,\theta)$ on $\theta$ is weak, so that 
\begin{equation}
\begin{array}{l}
\displaystyle
a(r) \approx \frac{\pi}{ {\rm i} \lambda_{\rm F}^2}  \times \\ \\
\displaystyle
\int_{R_0}^{+\infty} 
[\cos \chi(R)]
\exp \left\{ {\rm i} \left[ \varphi_{\rm a}(R) + \frac{\pi R^2}{2 \lambda_{\rm F}^2} \right] \right\}
J_0 \left( \frac{\pi Rr}{\lambda_{\rm F}^2} \right) R \, {\rm d}R.
\end{array}
\end{equation}
The change of variable $v=\pi R^2/(2 \lambda_{\rm F}^2)$ yields 
\begin{equation}
\begin{array}{l}
a(r) = \\ \\
\displaystyle
-{\rm i} \int_{v_0}^{+\infty} [\cos \chi(v)] J_0 \left( \frac{r}{\lambda_{\rm F}} \sqrt{2\pi v} \right) \exp \{ {\rm i} [\varphi_{\rm a}(v) + v ] \} \, {\rm d}v,
\label{eq_amplitude_airless_v}
\end{array}
\end{equation} 
where $v_0= \pi R_0^2/(2 \lambda_{\rm F}^2)$.
This is Eq.~\ref{eq_Sommerfeld}, where 
\begin{equation}
\begin{array}{l}
\displaystyle
f(v)= -{\rm i} [\cos \chi(v)] J_0 \left( \frac{r}{\lambda_{\rm F}} \sqrt{2\pi v} \right) \\ \\
g(v)=  \varphi_{\rm a}(v) + v,
\label{eq_f_g}
\end{array}
\end{equation} 
with $a= v_0$ and $b= +\infty$.

We first consider $f(b) = -{\rm i} [\cos \chi(b)] J_0 (r\sqrt{2\pi b}/\lambda_{\rm F})$ as $b$ approaches infinity.
Then $\cos \chi(b) = \Delta/l \approx \Delta/R$ approaches zero, as does 
$J_0 (r\sqrt{2\pi b}/\lambda_{\rm F})$ for $r \neq 0$ (while $J_0=1$ remains finite for $r=0$).
Consequently, the term $f(b) \exp[{\rm i} g(b)]/{\rm i} g'(b)$ vanishes in Eq.~\ref{eq_Sommerfeld} for $b=+\infty$.

This leaves only the term $f(a) \exp[{\rm i} g(a)]/{\rm i} g'(a)$ (where $a=v_0$).
We have $g'(v_0)= \varphi'_{\rm a}(v_0) + 1$, where here the derivation is made with respect to $v$. We have $\varphi'_{\rm a}(v_0)= ({\rm d}\varphi_{\rm a}/{\rm d}R)({\rm d}R/{\rm d}v)(R_0)$. From ${\rm d}R/{\rm d}v = \lambda_{\rm F}^2/(\pi R_0)$ and Eq.~\ref{eq_dphi}, we obtain $g'(v_0)=  1 - \omega_0 \Delta/R_0$, where $\omega_0$ is the deviation suffered in the geometrical optics regime by a ray that grazes the radius $R_0$. But $r_0=  R_0 - \omega_0 \Delta$, so that $g'(v_0)= r_0/R_0$. 
Moreover, from Fig.~\ref{fig_geo_occ}, $\cos \chi(v_0) \approx \Delta/\sqrt{\Delta^2 + R_0^2} \approx 1$ because $R_0 \ll \Delta$. 
From this, we finally obtain 
\begin{equation}
a(r) \approx  
\left( \frac{R_0}{r_0} \right)
\exp \left\{ {\rm i} \left[ \varphi_{\rm a}(R_0) + \frac{\pi R_0^2}{2 \lambda_{\rm F}^2} \right] \right\}
J_0 \left( \frac{\pi R_0 r}{\lambda_{\rm F}} \right)
\end{equation} 
and 
\begin{equation}
\phi_{\rm Diff} (r) = |a(r)|^2 =   
\left( \frac{R_0}{r_0} \right)^2
J_0^2 \left( \frac{\pi R_0 r}{\lambda_{\rm F}} \right).
\label{eq_CF_rR_0_r0}
\end{equation} 
We recall that this expression is valid for $r \ll \lambda_{\rm F}$, and only for a tenuous atmosphere.

\section{The stationary phase method}
\label{app_stationary_method}

If the function $g(v)$ becomes stationary in the interval $[a,b]$, i.e. if there exists a value $v_{\rm s} \in [a,b]$ where $g'(v_{\rm s})=0$, then Eq.~\ref{eq_Sommerfeld} cannot be used. In this case, assuming again that $f(v)$ varies much more slowly than $\exp[{\rm i} g(v)]$, most of the contribution to the integral  $\int_a^b f(v) \exp[{\rm i} g(v)] \, {\rm d}v$ comes from a neighborhood of $v_{\rm s}$, a situation known as the stationary phase condition. This neighborhood covers an interval of $v$ whose width is a few times $\sqrt{2\pi/|g''(v_{\rm s})|}$  (which is the width of the first Fresnel zone associated with $g$ near $v_{\rm s}$). If $v_{\rm s}$ is well inside $[a,b]$, in the sense that $v_{\rm s}$ is several times $\sqrt{2\pi/|g''(v_{\rm s})|}$ away from both bounds $a$ and $b$, then formally (\citealt{Dieudonne1980}) 
\begin{equation}
\int_{a}^{b} f(v) \exp[{\rm i} g(v)] \, {\rm d}v \approx 
(1+{\rm i}) f(v_{\rm s}) \sqrt{\frac{\pi}{g''(v_{\rm s})}} \exp[{\rm i} g(v_{\rm s})].
\label{eq_Fresnel_int}
\end{equation} 
We note that the modulus of the right-hand term of Eq.~\ref{eq_Fresnel_int} is the width of the first Fresnel zone 
associated with $g$ near $v_{\rm s}$, times $|f(v_{\rm s})|$.
If $g''(v_{\rm s}) > 0$ (local minimum for $g(v)$), 
\begin{equation}
\int_{a}^{b} f(v) \exp[{\rm i} g(v)] \, {\rm d}v = (1+{\rm i}) f(v_{\rm s}) \sqrt{\frac{\pi}{g''(v_{\rm s})}} \exp[{\rm i} g(v_{\rm s})]. 
\label{eq_Fresnel_integral_g"_positive}
\end{equation} 
If $g''(v_{\rm s}) < 0$ (local maximum for $g(v)$), we take by convention $\sqrt{g''}:= {\rm i} \sqrt{-g''}$, so that 
\begin{equation}
\int_{a}^{b} f(v) \exp[{\rm i} g(v)] \, {\rm d}v = -{\rm i}(1+{\rm i}) f(v_{\rm s}) \sqrt{\frac{\pi}{-g''(v_{\rm s})}} \exp[{\rm i} g(v_{\rm s})].
\label{eq_Fresnel_integral_g"_negative}
\end{equation} 
Thus, considering a local maximum of $g(v)$ instead of a local minimum introduces a $\pi/2$ phase shift in the evaluation of the integral, due to the prefactor $-{\rm i}$ in the second equation.

As $v_{\rm s}$ approaches one of the bounds, say $a$, Eq.~\ref{eq_Fresnel_int} is no more valid as an edge effect occurs. Assuming $g''(v_{\rm s}) > 0$ and $b-a \gg1/\sqrt{g''(v_{\rm s})}$, we have 
\begin{equation}
\begin{array}{ll}
\int_{a}^{b} f(v) \exp[{\rm i} g(v)] \, {\rm d}v \approx & \displaystyle \int_{a}^{v_{\rm s}} f(v) \exp[{\rm i} g(v)] \, {\rm d}v + \\ \\
                                                                     & \displaystyle \int_{v_{\rm s}}^{+\infty} f(v) \exp[{\rm i} g(v)] \, {\rm d}v
\end{array}
\label{eq_split_integral}
\end{equation} 
The second integral is just the half of the integral in Eq.~\ref{eq_Fresnel_int}. The first integral can be evaluated by expanding $g$ to the second order near $v_{\rm s}$, $g(v) \approx g(v_{\rm s}) + g''(v_{\rm s})(v-v_{\rm s})^2/2$. We define the Fresnel function as 
\begin{equation}
F_{\rm r}(\alpha):= \int_0^{\alpha} \exp \left( {\rm i} \frac{\pi  t^2}{2} \right) {\rm d}t,
\label{eq_Fresnel_func}
\end{equation} 
which yields 
\begin{equation}
\begin{array}{l}
\displaystyle
\int_{a}^{v_{\rm s}} f(v) \exp[{\rm i} g(v)]  \, {\rm d}v \approx \\ \\
\displaystyle
f(v_{\rm s}) \sqrt{\frac{\pi}{g''(v_{\rm s})}} \exp[{\rm i} g(v_{\rm s})] \, F_{\rm r} \left[ \sqrt{ \frac{g''(v_{\rm s})}{\pi}} (v_{\rm s} - a) \right],
\end{array}
\end{equation} 
so that 
\begin{equation}
\begin{array}{l}
\displaystyle
\int_{a}^{b} f(v) \exp[{\rm i} g(v)] \, {\rm d}v = (1+{\rm i})f(v_{\rm s}) \sqrt{\frac{\pi}{g''(v_{\rm s})}} \exp[{\rm i} g(v_{\rm s})] \times \\ \\
\displaystyle
\left\{ \frac{1}{2} + \frac{1}{1+{\rm i}} F_{\rm r} \left[ \sqrt{ \frac{g''(v_{\rm s})}{\pi}} (v_{\rm s} - a) \right] \right\}.
\end{array}
\label{eq_Fresnel_int_edge}
\end{equation}

The function $1/2 + F_{\rm r}(\alpha)/(1+{\rm i})$ is the classical normalized complex amplitude of a plane wave diffracted by a sharp opaque rectilinear edge, where $\alpha$ is the position of the observer (normalized to the Fresnel scale) relative to the limit of the geometrical shadow. The Sommerfeld lemma applied to $\int_{-\infty}^{\alpha} \exp({\rm i} \pi t^2/2) \, {\rm d}t$ and $\int_{\alpha}^{+\infty} \exp({\rm i} \pi t^2/2) \, {\rm d}t$ provides the following asymptotic expressions, 
\begin{eqnarray}
\alpha \rightarrow -\infty,~~ \frac{1}{2} + \frac{F_{\sf r} (\alpha)}{1+{\rm i}} \approx 
-\left( \frac{1+{\rm i}}{2\pi} \right) \frac{ e^{{\rm i} \frac{\pi}{2} \alpha^2}}{\alpha}  \label{eq_Fr_asymptot_minus_infty} \\ 
\nonumber \\
\alpha \rightarrow +\infty,~~ \frac{1}{2} + \frac{F_{\sf r} (\alpha)}{1+{\rm i}} \approx 
1 -\left( \frac{1+{\rm i}}{2\pi} \right) \frac{ e^{{\rm i} \frac{\pi}{2} \alpha^2}}{\alpha} \label{eq_Fr_asymptot_plus_infty},
\end{eqnarray} 
so that 
\begin{equation}
\alpha \rightarrow -\infty,~~ \left| \frac{1}{2} + \frac{F_{\sf r} (\alpha)}{1+{\rm i}} \right|^2 \approx \frac{1}{2 \pi^2 \alpha^2}
\label{eq_damping_Fresnel_fringes}
\end{equation}
\begin{equation}
\alpha \rightarrow +\infty,~~ \left| \frac{1}{2} + \frac{F_{\sf r} (\alpha)}{1+{\rm i}} \right|^2 \approx 
1 - \frac{\sqrt{2}}{\pi \alpha} \cos \left( \frac{\pi}{2} \alpha^2 + \frac{\pi}{4} \right) + \frac{1}{2\pi^2 \alpha^2}.
\label{eq_approx_Fresnel_fringes}
\end{equation} 
Eq.~\ref{eq_damping_Fresnel_fringes} shows that the flux decays like $1/\alpha^2$ inside the shadow, while Eq.~\ref{eq_approx_Fresnel_fringes} shows that the Fresnel fringe visibility damps like $1/\alpha$ outside the shadow. Numerical integrations show that the approximations in Eqs.~\ref{eq_damping_Fresnel_fringes} and \ref{eq_approx_Fresnel_fringes} agree with their respective exact values 
to within 1\% for $\alpha \lesssim -2.6$ and $\alpha \gtrsim +2.2$, respectively.

\section{Comparison of the Poisson and the Airy disk}
\label{app_Poisson_vs_Airy}

\begin{figure}[!t]
\centerline{\includegraphics[totalheight=50mm,trim=0 0 0 0]{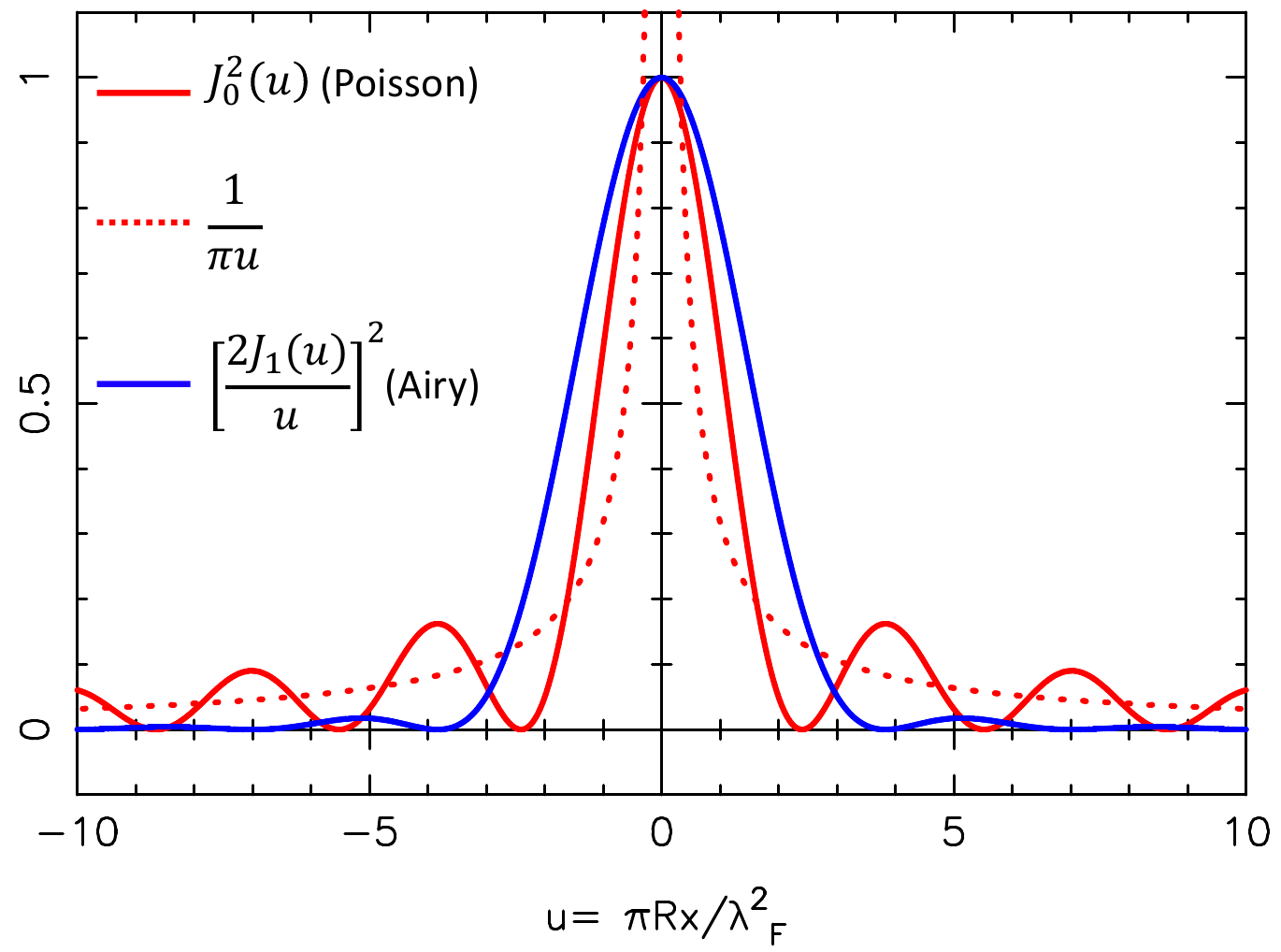}}
\centerline{\includegraphics[totalheight=50mm,trim=0 0 0 0]{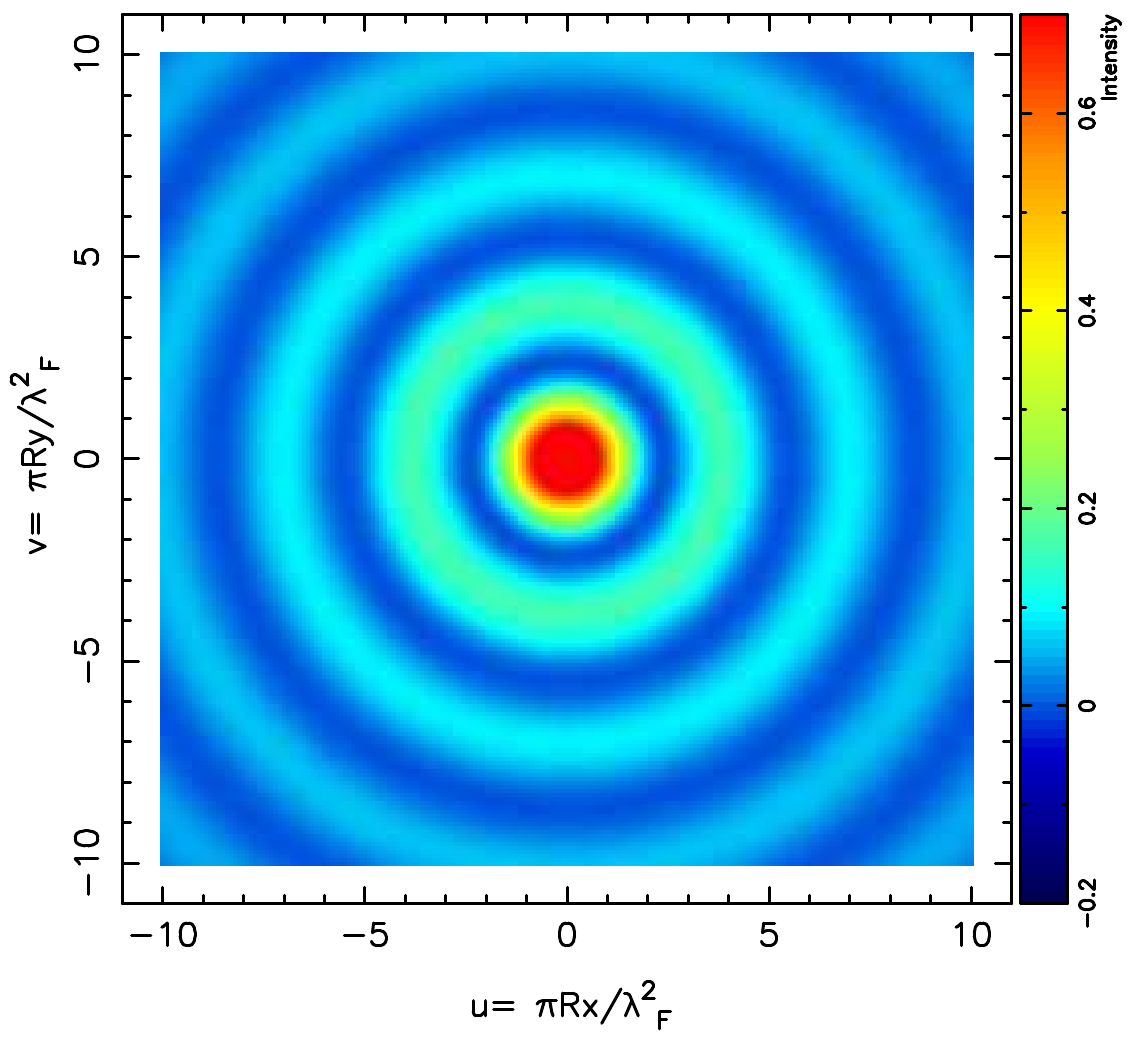}}
\centerline{\includegraphics[totalheight=50mm,trim=0 0 0 0]{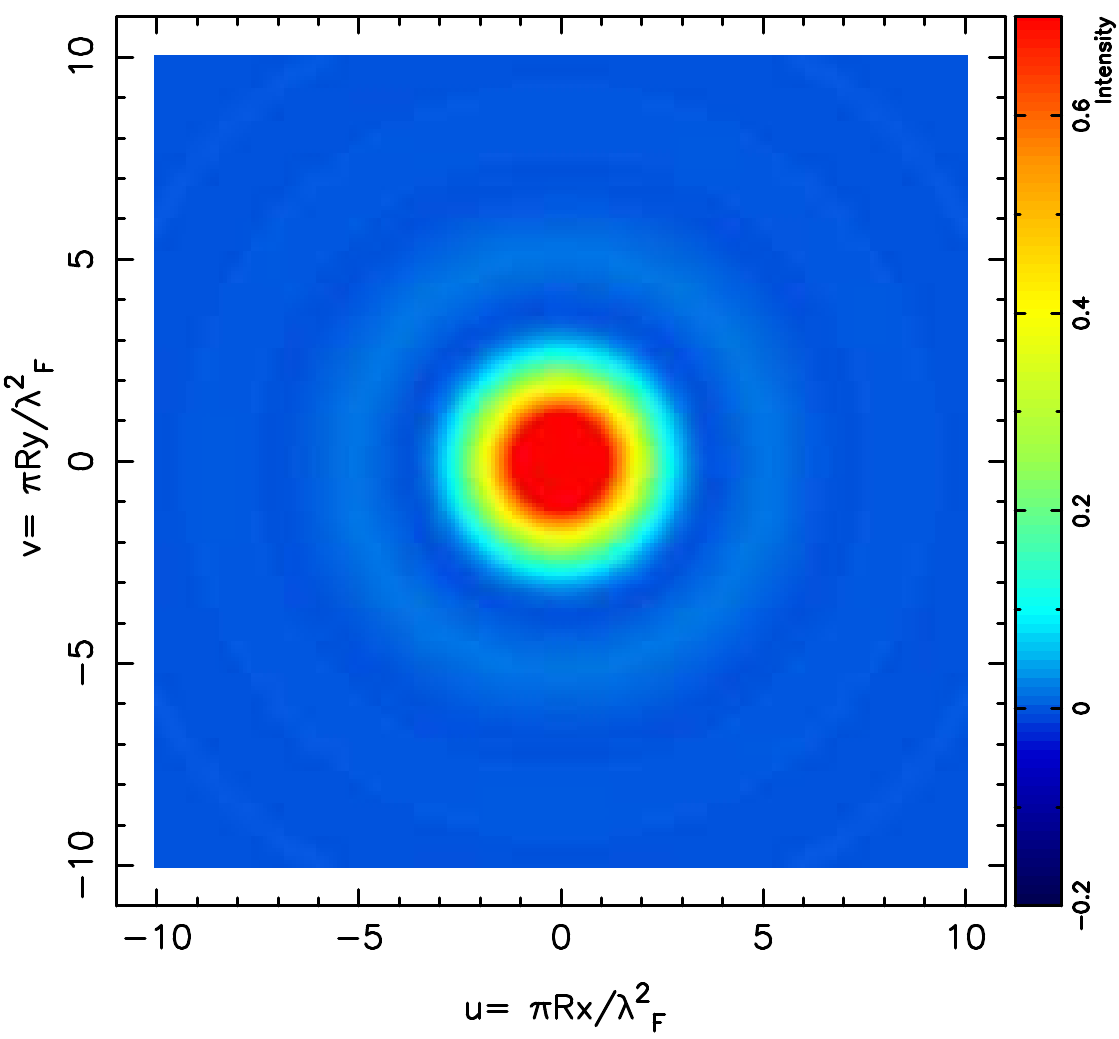}} 
\caption{\it Upper panel\rm:
Radial profile of the Poisson spot created by an opaque circular disk of radius $R$ is plotted in red, while the Airy disk profile created by a lens of same radius is plotted in blue. Both profiles are normalized to unity at $r=0$. 
The Poisson spot is described by the function $J_0^2(u)$ (Eq.~\ref{eq_Poisson_spot}), while the Airy disk is described by the function $[2J_1(u)/u]^2$ (Eq.~\ref{eq_Airy}).
The normalized geometrical optics approximation given by Eq.~\ref{eq_flash_geo_optics_iso} is plotted as a red dotted line which diverges to infinity at the origin. 
The Poisson spot is narrower than the corresponding Airy disk by about 60\% (Eqs.~\ref{eq_diam_Poisson} and \ref{eq_r1_Airy}). The fringes around the Poisson spot damp out more slowly ($\propto 1/u$) that those of the Airy disk ($\propto 1/u^3$); see Eq~\ref{eq_damping_CF_vs_Airy}. 
\it Middle panel\rm:
2D-map of the Poisson spot, corresponding to the red profile in the upper panel. 
\it Lower panel\rm:
2D-map of the Airy disk, corresponding to the blue profile in the upper panel.}
\label{fig_flash_Airy_1D}
\end{figure}

Here we compare the structure of the Poisson spot with the Airy disk produced by a lens with diameter $2R$ and focal lens $\Delta$. The Airy disk profile is classically given by (see e.g. \citealt{Elmore1969})
\begin{equation}
\phi_{\rm A}(r)= \left( \frac{\pi R^2}{\lambda_{\rm F}^2} \right)^2 \left[ \frac{J_1(u)}{u} \right]^2.
\label{eq_Airy}
\end{equation}
The first zero of $J_1$ being $u_1= 3.83...$, the first dark fringe occurs at the well known value
\begin{equation}
r_1= \frac{u_1 \lambda_{\rm F}^2}{\pi R} 
\approx  1.22  \frac{\lambda_{\rm F}^2}{R} 
= 1.22 \left(  \frac{\lambda}{2R} \right)  \Delta.
\label{eq_r1_Airy}
\end{equation}

The Poisson spot and Airy disk are displayed in Fig.~\ref{fig_flash_Airy_1D}. The comparison of Eqs.~\ref{eq_diam_Poisson} and \ref{eq_r1_Airy} shows that the central flash caused by a flash layer of radius $R$ is more peaked by about 60\% compared to the Airy disk caused by a lens of same radius. Moreover, the asymptotic form given in Eq.~\ref{eq_asymptotic} shows that
\begin{equation}
J^2_0(u) \propto \frac{1}{u} {\rm~~while~~} \left[ \frac{J_1(u)}{u} \right]^2 \propto \frac{1}{u^3},
\label{eq_damping_CF_vs_Airy}
\end{equation}
meaning that the central flash has more extended wings. In particular, the energy contained in the wings of the central flash is unbounded, contrarily to the energy contained in the wings of the Airy disk. In that sense, an object with an atmosphere is not as good an optical system  as a lens with aperture $2R$, since its point spread function is less sharp, as already noted by \cite{Hubbard1977}.

\section{Comparison of point-like and finite-size stars}
\label{app_point_vs_finite_size}

\begin{figure}[!h]
\centerline{\includegraphics[totalheight=75mm,trim=0 0 0 0]{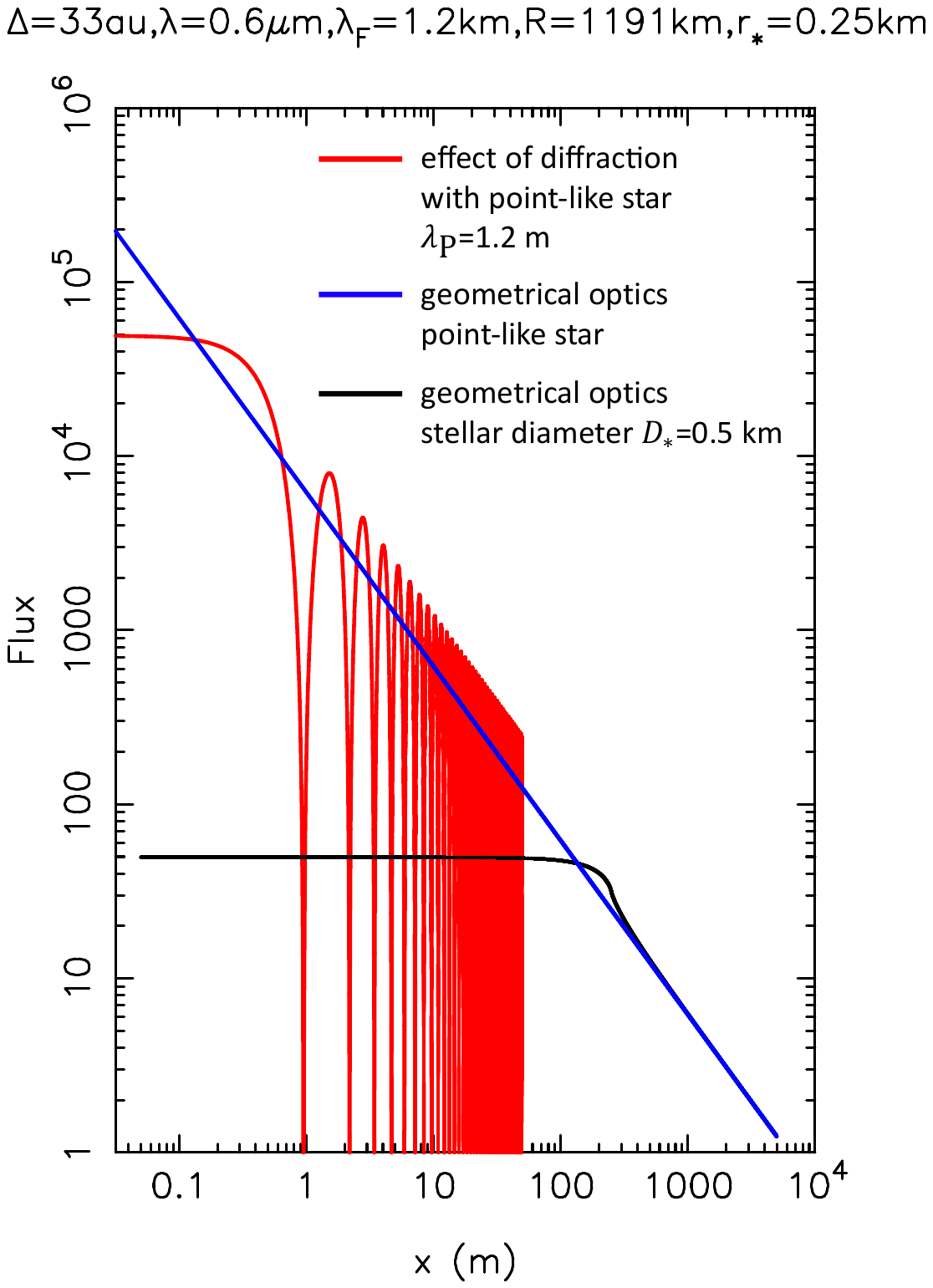}}
\caption{Log-log plot showing the flash profiles caused by a spherical and transparent Pluto's atmosphere, as derived from the 6 June 2020 occultation  \citep{Sicardy2021} and using the parameters listed in Table~\ref{tab_CF_Triton_Pluto}. 
Red curve: the flash caused by a point-like star and monochromatic wave, 
giving a finite diffraction peak and interference fringes (because Pluto's atmosphere is dense) with spacing $\lambda_{\rm P} \approx 1.2$~m (Table~\ref{tab_CF_Triton_Pluto}). 
The flash has then the same radial dependence as the Poisson curve shown in red curve in Fig.~\ref{fig_Fresnel_Poisson}; 
Black curve: the flash caused by a star with a diameter $D_*=0.5$~km projected at Pluto's distance
(Eqs.~\ref{eq_phi_r_gt_r*} and \ref{eq_phi_r_lt_r*});
Blue curve: the flash obtained using the geometrical optics approximation (Eq.~\ref{eq_flash_geo_optics}).}
\label{fig_log_log_diffra_stellar_diam}
\end{figure}

Because of the very large differences in both the widths and amplitudes of the flashes between the point-like and the finite-size star cases, it is preferable to use log-log axes to illustrate the differences between the two; see Fig.~\ref{fig_log_log_diffra_stellar_diam}. 

\section{Effects of limb darkening}
\label{app_limb_dark}

We consider a subdisk of radius $\rho$ nested inside the stellar disk of radius $r_*$. From Eqs.~\ref{eq_phi_r_gt_r*} and \ref{eq_phi_r_lt_r*}, the contribution of the subdisk to the surface areas of the two stellar images is
\begin{equation}
S(\rho)= C \rho f \left( \frac{\rho}{r} \right), 
\end{equation}
where 
\begin{equation}
\begin{array}{l}
C= 8R \Phi_\perp (0) \\ \\
f(s)=  E(1/s) {\rm~for~} s \geq 1 \\ \\
f(s)=  G(s) {\rm~for~} s \leq 1. \\
\end{array}
\end{equation}
Consequently, an annulus of width ${\rm d} \rho$ of the star provides a surface area of the images that amounts to
\begin{equation}
{\rm d} S=  C \frac{{\rm d}}{{\rm d \rho}} \left[\rho f \left( \frac{\rho}{r} \right) \right] {\rm d} \rho.
\end{equation}
The limb darkening of the stellar disk is described by a radiance profile $I(\rho/r_*)$. Weighing the surface area ${\rm d} S$ of each annulus by $I(\rho/r_*)$ and normalizing it to the total flux received from the unocculted star, we obtain the normalized central flash profile
\begin{equation}
\phi_{\rm LD}(r)= \phi_{\rm Star} (r) \frac{r_*}{2f(r_*/r)} 
\frac{\int_0^{r_*}   I(\rho/r_*) \frac{\rm d}{{\rm d}\rho}
\left[\rho f(\rho/r) \right]
{\rm d}\rho}{\int_0^{r_*} I(\rho/r_*)  \rho \, {\rm d}\rho},
\label{eq_phi_LD_gen}
\end{equation}
where $\phi_{\rm Star}(r)= C f(r_*/r)/(\pi r*)$ is the normalized flux that would be observed without limb darkening (Eqs.~\ref{eq_phi_r_gt_r*} and \ref{eq_phi_r_lt_r*}). Using the variable $u=\rho/r_*$ and integrating by parts the numerator of Eq.~\ref{eq_phi_LD_gen}, we obtain
\begin{equation}
\phi_{\rm LD}(r)= \frac{\phi_{\rm Star}(r)}{2\int_0^1 I(u) u \, {\rm d}u} 
\left[ I(1) - \frac{\int_0^1 f(r_*u/r) I'(u) u \, {\rm d} u}{f(r_*/r)} \right],
\label{eq_phi_LD_reduced}
\end{equation}
where $I'= {\rm d}I/{\rm d}u$. For a uniform stellar disk ($I'(u) \equiv 0$), we retrieve the expected result $\Phi_{\rm LD}(r)= \phi_{\rm Star}(r)$. 

In the limiting case $r \gg r_*$, we have $f(\rho/r)=G(\rho/r) \approx \pi \rho/(4r)$. Using this approximation in Eq.~\ref{eq_phi_LD_gen}, we obtain $\Phi_{\rm LD}(r) \approx \phi_{\rm Star}(r)$. Thus, the limb darkening has no effect on the flash profile for $r \gg r_*$. This is expected since in this case the surface elements of the stellar disk are uniformly distorted by the atmosphere.

The peak value of the flash is obtained by making $r=0$ in Eq.~\ref{eq_phi_LD_gen}. Then $f(\rho/r)=E(r/\rho)=E(0)=\pi/2$. Injecting this value in Eq.~\ref{eq_phi_LD_gen}, we get
\begin{equation}
\phi_{\rm LD}(0)= \phi_{\rm Star}(0) \frac{\int_0^1 I(u) \, {\rm d}u}{2\int_0^1 I(u)u \, {\rm d}u}.
\label{eq_phi_max_with_LD}
\end{equation}
Elementary calculations show that
\begin{equation}
\begin{array}{ll}
\int_0^1 I(u) \, {\rm d}u = & 2\int_0^1 I(u)u \, {\rm d}u + \\ \\
                                     & \int_0^{1/2} [I(u)-I(1-u)](1-2u) \, {\rm d}u. \\
\end{array}
\end{equation}
In the last integral, we have $1-2u \geq 0$. Moreover, the limb darkening causes a monotonic decrease of 
$I(u)$ between 0 and 1, so that $I(u) \geq I(1-u)$ for $u \in [0,1/2]$. Consequently, $\int_0^1 I(u) \, {\rm d}u \geq 2\int_0^1 I(u)u \, {\rm d}u$ and $\phi_{\rm LD}(0) \geq \phi_{\rm Star}(0)$. In other words, for a given stellar radius $r_*$, the limb darkening increases the height of the flash compared to the case without limb darkening. This is expected since the limb darkening causes a concentration of light near the star center, reducing its effective radius.

The various integrals appearing in Eqs.~\ref{eq_phi_LD_reduced}-\ref{eq_phi_max_with_LD} can be evaluated  by using an expression of the form \citep{Claret2000,Claret2011}
\begin{equation}
I(u)= 1 - \sum_{k=1}^4 c_k  (1 - \mu^{k/2}), {\rm~where~} \mu = \sqrt{1-u^2}.
\label{eq_limb_darkening}
\end{equation}
This form introduces infinite slopes of $I(u)$ at $u=1$, and a singularity for $I'(u)$ at that point. In order to avoid the resulting numerical difficulties when evaluating the integrals in Eqs.~\ref{eq_phi_LD_reduced} and  \ref{eq_phi_max_with_LD}, we can use  the change of variable $v=(1-u)^{1/4}$, so that
\begin{equation}
\begin{array}{l}
\displaystyle 
\int_0^1 f \left(\frac{r_*}{r}u \right) I'(u) u \, {\rm d}u = \\ \\
\displaystyle
\sum_{k=1}^4 2k c_k \int_0^1  f \left[ \frac{r_*}{r} (1-v^4) \right] (1-v^4)^2 (2-v^4)^{\frac{k}{4}-1} v^{k-1} \, {\rm d}v.
\end{array}
\label{eq_int_f_I'_u_du} 
\end{equation}
Likewise, we can use the change of variable $v=(1-u)^{5/8}$, to obtain
\begin{equation}
\begin{array}{ll}
\displaystyle
\int_0^1 I(u) \, {\rm d}u= & \displaystyle \left( 1 - \sum_{k=1}^{4} c_k \right) + \\ \\
& \displaystyle \frac{8}{5}  \sum_{k=1}^{4} c_k \int_0^1 (2- v^{\frac{8}{5}})^{\frac{k}{4}} v^{\frac{2k+3}{5}} \, {\rm d}v,
\end{array}
\label{eq_int_I_du} 
\end{equation}
\begin{equation}
\begin{array}{ll}
\displaystyle
\int_0^1 I(u) u \, {\rm d}u= & \displaystyle \frac{1}{2} \left( 1 - \sum_{k=1}^{4} c_k \right) + \\ \\
\displaystyle
& \displaystyle \frac{8}{5}  \sum_{k=1}^{4} c_k \int_0^1 (2- v^{\frac{8}{5}})^{\frac{k}{4}} (1-v^{\frac{8}{5}}) v^{\frac{2k+3}{5}} \, {\rm d}v.
\end{array}
\label{eq_int_I_u_du} 
\end{equation}

\section{Summary of the central flash expressions}
\label{app_table_general}

The Table~\ref{tab_CF_formulae} summarizes the various expressions of the central flashes obtained in this paper, splitting the case of a monochromatic wave caused by a point-like star, and the case of a star with finite apparent size in the geometrical optics approximation.
\begin{table*}[!b]
\caption{Parameters of the central flashes in the immediate vicinity of the shadow center. }
\label{tab_CF_formulae}
\renewcommand{\arraystretch}{1} 
 \begin{tabular}{lll}
 \hline
 \hline
\multicolumn{3}{c}{Case with diffraction by an occulter with dense atmosphere, point-like star and monochromatic wave} \\
 \hline
 & General case & Isothermal atmosphere, \\
 &                        & near centrality ($r \ll R_{\rm CF}$) \\
 \hline
 \\
 Radial profile, $\displaystyle \phi_{\rm Diff} \left(r \lesssim \frac{\lambda_{\rm F}^2}{R_{\rm CF}} \right)$ & 
 $\displaystyle 2\pi^2 \left( \frac{R_{\rm CF}}{\lambda_{\rm F}} \right)^2 J_0^2 \left( \frac{\pi R_{\rm CF} r}{\lambda_{\rm F}^2} \right)  \phi_\perp(0)$ &
 $\displaystyle
 2\pi^2 \left( \frac{R_{\rm CF} H}{\lambda_{\rm F}^2} \right)  J_0^2 \left( \frac{\pi R_{\rm CF} r}{\lambda_{\rm F}^2} \right)$ \\ 
 \\
 Peak value,  $\phi_{\rm Diff}(0)$ & 
 $\displaystyle 2\pi^2 \left( \frac{R_{\rm CF}}{\lambda_{\rm F}} \right)^2 \phi_\perp(0)$ &
 $\displaystyle 2\pi^2 \left( \frac{R_{\rm CF} H}{\lambda_{\rm F}^2} \right)$ \\ \\
 Diameter of first dark fringe & $\displaystyle 1.53 \left( \frac{\lambda_{\rm F}^2}{R_{\rm CF}} \right)$ & $\displaystyle 1.53 \left( \frac{\lambda_{\rm F}^2}{R_{\rm CF}} \right)$ \\
 \\
 Asymptotic form $\displaystyle \phi_{\rm Diff} \left(r \gtrsim \frac{\lambda_{\rm F}^2}{R_{\rm CF}} \right)$ &  
 $\displaystyle \phi_1 + \phi_2 + 2\sqrt{\phi_1 \phi_2} \sin \left( \varphi_2 -  \varphi_1 \right)$ &
 $\displaystyle \frac{2H}{r} + \frac{2H}{r} \sin \left( \frac{2\pi R_{\rm CF} r}{\lambda_{\rm F}^2} \right)$ \\
 \\
 Poisson fringe spacing, $\lambda_{\rm P}$ & $\displaystyle \frac{\lambda_{\rm F}^2}{R_{\rm CF}}$ &  $\displaystyle \frac{\lambda_{\rm F}^2}{R_{\rm CF}}$ \\ 
 \\
 Flash layer thickness, $\Delta R_{\rm Diff}$ & 
 $\displaystyle  \sim \lambda_{\rm F} \sqrt{\phi_\perp(0)}$ & 
 $\displaystyle \sim \lambda_{\rm F} \sqrt{\frac{H}{R_{\rm CF}}}$ \\ \\
 \hline
 \multicolumn{3}{c}{Case with finite stellar diameter in the geometrical optics approximation and uniform stellar disk} \\
\hline
 \\
 Radial profile, $ \phi_{\rm Star}(r \geq r_*)$ &  
 $\displaystyle \left( \frac{8R}{\pi r_*}\right) G\left(\frac{r_*}{r}\right) \phi_\perp(0)$ & 
 $\displaystyle \left( \frac{8H}{\pi r_*}\right) G\left(\frac{r_*}{r}\right)$ \\ 
 \\
 Radial profile, $\phi_{\rm Star}(r \leq r_*)$ & $\displaystyle
  \left( \frac{8R}{\pi r_*}\right) E\left(\frac{r}{r_*}\right) \phi_\perp(0)$ & 
 $\displaystyle \left( \frac{8H}{\pi r_*}\right) E\left(\frac{r}{r_*}\right)$ \\
  \\
 Flash width, ${\rm FWHM}(\phi_{\rm Star})$ & $1.14D_*$ & $1.14D_*$ \\
 \\
 \hline
\end{tabular}
\tablefoot{See the definitions in Table~\ref{tab_definitions}. The effects due to limb darkening are described in Appendix~\ref{app_limb_dark}}
\end{table*}

\end{appendix}

\end{document}